\documentclass[
twocolumn,
preprintnumbers,superscriptaddress,showpacs,supberscriptaddress,
nofootinbib,longbibliography]{revtex4-2}

\usepackage[%dvipdfm, 
bookmarks=true,colorlinks,linkcolor=blue,urlcolor=cyan,citecolor=red]{hyperref}

\usepackage{graphicx}
\usepackage{latexsym}
\usepackage{amsfonts}
\usepackage{amssymb}
\usepackage{amsmath}
\usepackage{bm}
\usepackage{mathrsfs}
\usepackage{slashed}
\usepackage{ascmac}
\usepackage{comment}

\usepackage{dcolumn}% Align table columns on decimal point

\allowdisplaybreaks[3] 

\usepackage[utf8]{inputenc} %to your latex preamble

\newcommand{\cout}[1]{ \if 0 {#1} \fi }
\newcommand{\nn}{\nonumber}
\renewcommand{\=}{&=&}
\newcommand{\nnb}{\nonumber \\}
\newcommand{\pd}{\partial}

\renewcommand{\r}{\right}

\renewcommand{\a}{\alpha}
\renewcommand{\b}{\beta}

\newcommand{\m}{\mu}
\newcommand{\n}{\nu}
\renewcommand{\r}{\rho}
\newcommand{\s}{\sigma}

\newcommand{\Gam}{ \Gamma }
\newcommand{\gam}{ \gamma }

\newcommand{\bu}{{\bm u}}

\newcommand{\order}{{\mathcal O}}

\newcommand{\para}{ \parallel}

\newcommand{\zero}{{(0)}}
\newcommand{\one}{{(1)}}

\usepackage{color} % For blue in-text comments and additions
\usepackage[normalem]{ulem}  % \sout{old text} for strikeout
\newcommand{\com}[1]{[{\color[rgb]{0,0,1}{#1}}]}

\newcommand{\gray}[1]{{\color[rgb]{0.7,0.7,0.7}{#1}}}

\renewcommand{\P}{\tilde {\mathcal F}}

\begin{document}

%\title{Complete solutions for the linear waves in first-order spin magnetohydrodynamics
%\\

\title{
Anisotropic linear waves and breakdown of the momentum expansion 
\\
in spin magnetohydrodynamics
}

\author{Zhe Fang}
\email{12345071@zju.edu.cn}
\affiliation{Zhejiang Institute of Modern Physics, Department of Physics, Zhejiang University, Hangzhou, Zhejiang 310027, China}

\author{Koichi Hattori}
\email{koichi.hattori@zju.edu.cn}
\affiliation{Zhejiang Institute of Modern Physics, Department of Physics, Zhejiang University, Hangzhou, Zhejiang 310027, China}
\affiliation{Research Center for Nuclear Physics, Osaka University, 
10-1 Mihogaoka, Ibaraki, Osaka 567-0047, Japan}

\author{Jin Hu}
\email{hu-j23@fzu.edu.cn}
\affiliation{Department of Physics, Fuzhou University, Fujian 350116, China}

%\preprint{YITP-xxx}
\date{\today}

\begin{abstract}
We formulate spin magnetohydrodynamics (MHD) by including the magnetic-flux and total angular momentum conservation laws in the hydrodynamic framework.  
To specify the local angular momentum conservation, 
we choose the totally antisymmetric spin current. 
The entropy-current analysis allows for ten dissipative first-order transport coefficients including anisotropic spin relaxation rates and the conversion rate between a vorticity (shear) to a symmetric stress (antisymmetric torque), {as well as anisotropic viscosities and resistivities}. By employing the linear-mode analysis, we solve the first-order spin MHD equations to determine the dispersion relations with the complete information of anisotropy retained. 
%Our analysis indicates the existence of six modes, two Alfven-wave modes and four magneto-sonic modes in the MHD sector and three damping modes in the spin sector. 
Our analytic solutions indicate that the small-momentum expansion is spoiled by blow up of the higher-order terms when the angle between the momentum and the magnetic field approaches the right angle. 
This also reveals the existence of another expansion parameter, and, in light of it, we provide solutions in an alternative series expression beyond the critical angle.  
We confirm that these two series expansions work well 
in the appropriate angle ranges  
as compared with numerical results. 
{Building on our findings regarding the breakdown of the small-momentum expansion in first-order theory, we proceed to discussing how these first-order solutions are modified when we include the relaxation dynamics for dissipative modes with the Israel-Stewart framework. 
We find that, due to the presence of the critical behavior in the first-order solutions, there remains a diffusive window even after the relaxation dynamics is introduced. 
}
%leading to a deviation between our general solution and the numerical results obtained from our equations. We explain the deviation and also find a valid expansion at this range.
%\gray{We formulate spin magnetohydrodynamics (MHD) based on the magnetic-flux conservation and the totally antisymmetric spin current and solve the obtained hydrodynamic equations within the linear-mode analysis. 
%The entropy-current analysis allows for 
%ten dissipative transport coefficients in the first-order constitutive equations. 
%We find analytic solutions for an arbitrary angle of the wave vector inclined from the direction of a magnetic field with all the first-order dissipative corrections, which have been missing for long in the literature even without the spin degrees of freedom. 
%The solutions are composed of the five and four coupled modes in spin MHD and the four and two coupled modes with seven dissipative transport coefficients in MHD without spin. We emphasize that, at the right angle to the magnetic-field direction, 
%the naive solutions, which are conventionally obtained from the MHD equations simplified in this limit of the angle in the literature, do not correctly capture the mixing among the coupled modes, and do not agree with the limit taken in the general analytic solutions because of the non-commutative nature between the limit of the angle and the diagonalization. }
\end{abstract}

\maketitle
%\tableofcontents

\section{Introduction}\label{s1}

In the realm of nuclear physics, the study of spin polarization in relativistic nucleus-nucleus collisions has unveiled a fascinating exploration domain. 
High-energy collisions generate vortical fields within the quark-gluon plasma (QGP) through the initial orbital angular momentum, leading to the global spin polarization of $\Lambda$ and $\bar \Lambda$ hyperons and the alignment of vector mesons via spin-orbit couplings \cite{Liang:2004ph,Liang:2004xn,Betz:2007kg,Becattini:2007sr,Gao:2007bc,Becattini:2016gvu}.
The polarization of hyperons was observed in experiments by STAR at RHIC and ALICE at the LHC \cite{STAR:2017ckg, STAR:2018gyt, mohanty_spin_2021,STAR:2019erd,STAR:2020xbm}, while the alignment of vector mesons was also observed in recent years \cite{ALICE:2022dyy,Micheletti:2023qlh}, representing a pivotal moment in understanding the QCD matter under extreme conditions.

Theoretical investigations have also been devoted to developing relativistic frameworks for spin dynamics in QGP.  Early theoretical projections suggested that the spin polarization of particles would be affected by the thermal vorticity tensor at the freeze-out surface, intricately associated with the fluid's angular velocity and acceleration.
Various theoretical models are adopted to include the thermal vorticity and have successfully described the global polarization using the modified Cooper-Frye formula \cite{Becattini:2007sr,Fang:2016vpj,Karpenko:2016jyx,Shi:2017wpk,Fu:2020oxj,Liu:2021nyg}. 
Nevertheless, the experimental results on the local $\Lambda$  spin polarization pose a spin sign puzzle that cannot be solely explained by thermal effects \cite{Becattini:2021iol, Wu:2019eyi,Becattini:2024uha}, i.e., theoretical calculations give the opposite azimuthal angle dependence of $ \Lambda $ polarization to the experimental results.
 This deviation may be attributed to the non-equilibrium evolution of spin degrees of freedom, inspiring theoretical research on hydrodynamic descriptions that include spin effects. 
 Substantial theoretical studies have been conducted on the relativistic spin hydrodynamics based on the entropy analysis or the second law of thermodynamics  \cite{Hattori:2019lfp,Fukushima:2020ucl,Gallegos:2021bzp,Li:2020eon,Hu:2021lnx,Hu:2022azy,Singh:2022ltu,Cao:2022aku,Daher:2022wzf,Sarwar:2022yzs,Kiamari:2023fbe,Xie:2023gbo,Ren:2024pur}, quantum kinetic theories for fermions \cite{Florkowski:2017ruc,Weickgenannt:2020aaf,Yang:2020hri, Bhadury:2020cop,Wang:2020pej,Sheng:2021kfc,Muller:2021hpe, Kumar:2022ylt, Kumar:2023ghs,Peng:2021ago,Hu:2021pwh,Hu:2022lpi,Hu:2022xjn,Hongo:2022izs,Hidaka:2023oze,Weickgenannt:2022zxs,Weickgenannt:2023btk,Bhadury:2024ckc,Fang:2024vds, Lin:2024cxo},  
 holographic approaches \cite{Garbiso:2020puw,Gallegos:2020otk,Hashimoto:2013bna}, effective action approach \cite{Montenegro:2017rbu,Montenegro:2020paq}, and the statistical density operator methods \cite{Becattini:2009wh,Becattini:2012pp,Becattini:2018duy,Hongo:2021ona, Hu:2021lnx,Hu:2022azy,Tiwari:2024trl}.  
{
Different from hydrodynamic modes, spin density is not a strictly conserved charge since the spatial rotational symmetry only guarantees the conservation of the total angular momentum including the orbital part. 
Therefore, there is not a {\it a priori} justification of spin hydrodynamics as a (quasi-)hydrodynamics, which should be instead justified {\it posteriori} by evaluating the spin relaxation rate. 
%is small enough compared to the momentum relaxation rate.  
Evaluations of the spin relaxation rate have been addressed recently in Refs.~\cite{Hongo:2021ona,Hongo:2022izs,Hidaka:2023oze,Wagner:2024fry,Wagner:2024fhf} for relativistic heavy-ion collisions. 
}
% The spin relaxation rate was also evaluated with the Kubo formula \cite{Hongo:2022izs}. 
% Note that in the framework of spin hydrodynamics, the conversion between orbital angular momentum and spin angular momentum needs to be considered, and therefore spin density is treated as a non-conserved quantity, whose evolution draws extensive research interests and can be described in a hydrodynamic way. This offers an intriguing pathway for investigating the thermodynamic behavior in extreme collision scenarios. In addition to hydrodynamic framework that we are focusing on, quantum kinetic theory is also viewed as good candidate to describe spin dynamic evolution, see the related developments in this aspect \cite{Weickgenannt:2020aaf,Yang:2020hri,Wang:2020pej,Sheng:2021kfc,Fang:2024vds}. 
Also, the shear-induced polarization \cite{Liu:2021uhn} has been proposed as a source triggering the local spin polarization. Including them in hydrodynamic simulations seems to qualitatively resolve the sign issue \cite{Becattini:2021suc,Fu:2021pok} (see also Refs.~\cite{Lin:2022tma,Lin:2024zik}).

Besides the experimental observation of the spin polarization, strong magnetic fields are expected to be generated in the non-central collisions (see, e.g., Refs.~\cite{Kharzeev:2015znc, Skokov:2016yrj, Hattori:2016emy,Hernandez:2017mch} for reviews and references therein). 
%\cite{adam_global_2018, yan_dynamical_2021}. 
These magnetic fields are thought to be transient as they will fade out rapidly in the heavy-ion collision, but may result in some important phenomena like chiral magnetic effect \cite{Fukushima:2008xe, Kharzeev:1998kz, Kharzeev:2004ey, Kharzeev:2007jp} and chiral magnetic wave \cite{Kharzeev:2010gd}. 
{Recently, there have been developments in the numerical simulation of relativistic magneto-hydrodynamics(MHD) \cite{Nakamura:2022idq,Nakamura:2022wqr,Nakamura:2022ssn,Mayer:2024dze,Mayer:2024kkv}.
}
Furthermore, the strong magnetic field may also have a significant impact on the spin polarization. Unlike rotational effects, a magnetic field induces opposite spin polarization of particles and antiparticles as they carry opposite charges. 
{Though there is still no experimental certainty, the deviations in the polarization of  $\Lambda$ and $\bar \Lambda$ in heavy-ion collisions have drawn a lot of attention recently \cite{STAR:2018gyt}, which may be accounted by the effect of the magnetic field.}
%This could account for the observed deviations in the polarization of  $\Lambda$ and $\bar \Lambda$ in heavy-ion collisions \cite{STAR:2018gyt}. 
{Recently, magnetic-field effects were discussed as sources of not only the splitting between $\Lambda$ and $\bar \Lambda$ \cite{Xu:2022hql, Peng:2022cya,Buzzegoli:2022qrr} but also the local spin polarization \cite{Sun:2024isb}. 
}
The interplay between the magnetic fields and spin dynamics in the QGP draws considerable interest.

In this study, we explore formulation and linear-mode analysis of the relativistic spin MHD, extending our study on the first-order MHD \cite{Fang:2024skm}. 
%\com{Do not abbreviate verbs in professional writing.} \gray{and it is promising to add also the spin into the simulation.} 
%\com{In what sense, is it promising?}
We plug spin dynamics into relativistic MHD that has been formulated in recent years {with anisotropic viscosities and resistivities} {
\cite{Grozdanov:2016tdf,Pu:2016ayh,Hattori:2017usa,Hernandez:2017mch, Denicol:2018rbw,Denicol:2019iyh,Hongo:2021ona,Armas:2022wvb,Panda:2021pvq} (see Ref.~\cite{Hattori:2022hyo} for a review). 
}
%\gray{based on the Maxwell equations \cite{Grozdanov:2016tdf,Hattori:2017usa,Hongo:2021ona,Armas:2022wvb,Hernandez:2017mch,Pu:2016ayh} and Boltzmann equations \cite{Panda:2021pvq,Denicol:2018rbw} in recent years (see Ref.~\cite{Hattori:2022hyo} for a review). }
%\com{This sorting is not precise. Here, we may not need to classify the methods to avoid complication.}
We will find anisotropic rotational viscosities and a new transport coefficient that converts the vorticity to a symmetric stress and the shear to an antisymmetric torque. 
We also solve the system of hydrodynamic equations for the linear perturbation in the vicinity of an equilibrium configuration. 
Note that, while the linear-mode analyses have been applied to MHD  \cite{Grozdanov:2016tdf,Biswas:2020rps,Armas:2022wvb,Fang:2024skm} and spin hydrodynamics  \cite{Hattori:2019lfp,Hongo:2021ona,Hu:2022azy,Xie:2023gbo,Daher:2022wzf,Sarwar:2022yzs}, the linear waves in the spin MHD has not been addressed, which is exactly what we will present in this work. 
We also note that spin MHD was recently discussed with the kinetic theory approach \cite{Bhadury:2022ulr,Bhadury:2024whs}.
{
In this paper, we also introduce relaxation dynamics with the Israel-Stewart (IS) framework. 
This offers an application of the IS framework to anisotropic systems. 
We will find that there remains a diffusive window, of which the boundary is specified by a direction of momentum as well as its magnitude. 
}

\cout{
\gray{Furthermore, the influence of other factors such as vortical \cite{Kiamari:2023fbe} and electric fields \cite{Singh:2022ltu} on the system is also being considered in some recent works.}\com{delete ? $\to$ Combined in the spin hydro in the above.}
}

Notably, when solving the linearized equations, we retain the full anisotropy of the solutions parametrized by the general angle $\theta$ between the momentum and magnetic field, which is one of the important properties inherent in a magnetized medium. Relying on the anisotropic analytical solutions, we find that the momentum expansion breaks down when $\theta$ approaches $\pi/2$, because diverging powers of $1/\cos\theta$ emerge in the higher-order terms in the momentum expansion. 
%This concerns with the mathematical issue of the sequence of taking various limits. 
This issue stems from non-commutative limits of small values of the momentum and the cosine. 
In light of this, we provide an alternative solution that covers the regime where the momentum expansion breaks down. 
We confirm these two series expressions work well in the appropriate regimes by comparing them with the numerical results. Our study may have broader implications for solving the perturbation problems involving multiple small parameters.

In Sec.\ref{s2}, we provide the derivation details of the hydrodynamic equations for spin MHD. In Sec.\ref{s3}, we focus on the linear-mode analysis near an equilibrium state. 
We elaborate on the solutions for the linear waves in Sec.\ref{s4}. In this part, a pair of the Alfven waves and two pairs of the magneto-sonic waves are found in the  MHD sector, while three damping modes are found in the spin sector. 
{We provide a remedy for the breakdown of the momentum expansion. }
A comparison with our previous work on MHD \cite{Fang:2024skm} is also given. We give a summary in Sec.\ref{s5}.

%A disagreement arise when we compare the limit of solution at a specific angle when the wave vector is perpendicular to the magnetic field. We explain it stems from the invalidity of $k$ expansion at this point, and we offer the other approach to achieve the appropriate expansion. 

In this paper, we use the mostly plus metric convention ${\eta ^{\mu \nu }} = \text{diag}( - 1,1,1,1)$ and the completely antisymmetric tensor with the convention $\epsilon^{0123} =  + 1$. 
Then, the fluid velocity ${u^\mu }$ is normalized 
as ${u^\mu }{u_\mu } = - 1$. 
We define the projection operator ${\Delta ^{\mu \nu }} = {\eta ^{\mu \nu }} + {u^\mu }{u^\nu }$ such that $u_\mu \Delta^{\mu\nu}=0 $.

\section{Formulation of spin magnetohydrodynamics}\label{s2}

\subsection{Conservation laws}
Our investigation commences by introducing a finite angular-momentum density into the framework of magnetohydrodynamics. We focus on neutral systems in the absence of vector or axial charges.   
The conservation laws deduced from the translational, 
rotational and one-form symmetries read 
\cite{Grozdanov:2016tdf,Hongo:2020qpv,Hattori:2022hyo}  
\begin{equation}
{\partial _\mu }{\Theta ^{\mu \nu }} = 0 , \quad 
{\partial _\mu }{J^{\mu \a\b }} = 0 , \quad 
{\partial _\mu }{{\tilde F}^{\mu \nu }} = 0 , 
\label{con4}
\end{equation}
where $ \Theta ^{\mu \nu }$, $ J^{\mu \a\b }$, and ${\tilde F}^{\mu \nu } $ are the energy-momentum tensor, 
the angular momentum tensor, and 
the dual electromagnetic field strength tensor, respectively.

The angular momentum tensor $J^{\mu\a\b} $ can be decomposed 
into the contributions of {the spin current} $\Sigma^{\mu\a\b} $ 
and the orbital angular momentum as 
\begin{eqnarray}
J^{\mu\a\b} = \Sigma^{\mu\a\b}
+ x^\a \Theta^{\mu\b} - x^\b \Theta^{\mu\a}
\, .
\end{eqnarray}
Then, the total angular momentum conservation in Eq.~(\ref{con4}) is cast into another form 
\begin{eqnarray}
\label{eq:spin-cons}
\partial _\mu \Sigma^{\mu \a\b } =  - 2 \Theta^{[\a\b]}
\, .
\end{eqnarray}
This equation indicates that there is no symmetry that guarantees the conservation of spin in relativistic systems. 
The antisymmetric part of the energy-momentum tensor 
quantifies the torque exerting on a fluid cell 
that possesses a nonzero average spin. 
Since the spin density is not a conserved quantity, 
spin hydrodynamics should be understood as an extended 
hydrodynamic framework with a quasihydrodynamic mode 
as previously formulated in the literature \cite{Hattori:2019lfp,Fukushima:2020ucl,Hongo:2021ona,Cao:2022aku}. 
The temporal components of these currents 
provide the density of corresponding conserved charges 
\begin{eqnarray}
\label{eq:charges}
e =  u_\mu u_\nu {\Theta ^{\mu \nu }} , \quad 
S^{\a\b} = - u_\mu \Sigma^{\mu \a\b } , \quad 
B^\mu  = {-}  {\tilde F}^{\mu \nu } u_\nu . 
\end{eqnarray}
Note that there is an ambiguity of the pseudo-gauge choice 
where the local conservation laws of 
the energy-momentum tensor 
and the angular momentum tensor are preserved 
under the simultaneous shifts of ${\Theta ^{\mu \nu }} $ 
and $J^{\mu\a\b} $ transformations~\cite{Belinfante1939,Belinfante1940,Rosenfeld1940} 
(see, e.g., Refs.~\cite{Becattini:2018duy, Hattori:2019lfp, Li:2020eon, Fukushima:2020ucl} for recent discussions). 
Because of this ambiguity, the separation between 
the spin and orbital angular momentum is not unique 
and depends on a choice of the pseudo-gauge. 
In the end, the definitions of the spin and orbital angular momentum may obey the measurement processes. 
We assume that $\Sigma^{\mu\alpha\beta}$ is 
a totally antisymmetric tensor.  
In this pseudo-gauge, we find an identity 
$u_\mu S^{\mu\nu} = 0 $ from Eq.~(\ref{eq:charges}), 
indicating that 
the spin density only has three spatial components. 
Accordingly, three components of Eq.~(\ref{eq:spin-cons}) 
are not dynamical equations, but 
serve as constraint equations \cite{Cao:2022aku}
\begin{eqnarray}
\label{eq:const}
u_\a \Theta^{[\a\b]} 
= - \frac12 u_\a \pd_\mu \Sigma^{\mu\a\b} 
.
\end{eqnarray}
As a consequence, the temporal component of 
$\Theta^{[\a\b]}$ is determined 
by the spatial derivative of the spin density.

Next, the conserved quantities (\ref{eq:charges}) 
are supposed to satisfy the first law of thermodynamics:
\begin{subequations}
\label{con5}
\begin{eqnarray} 
Ts \= e + {p } - {\omega _{\mu \nu }}{S^{\mu \nu }} - {H_\mu }{B^\mu } , \\
Tds \= de - {\omega _{\mu \nu }}d{S^{\mu \nu }} - {H_\mu }d{B^\mu } ,\\
TDs \= De - {\omega _{\mu \nu }}D{S^{\mu \nu }} - {H_\mu }D{B^\mu },
\end{eqnarray}
\end{subequations}
The temporal derivative $D$ is defined as $u^\mu\partial_\mu$.  
We introduced the spin potential $ \omega_{\m\n}$ 
and the magnetic field $H_\mu $ as the Lagrange multipliers 
for the spin density $S^{\mu\nu} $ and the magnetic flux $B^\mu $, respectively. We assume the linear relations between the conserved charges 
and the Lagrange multipliers, i.e.,  
$H^\mu= B^\mu/\mu_m $ and $\omega^{\mu\nu}=S^{\mu\nu}/\chi$ 
with the magnetic permeability $ \mu_m$ 
and the spin susceptibility $\chi $ being constants in spacetime.

%To cast the conservation equations into a formulation that describe the relation between thermodynamic variables and conserved charges, we employ a derivative expansion to the conserved current. At zeroth order, the relevant tensors available can be utilized to construct the corresponding terms as

{To close the equations of motion, one needs to 
express the spatial components of the conserved currents 
by the conserved charges. 
This is implemented on the basis of a derivative expansion.
}
Up to the first-order in derivative, 
one can write the conserved currents as 
\begin{equation}
    \begin{array}{l}
{\Theta ^{\mu \nu }} = e{u^\mu }{u^\nu } + {p_\parallel }{b^\mu }{b^\nu } + {p_ \bot }{\Xi ^{\mu \nu }} 
%\textcolor{red}{+ u^{[\mu}q^{\nu]}}
+ \Theta _{(1)}^{\mu \nu }, \\
{\Sigma ^{\mu \alpha \beta }} = u^\mu S^{\alpha\beta} - u^\alpha S^{\mu\beta}+u^\beta S^{\mu\alpha}+ \Sigma _{(1)}^{\mu \alpha \beta } ,\\
{{\tilde F}^{\mu \nu }} = {B^\mu }{u^\nu }  - {B^\nu }{u^\mu } + \tilde F_{(1)}^{\mu \nu },
\end{array}\label{con6}
\end{equation}
where we introduced a unit vector ${b^\mu } = {B^\mu }/\sqrt {{B^\nu }{B_\nu }}$ and the projection operator ${\Xi ^{\mu \nu }} = {\Delta ^{\mu \nu }} - {b^\mu }{b^\nu }$ 
such that $b_\mu \Xi ^{\mu \nu }=0= u_\mu \Xi ^{\mu \nu }$. 
{
The unknown parameters $ p_{\para,\perp}$ are identified with 
thermodynamic pressure below. 
The above explicitly written tensor structures exhaust 
possible tensor structures respecting the discrete symmetries in the zeroth order in derivative. 
The last terms on the right-hand side of Eq.(\ref{con6}) with the subscript $(1)$ denote the first-order corrections 
that will be constrained by the entropy-current analysis below. 
}

\subsection{Entropy-current analysis}

{
The entropy-current analysis provides constraints on the possible first-order tensor structures 
in Eq.~(\ref{con6}). 
}
% the conserved charges and the viscosity parameters.
Up to the first order, the entropy current is given in the form $s^\mu = su^\mu + s_\one^\mu$, where $s_\one^\mu\sim \order(\partial)$. 
By making use of the first law of thermodynamics (\ref{con5})
and the equations of motion (\ref{con4}) 
with Eq.~(\ref{con6}), one finds that 
\begin{eqnarray}  \label{con9}
\pd_\mu s^\mu 
%\= \pd_\mu ( u^\mu s + s_\one^\mu) 
\= s \pd_\mu u^\mu + D s + \pd_\mu s_\one^\mu 
\\
\= \beta  ( Ts - \epsilon - p_\perp  
+ \omega_{\a\b}  S^{\alpha\beta} + H_\mu B^\mu ) \pd_\mu u^\mu
\nnb
&&- \beta \big[\, (p_\para - p_\perp)  b^\mu b^\nu 
+ B  b^\mu H^\nu 
+ 2  S^{\m\a} \omega^\n_{\ \, \a} \, \big] \pd_\mu u_\nu 
\nnb
&&- \Theta^{\mu\nu}_\one (\pd_\mu\beta _\nu - 2\beta \omega_{\mu\nu} )
\nnb
&&- \Sigma^{\mu\a\b}_\one  \pd_\mu (\beta \omega_{\a\b} )
+  \tilde F^{\mu\nu}_\one \pd_\mu (\beta H_\nu )
 \nnb
 &&+ \pd_\mu (s_\one^\mu + \beta u_\nu \Theta^{\mu\nu}_\one
 + \beta \omega_{\a\b} \Sigma^{\mu\a\b}_\one 
 - \beta H_\nu  \tilde F^{\mu\nu}_\one )
 \nn
 ,
\end{eqnarray} 
%\begin{eqnarray}
%\label{con9}
%{\partial _\mu }{s^\mu } 
%\= \beta  ( Ts - \epsilon - p_\perp  
%+ \omega_{\a\b}  S^{\alpha\beta} + H_\mu B^\mu ) \pd_\mu u^\mu
%\\
%&& 
%- \beta  \{ ({p_\parallel } - {p_ \bot })  {b^\nu }
%+  B   {H^\nu }  \}   b_\mu \partial ^\mu  u_\nu 
%\nnb
%&&
%- \Theta^{\mu\nu}_\one (\pd_\mu\beta _\nu-2\beta \omega_{\mu\nu} )
%\nnb
%&&
%- \Sigma^{\mu\a\b}_\one  \pd_\mu (\beta \omega_{\a\b} )
%+  \tilde F^{\mu\nu}_\one \pd_\mu (\beta H_\nu )
% \nnb
% &&
% + \pd_\mu (s_\one^\mu + \beta u_\nu \Theta^{\mu\nu}_\one
% + \beta \omega_{\a\b} \Sigma^{\mu\a\b}_\one 
% - \beta H_\nu  \tilde F^{\mu\nu}_\one )
%. \nn
%\end{eqnarray}
%
%\begin{eqnarray}
%{\partial _\mu }{s^\mu } \= \partial_\mu(s u^\mu-\Theta_{(1)}^{\mu\nu}\beta_\nu
%{ -  \beta {H_\nu } \tilde F_{(1)}^{\mu \nu } })
%\nnb
%&&
%%- \beta {H_\mu }({u^\mu }{\partial _\nu }{B^\nu } + {B^\nu }{\partial _\nu }{u^\mu } + {\partial _\mu }\tilde F_{(1)}^{\mu \nu })
%{+  u_\nu  b_\mu \{
%\beta ({p_\parallel } - {p_ \bot }) {\partial^\mu }{b^\nu }
%+  B \partial ^\mu ( \beta {H^\nu }) \} }
%\gray{- \beta {H_\mu }({u^\mu }{\partial _\nu }{B^\nu } + {B^\nu }{\partial _\nu }{u^\mu }  )} \}
%\nnb 
%&&
%%\textcolor{red}{- u^{[\mu}q^{\nu]}(\partial_\mu \beta_\nu-2\beta\omega_{\mu\nu})} 
%- \Theta _{(1)}^{\mu \nu } 
%({\partial _\mu }{\beta _\nu } - 2\beta {\omega _{\mu \nu }})
%{ +  \tilde F_{(1)}^{\mu \nu } \pd_\mu (\beta {H_\nu }) }
%.
%\label{con9}
%\end{eqnarray}
where we used $ u_\mu \omega^{\mu\nu} = 0 = u_\mu S^{\mu\nu} $ 
in the totally antisymmetric pseudo-gauge mentioned above. 
In a {local equilibrium state} where $\Theta^{\mu\nu}_\one (\pd_\mu\beta _\nu - 2\beta \omega_{\mu\nu} ) \to 0 $, 
the spin potential is determined by the thermal vorticity 
$ \beta \omega_{\mu\nu} = \frac12 \pd_{[\mu}\beta _{\nu]}$. 
Therefore, in a weak vorticity $\pd_{[\mu}\beta _{\nu]} \sim \order(\pd^1)$, the magnitude of the spin potential is 
$ \beta \omega_{\mu\nu} \sim \pd_{[\mu}\beta _{\nu]} 
\sim \order(\pd^1)$. 
Accordingly, we drop the term proportional to $\pd_\mu \Sigma^{\mu\a\b}_\one $ which is a third-order correction. 
When $ S^{\mu\nu} \sim \omega^{\mu\nu}$, 
the term proportional to $ S^{\m\a} \omega^\n_{\ \, \a} $ 
is also a third-order correction and can be dropped; 
Otherwise, this term provides a pressure anisotropy 
induced by spin.

{At the leading-order in derivative, 
the entropy production should vanish. 
According to this condition, we find that}
\begin{subequations}
\label{con11}
\begin{eqnarray} 
&& Ts = \epsilon + p_\perp  
- \omega_{\a\b}  S^{\alpha\beta} - H_\mu B^\mu ,
\\
&&
({p_\parallel } - {p_ \bot }){b^\nu } + {H^\nu }B = 0
 .
\end{eqnarray}
\end{subequations}
{
One can identify $ p_\perp $ with the thermodynamic pressure 
in Eq.~(\ref{con5}) according to the first condition. 
The second condition shows that the pressure becomes anisotropic 
in the presence of a magnetic field. 
}

For the first-order corrections to be semi-positive definite, 
we require that 
individual contributions of distinct thermodynamic forces 
take semi-positive values, i.e., 
\begin{subequations}
\label{eq:first-order}
\begin{eqnarray} 
&& 
- \Theta _{(1)}^{\mu \nu } ({\partial _\mu }{\beta _\nu } - 2\beta {\omega _{\mu \nu }}) \geq 0,
\\
&& 
  \tilde F_{(1)}^{\mu \nu } \pd_\mu (\beta {H_\nu }) \geq 0 
  .
\end{eqnarray}
\end{subequations}
To satisfy the inequality, the energy-momentum tensor should be decomposed as 
\begin{eqnarray} 
\Theta_{(1)}^{\m\n} \= 
 2h^{(\mu} u^{\nu)} + 2 q^{[\mu} u^{\nu]} 
\label{con12}\\
&&
- T  \left(
\begin{array}{l}
1
\\
1
\\
\end{array}
\right)\left(
\begin{array}{ll}
\eta^{\m\n\r\s} & \xi^{\m\n\r\s}
\\
\xi^{\prime \m\n\r\s} & \gam^{\m\n\r\s}
%2 q^{[\mu} u^{\nu]} + \phi^{\mu\nu}
\end{array}
\right)
 \left(
\begin{array}{l}
\pd_{(\r} \beta_{\s)}  
\\
\pd_{[\r} \beta_{\s]} - 2\beta  \omega_{\r\s}
\end{array}
\right) 
\nn
.
\end{eqnarray}
{The temporal components with $u^\mu $} are the conventional heat current $h^\mu $ and the boost heat current $q^\mu $ introduced in Ref.~\cite{Hattori:2019lfp}. 
We choose to work with the Landau frame, so that 
the heat current is vanishing $h^{\mu} =0 $. 
Also, plugging the expansion (\ref{con12}) 
into the constraint equation (\ref{eq:const}), 
one finds that the boost heat current is completely 
determined by the spin density as 
\begin{eqnarray}
q^\mu 
%\= u_\a \Theta^{[\a\b]} _\one
= - \frac12 u_\a \pd_\mu \Sigma^{\mu\a\b} _\zero
= - \frac12  \Delta_\rho^\b \Delta_\a^\gam \pd_\gam S^{\alpha\rho}  
%- \frac12 u_\a \pd_\mu \Sigma _{(1)}^{\mu \alpha \beta } 
.
\end{eqnarray}
In our power counting $ S^{\mu\nu} \sim \order(\pd^1)$, 
the boost heat current turns out to be 
the second-order quantity $q^\mu \sim \order(\pd^2) $. 
Therefore, it can be dropped 
in the current working accuracy. 
Under those conditions, one can take 
\begin{eqnarray}
h^\mu = 0 =q^\mu .
\end{eqnarray}

In Eq.~(\ref{con12}), the viscous tensors are presented in the matrix form. 
While the symmetric component $\eta^{\m\n\r\s} $ provides 
the viscosities in MHD \cite{Grozdanov:2016tdf,Hongo:2020qpv,Hattori:2022hyo}, 
the antisymmetric part $\gam^{\m\n\r\s} $ provides a torque when there is a deviation between rotations of a fluid cell and the surrounding fluid \cite{Hattori:2019lfp}. 
The existence of the cross terms $ \xi^{\m\n\r\s}$ and $\xi^{\prime\m\n\r\s} $ was pointed out in Ref.~\cite{Cao:2022aku}. 
Onsager's reciprocal relation states that 
$ \eta^{\m\n\r\s}(b^\mu) = \eta^{\r\s\m\n} (-b^\mu) $, 
$ \gam^{\m\n\r\s}(b^\mu) = \gam^{\prime\r\s\m\n} (-b^\mu) $, 
$ \xi^{\m\n\r\s}(b^\mu) = \xi^{\r\s\m\n} (-b^\mu) $, 
and $ \xi^{\m\n\r\s}(b^\mu) = \xi^{\prime\r\s\m\n} (-b^\mu) $. 
These vectors and tensors are all transverse to $ u_\mu$.

The general structures of the viscous tensors are constructed as 
\begin{subequations}
\label{con16}
\begin{eqnarray}  
{\eta ^{\mu \nu \rho \sigma }} 
%\= {\zeta _ \bot }{\Xi ^{\mu \nu }}{\Xi ^{\rho \sigma }} + {\zeta _\parallel }{b^\mu }{b^\nu }{b^\rho }{b^\sigma }
%+ {\zeta _ \times }({b^\mu }{b^\nu }{\Xi ^{\rho \sigma }} + {\Xi ^{\mu \nu }}{b^\rho }{b^\sigma })
%\nnb
\= 
\begin{pmatrix}
b^\mu b^\nu & \Xi^{\mu\nu}
\end{pmatrix}
\begin{pmatrix}
\zeta_\para & \zeta_\times \\
\zeta'_\times & \zeta_\perp
\end{pmatrix}
\begin{pmatrix}
b^\rho b^\sigma \\ \Xi^{\rho\sigma}
\end{pmatrix}
\nnb
&&
+ 2{\eta _\parallel }({b^\mu }{\Xi ^{\nu (\rho }}{b^{\sigma )}} + {b^\nu }{\Xi ^{\mu (\rho }}{b^{\sigma )}})
\\
&&
 + {\eta _ \bot }({\Xi ^{\mu \rho }}{\Xi ^{\nu \sigma }} + {\Xi ^{\mu \sigma }}{\Xi ^{\nu \rho }} - {\Xi ^{\mu \nu }}{\Xi ^{\rho \sigma }}) 
 ,
\nnb 
{\gamma ^{\mu \nu \rho \sigma }} 
\= {\gamma _ \bot }({\Xi ^{\mu \rho }}{\Xi ^{\nu \sigma }} - {\Xi ^{\mu \sigma }}{\Xi ^{\nu \rho }}) 
\nnb
&&
- 2{\gamma _\parallel }({b^\mu }{\Xi ^{\nu [\rho }}{b^{\sigma ]}} - {b^\nu }{\Xi ^{\mu [\rho }}{b^{\sigma ]}})
,
\\
{\xi ^{\mu \nu \rho \sigma }} 
\= 2{\xi _\parallel }({b^\mu }{\Xi ^{\nu [\rho }}{b^{\sigma ]}} + {b^\nu }{\Xi ^{\mu [\rho }}{b^{\sigma ]}} 
\nnb
&& \quad \quad 
+ {b^\mu }{\Xi ^{\nu (\rho }}{b^{\sigma )}} - {b^\nu }{\Xi ^{\mu (\rho }}{b^{\sigma )}}), 
\end{eqnarray}
\end{subequations}
where $\zeta_\times = \zeta_\times' $ by virtue of Onsager's reciprocal relation. 
These viscous tensors are analogs of those in the presence of 
a strong vorticity \cite{Cao:2022aku}. 
{
The transport coefficients, $\gam_\perp, \, \gam_\para , \, \xi_\para$, in the antisymmetric part control the spin relaxation rate in Eq.~(\ref{eq:spin-cons}).\footnote{
One can confirm this in the solutions for the linear-mode analysis in a later section [see, e.g., Eqs.~(\ref{quantic NLO solution-set1}) and (\ref{quartic NLO solution})].} 
Therefore, spin density persists near, {\it albeit not strictly in}, an equilibrium and the current framework works as a quasi-hydrodynamics when these transport coefficients are small enough. 
%Evaluation of the spin relaxation rate for relativistic-heavy-ion collisions have been addressed recently in Refs.~\cite{Hongo:2021ona,Hongo:2022izs,Hidaka:2023oze,Wagner:2024fry,Wagner:2024fhf}. 
}
{
The symmetric viscous tensor $\eta^{\mu\nu\rho\sigma}$ reduces to the familiar viscous tensor in the isotropic limit $\zeta_\para = \zeta_\perp =\zeta_\times$ and $\eta_\para = \eta_\perp$. 
The antisymmetric viscous tensor $\gamma^{\mu\nu\rho\sigma}$ also reduces to that without a magnetic field \cite{Hattori:2019lfp} in the isotropic limit $\gamma_\para = \gamma_\perp$. 
The cross term $\xi^{\mu\nu\rho\sigma}$ does not exist without $b^\mu$. 
Therefore, the most general tensor structure in Eq.~(14) works for an arbitrary magnitude of a magnetic field. 
}

%One clear difference stems from the charge-conjugation 
%properties of a magnetic field and a vorticity. 
%In the charge-neutral systems, there are no Hall terms, while Hall-like terms induced by the Coriolis force were found in Ref.~\cite{Cao:2022aku}. 

%The coefficients in (\ref{con16}) can be determined by finite-temperature field theory \cite{anchishkin_single-particle_2022}\cite{huovinen_chemical_2008}\cite{russkikh_dynamical_2007}. In this article, we will not discuss how to determine these coefficients. One can refer to a letter\cite{hattori_fate_2019} to understand their physical meaning.

As for the dual field strength tensor, 
the first-order correction should be proportional to 
the thermodynamic force as 
\begin{equation}
\tilde F_{(1)}^{\mu \nu } = - T{\rho ^{\mu \nu \rho \sigma }}{\partial_ {[\rho }}(\beta {H_{\sigma ]}})
. \label{don16}
\end{equation}
To ensure Eq.~(\ref{eq:first-order}) is satisfied, the general form of the fourth-rank tensor $\rho ^{\mu \nu \rho \sigma } $
can be constructed as \cite{Grozdanov:2016tdf,Hongo:2020qpv,Hattori:2022hyo}
\begin{equation}
    {\rho ^{\mu \nu \rho \sigma }} = - 2{\rho_ \bot }({b^\mu }{\Xi ^{\nu [\rho }}{b^{\sigma ]}} - {b^\nu }{\Xi ^{\mu [\rho }}{b^{\sigma ]}})  +  2{\rho_\parallel }{\Xi ^{\mu [\rho }}{\Xi ^{\sigma ]\nu }}.\label{add1}
\end{equation}
This tensor provides the constitutive equation for an induced electric field and is a resistivity tensor \cite{Grozdanov:2016tdf, Hattori:2017usa, Hattori:2022hyo}. 
There is no Hall term in neutral systems.

The semi-positivity of the entropy current requires that 
all the transport coefficients be semi-positive except for $\zeta_\times $ 
and $\xi _\parallel$ in 
the off-diagonal components of matrices. 
These transport coefficients are instead constrained by semi-positivity of the matrices, i.e., the eigenvalues. 
The coefficients should satisfy the inequalities 
\begin{eqnarray}
\label{inequality}
&&  {\zeta _ \bot } \ge 0, \quad 
{\zeta _\parallel } \ge 0, \quad 
{\zeta _\parallel }{\zeta _ \bot } \ge \zeta _ \times ^2, \quad 
 {\eta _\parallel } \ge 0,  \quad 
{\eta _ \bot } \ge 0, \quad 
\nnb
 && 
 {\gamma _\parallel } \ge 0, \quad 
{\gamma _ \bot } \ge 0, \quad 
{\gamma _\parallel }{\eta _\parallel } \ge {\xi _\parallel^2 }
, \quad 
\label{con20} 
\\
&&
\rho_\perp \ge 0, \quad \rho_\parallel\ge 0
. \nn
\end{eqnarray}

\section{Linear-mode analysis}\label{s3}

In this section, we derive the linearized first-order hydrodynamic equations 
for small perturbations near an equilibrium state. 
This is often called the linear-mode analysis.  
We apply small perturbations on top of the equilibrium values 
$u^\mu  = (1,0,0,0)$, $B^\mu  = (0,0,0,B)$, 
and $ S^{\mu \nu } = 0 $. 
Here, without loss of generality, we took the direction of the magnetic field along the $ z$ axis in equilibrium. 
The conserved charges are displaced from 
their equilibrium values as 
\begin{eqnarray}
&& e \to e+ \delta e (x), \quad 
 {u^\mu } \to u^\mu  + \delta {u^\mu }(x) ,
\label{don01}
 \\
 &&
{B^\mu } \to B^\mu  + \delta {B^\mu } (x), \quad 
%{\mkern 1mu} B = {  B} + \delta B\\
{S^{\mu \nu }} \to { S^{\mu \nu }} + \delta {S^{\mu \nu }} (x) 
\nn
%{\mkern 1mu} {\omega ^{\mu \nu }}(x) = 0 + \delta {\omega ^{\mu \nu }}
.
\end{eqnarray}
We will linearize the hydrodynamic equations with respect to these perturbations. 
%The perturbation $\delta {B^\mu } $ can orient in other directions. 
Since $S^{\mu\nu}$ has only three spatial components 
due to our pseudo-gauge choice, it is more convenient 
to extract the spin density as a spatial vector 
\begin{eqnarray}
\label{eq:sigma}
\sigma^\mu=-\frac 12\epsilon^{\mu\nu\rho\sigma} u_\nu S_{\rho\sigma} . 
\end{eqnarray}
Thus, the last decomposition in Eq.~(\ref{don01}) is 
alternatively written as $\sigma^\mu = 0 + \delta \sigma^\mu $. 

For simplicity, we assume that the contributions of 
the matter and magnetic components to the equilibrium 
energy density and pressure can be separated as 
\begin{eqnarray}
 p_\perp=P+\frac{B^2}{2\mu_m}, \quad 
e=\epsilon+\frac{B^2}{2\mu_m} ,
\label{relations}
\end{eqnarray}
where $P$ and $\epsilon$ are the equilibrium energy density and pressure from the matter contribution.

\cout{
In \ref{relations}, one can assume that the x-y plane is isotropic, so we prescribe that ${\chi ^{xz}} = {\chi ^{yz}} \equiv \chi $, and define $\chi^{xy}=\chi^z$. Note that $\delta {\sigma _0} = \frac{1}{2}{\epsilon^{0\nu \rho \sigma }}({  S_{\rho \sigma }}\delta {u_\nu } +   {u_\nu }\delta {S_{\rho \sigma }}) = 0$, so we can use $\sigma^i$ to describe the spin. And we assume the magneto-conductivity is isotropic in x,y,z direction.

\com{Within the linear order in B, it may be isotropic.}
\com{What is magneto-conductivity?}

}

%Note that not all of these variables are necessary, for example, $\delta {u} = 0$. And not all the variables in (\ref{don01}) are linearly independent. In fact, there’re only 9 linear independent variables.

\subsection{Equations for the energy-momentum tensor}\label{s3.1}

The conservation law of the energy-momentum tensor in Eq.~(\ref{con4}) can be projected as 
\begin{eqnarray} 
  {u_\nu }{\partial _\mu }{\Theta ^{\mu \nu }} = 0 , \quad 
  \Xi _\nu ^\rho {\partial _\mu }{\Theta ^{\mu \nu }} = 0 , \quad 
  {b_\nu }{\partial _\mu }{\Theta ^{\mu \nu }} = 0  .
\label{don04}
\end{eqnarray}
Plugging Eq.~(\ref{con6}) into the above 
and focusing on the linear order in perturbations, 
%\begin{flalign}
%    \begin{array}{l}
%    De =  \\
%    - (e + {p_ \bot })\theta  + ({p_\parallel } - {p_ \bot }){u_\nu }{\partial _\mu }({b^\mu }{b^\nu }) + {u_\nu }{\partial _\mu }\Theta _{(1)}^{\mu \nu },\label{don05}
%\end{array}
%\end{flalign}
%
%\begin{flalign}
%    \begin{array}{l}
%0 = \Delta _\nu ^\rho {\partial _\mu }{\Theta ^{\mu \nu }} \\= \Delta _\nu ^\rho {\partial _\mu }(e{u^\mu }{u^\nu } + {p_\parallel }{b^\mu }{b^\nu } + {p_ \bot }{\Xi ^{\mu \nu }} + \Theta _{(1)}^{\mu \nu })\\
% = (e + {p_ \bot }){u^\mu }{\partial _\mu }{u^\rho } + {\Xi ^{\mu \rho }}{\partial _\mu }{p_ \bot } + {b^\mu }{b^\rho }{\partial _\mu }{p_\parallel }\\
% + ({p_\parallel } - {p_ \bot })\Delta _\nu ^\rho {\partial _\mu }({b^\mu }{b^\nu }) + \Delta _\nu ^\rho {\partial _\mu }\Theta _{(1)}^{\mu \nu }.
%\end{array}\label{don06}
%\end{flalign}  	
one arrives at the equations 
\begin{subequations}
\label{don13}
\begin{eqnarray} 
0 \=  \partial_0 \delta \epsilon+\frac{B}{\mu_m}\partial_0\delta B
\nnb
&+&h\partial_{\perp\mu} \delta u^\mu_\perp+ (h-\frac{B^2}{\mu_m})  \partial_z \delta u_z,
\\
0\=h{\partial _0 }{\delta u_x } + c_s^2{\partial _x }{\delta \epsilon } -\frac{B}{\mu_m} {\partial_z }\delta {B_x}+\frac{B}{\mu_m}\partial_x\delta B_z
\nnb
&-&[(\zeta_\perp+\eta_\perp)\partial_x^2+(\eta_\perp+\gamma_\perp)\partial_y^2+(\eta_\parallel+\gamma_\parallel-2\xi_\parallel)\partial_z^2]\delta u_x
\nnb
&-&(\zeta_\perp-\gamma_\perp)\partial_x\partial_y\delta u_y-(\zeta_\times+\eta_\parallel-\gamma_\parallel)\partial_z\partial_x\delta u_z
\nnb
&-& \frac 4\chi(\gamma_\parallel-\xi_\parallel)\partial_z\delta\sigma_{y}
+\frac {4}{\chi}\gamma_\perp\partial_y\delta \sigma_{z} 
\\
0\=h{\partial _0 }{\delta u_y } + c_s^2{\partial _y }{\delta \epsilon } -\frac{B}{\mu_m} {\partial_z }\delta {B_y}+\frac{B}{\mu_m}\partial_y\delta B_z
\nnb
&-&[(\zeta_\perp+\eta_\perp)\partial_y^2+(\eta_\perp+\gamma_\perp)\partial_x^2+(\eta_\parallel+\gamma_\parallel-2\xi_\parallel)\partial_z^2]\delta u_y
\nnb
&-&(\zeta_\perp-\gamma_\perp)\partial_x\partial_y\delta u_x-(\zeta_\times+\eta_\parallel-\gamma_\parallel)\partial_z\partial_y\delta u_z
\nnb
&+& \frac{4}{\chi}(\gamma_\parallel-\xi_\parallel)\partial_z\delta\sigma_{x} 
-\frac{4}{\chi}\gamma_\perp\partial_x\delta \sigma_{z}
,
\\
0\=c_s^2{\partial _z }\delta \epsilon
+(h-\frac{B^2}{\mu_m}){\partial _0 }{\delta u_z } 
\nnb
&-&(\zeta_\times+\eta_\parallel-\gamma_\parallel)\partial_z(\partial_x\delta u_x+\partial_y\delta u_y)
\nnb
&-&(\zeta_\parallel\partial_z^2+(\eta_\parallel+\gamma_\parallel+2\xi_\parallel)(\partial_x^2+\partial_y^2))\delta u_z
\nnb
&+&\frac{4}{\chi}(\xi_\parallel+\gamma_\parallel)(\partial_x\delta\sigma_{y}-\partial_y\delta\sigma_{x}),
\end{eqnarray}  
\end{subequations}
%\com{All the x,y,z indices were lowered.}
where we defined $  \delta u^\mu_\perp 
= (0,\delta u_x,\delta u_y,0)$ 
and the enthalpy 
\begin{eqnarray}
     h =   e +   p =
  \epsilon +   P +   B^2 /\mu_m .
  \label{eq:enthalpy}
\end{eqnarray} 
Here, $c_s^2= \delta P/ \delta \epsilon$ is the squared sound velocity.

\subsection{Equations for the magnetic field}\label{s3.2}

The equations for $\tilde F^{\mu\nu} $ can be 
projected and linearized in the same manner. 
The projected conservation law reads 
\begin{eqnarray} 
  {u_\nu }{\partial _\mu }{\tilde F^{\mu \nu }} = 0 , \quad 
  \Xi _\nu ^\rho {\partial _\mu }{\tilde F ^{\mu \nu }} = 0 , \quad 
  {b_\nu }{\partial _\mu }{\tilde F^{\mu \nu }} = 0  .
\label{eq:F-prj}
\end{eqnarray}
Note that the first equation contains neither a time derivative 
nor derivative corrections. 
This equation is the Gauss law constraint 
as explicitly shown below. 
Note also that the set of equations (\ref{eq:F-prj}) involves a redundancy due to an identity 
\begin{eqnarray}
\label{eq:sum-rule}
0 = \pd_\mu \pd_\nu \tilde F^{\mu\nu}
= (  \Xi_{\alpha\beta} 
-   u_\a   u_\b
+   b_\a    b_\b) \pd^\a \pd_\mu  \tilde F^{\mu\b}
. 
\end{eqnarray}
The left identity is satisfied by any antisymmetric tensor. 
The right identity serves as a sum-rule constraint 
for the set of equations (\ref{eq:F-prj}) 
as explicitly confirmed below. 
Therefore, we are left with two independent dynamical equations 
and the corresponding two spatial components of the magnetic field 
that satisfies $ u_\mu  B^\mu = 0 $ 
and $ \pd_\mu \delta B^\mu=0 $. 
The former equality reads $   B\delta {u_z} - \delta {B^0} = 0$, which is used to obtain the following equations.

%\begin{equation}
%    0 = (  u^\mu  + \delta {u^\mu })( {  B_\mu } + \delta {B_\mu })  = {  B_\mu }\delta {u^\mu } +   u_\mu \delta {B^\mu } =   B\delta {u^z} - \delta {B^0}.\label{don17}
%\end{equation}	  	
%Thus, $\delta {B^0} =  B \delta {u^z}$ and $\delta {B_0} =  -  B \delta {u_z}$. 

% Where the $F_{(1)}^{\mu\nu}$ is shown in (\ref{add1}). There should be 4 equations in (\ref{don15}), but formally, there's 5 variables attached with the magnetic ($\delta B$ and $\delta {b^\mu}$), thus the 5 variables should be linear dependent. Since ${u^\mu }{B_\mu } = 0$, one gets:

The explicit forms of the linearized equations are obtained as 
\begin{subequations}\label{don26}
\begin{eqnarray} 
0 \= \partial_i\delta B^i , \\
0 \= B\partial_z \delta u_x-\partial_0 \delta B_x
\nnb
&& 
- \rho'_\perp T \big[ \, \partial_z\partial_x \delta (\beta B_z)-\partial_z^2 \delta(\beta B_x) \, \big]
\nnb
&&
+ \rho'_\parallel T\big[ \,\partial_\bot^2 \delta(\beta B_x)
-\partial_x \partial_{\perp\mu}\delta (\beta B_\perp^\mu) \, \big] ,
\\
0 \= B\partial_z \delta u_y-\partial_0 \delta B_y
\nnb
&& 
- \rho'_\perp T\big[ \,\partial_z\partial_y \delta (\beta B_z)-\partial_z^2 \delta(\beta B_y)\, \big]
\nnb
&&
+ \rho'_\parallel T \big[ \,\partial_\bot^2 \delta(\beta B_y)-\partial_y \partial_{\perp\mu}\delta (\beta B_\perp^\mu)\, \big] ,
\\ 
0 \= -B\partial_{\perp\mu} \delta u^\mu_\perp-\partial_0 \delta B_z
\nnb
&& + \rho'_\perp T\big[ \,\partial_\perp^2 \delta(\beta B_z)-\partial_z\partial_{\perp\mu} \delta (\beta B_\perp^\mu)\, \big]
,
\end{eqnarray}
\end{subequations}
where we defined 
\begin{eqnarray}
    \rho'_\para = \frac{ \rho_\para}{\mu_m} , \quad 
     \rho'_\perp = \frac{ \rho_\perp}{\mu_m} .
\end{eqnarray}
The first equation is nothing but the Gauss law and one can also explicitly confirm the sum rule (\ref{eq:sum-rule}). 
Thus, there are only two independent dynamical equations in the above. 

%degrees of freedom in $\delta B^i $ due to the Gauss law and the sum rule (\ref{eq:sum-rule}).

%\com{All the x,y,z indices were lowered.}
%\begin{equation}
%    \begin{array}{l}
%0 = \partial^i\delta B^i\\
%0=B\partial_z \delta u^x-\partial_0 \delta B^x-\frac{\rho_\perp T}{\mu_m}[\partial_z\partial_x \delta (\beta B_z)-\partial_z^2 \delta(\beta B_x)]\\
%+\frac{\rho_\parallel T}{\mu_m}[\partial_\bot^2 \delta(\beta B_x)-\partial_x \partial_\perp\delta (\beta B_\perp)]\\
%0=B\partial_z \delta u^y-\partial_0 \delta B^y-\frac{\rho_\perp T}{\mu_m}[\partial_z\partial_y \delta (\beta B_z)-\partial_z^2 \delta(\beta B_y)]\\
%+\frac{\rho_\parallel T}{\mu_m}[\partial_\bot^2 \delta(\beta B_y)-\partial_y \partial_\perp\delta (\beta B_\perp)]\\
%0=-B\partial_\perp \delta u^\perp-\partial_0 \delta B^z+\frac{\rho_\perp T}{\mu_m}[\partial_\perp^2 \delta(\beta B_z)-\partial_z\partial_\perp \delta (\beta B_\perp)].
%\end{array}\label{don26}
%\end{equation}		  	
%
%And the $\delta(\beta B_i)$ can be expressed with $\delta B_i,\delta \epsilon$:
%\begin{equation}
%    \begin{array}{l}
%       \delta(\beta B^i)=B^i\delta \beta+\beta\delta B^i=-\frac{c_s^2 \beta B^i }{  h-  B^2/\mu_m}\delta\epsilon+\beta\delta B^i
%    \end{array}
%\end{equation}

\cout{
As mentioned above, 
the first equation is nothing but the Gauss law, 
indicating that the Fourier component of the perturbation $\delta B^i $ is always perpendicular to its momentum $k^i$ which will be introduced in Eq.~(\ref{eq:mode-exp}). 
Also, the last three equations are not independent as they satisfy the sum rule (\ref{eq:sum-rule}). 
Thus, there are only two independent degrees of freedom in $\delta B^i $.

\gray{
When solving the equations, one can simplify the equations 
{
by choosing two components transverse to the momentum. 
}
For example, when $k_z = 0$, $\delta B_y$ and $\delta B_z$ are independent variables; 
In this case, we should keep the last two equations. 
When $k_z\neq 0$, $\delta B_x$ and $\delta B_y$ are independent variables. Then, one should keep the second and third equations.}

}

The derivative of $ \delta \b$ can be expressed with 
that of $\delta \epsilon $ with the help of 
a relation obtained from 
the thermodynamic relation Eq.~(\ref{con5}), that is,  
\begin{eqnarray}
\label{eq:delta-b}
\delta\beta = -\frac{c_s^2 \beta}{h-  B^2/\mu_m}\delta\epsilon .
\end{eqnarray}

\subsection{Equations for the spin angular momentum}\label{s3.3}

\cout{
We then study the equation determined by the conserved equation ${\partial _\mu }{J^{\mu \alpha \beta }} = 0$, it implies the conservation of the angular momentum. Since we assumed that the $\Sigma^{\mu\alpha\beta}$ is full asymmetry, we will have an extra constrain\cite{frenkel_elektrodynamik_1926}: $S^{\mu\nu}u_\mu=0$, which lead to:
\begin{equation}
    \begin{array}{l}
         0=\delta (S^{\mu\nu}u_\mu)=  u_\mu \delta S^{\mu\nu}=-\delta S^{0\nu} 
    \end{array}
\end{equation}
\com{This should be discussed earlier since $\sigma^\mu $ 
is used in the equations from the energy-momentum conservation.}

}

%Make a perturbation on the equation (\ref{spin}) ,we get:
%\begin{equation}
%    \partial_0 \delta S^{ij}=-2\delta \Theta_{(1a)}^{ij}.
%\end{equation}	  	

As mentioned around Eq.~(\ref{eq:sigma}), 
the spin density only has three spatial degrees of freedom 
in the totally antisymmetric pseudo-gauge. 
The corresponding three equations are obtained by the projections 
\begin{eqnarray} 
&&
\Xi^1_\a b_\b (\partial _\mu \Sigma^{\mu \a\b } + 2 \Theta^{[\a\b]} ) = 0 ,
\nnb
&&
\Xi^2_\a b_\b (\partial _\mu \Sigma^{\mu \a\b } + 2 \Theta^{[\a\b]} ) = 0, 
\\
&&
\Xi^1_\a \Xi^2_\b (\partial _\mu \Sigma^{\mu \a\b } + 2 \Theta^{[\a\b]} ) = 0 .
\nn
\end{eqnarray}
One can obtain the linearized equations 
\begin{subequations}
\label{add3} 
\begin{eqnarray} 
0\= 
( %\frac{8\gamma_\parallel}{\chi}
\Gam_\para+\partial_0)\delta\sigma_x
\\
&&
+ 2 \big[ \,  \gamma_\parallel (\partial_y\delta u_z-\partial_z\delta u_y) 
+ \xi_\parallel (\partial_y\delta u_z+\partial_z\delta u_y) \, \big]
,
\nnb
0\=
(%\frac{8\gamma_\parallel}{\chi}
\Gam_\para+\partial_0)\delta\sigma_y
\\
&&
+ 2\big[ \, \gamma_\parallel (\partial_z\delta u_x - \partial_x\delta u_z )
-  \xi_\parallel (\partial_z\delta u_x + \partial_x\delta u_z  ) \, \big] ,
\nnb
0\=(%\frac{8\gamma_\perp}{\chi^z}
\Gam_\perp +\partial_0)\delta\sigma_z
+ 2\big[ \,\gamma_\perp(\partial_x\delta  u_y-\partial_y\delta u_x) \, \big]
,
\end{eqnarray}
\end{subequations}
where we defined 
\begin{eqnarray}
\Gam_\para =\frac{8\gamma_\parallel}{\chi} , \quad 
\Gam_\perp =\frac{8\gamma_\perp}{\chi}
.
\end{eqnarray}

\subsection{Summary of linearized equations}\label{s3.4}
We have obtained the linearized equations for spin MHD, and it is useful to note that the equations obtained above can be divided into two groups. One of them contains the variables $ (\delta {u_y},\delta {B_y},\delta {\sigma _x},\delta {\sigma _z})$, while the other contains $( \delta \epsilon,\delta {u_x},\delta {u_z},\delta {B_x},\delta {\sigma _y})$. 
The former group is summarized as 
\begin{subequations}\label{f2}
    \begin{eqnarray}
0\=h{\partial _0 }{\delta u_y } - 
%\frac{B}{\mu_m}
 \frac{h}{B} v_A^2  {\partial_z }\delta {B_y}
\\
&&
-[(\eta_\perp+\gamma_\perp)\partial_x^2+(\eta_\parallel+\gamma_\parallel-2\xi_\parallel)\partial_z^2]\delta u_y
\nnb
&&
- \frac 12 \Gam_\perp %\frac{4}{\chi}\gamma_\perp
\partial_x \delta \sigma_{z}
+\frac{4}{\chi}(\gamma_\parallel-\xi_\parallel)\partial_z\delta\sigma_{x},
\nnb
0\=B\partial_z \delta u_y-\partial_0 \delta B_y+ \rho'_\perp \partial_z^2 \delta B_y+\rho'_\parallel\partial_x^2 \delta B_y,
\\
0\=(\Gam_\para+\partial_0)\delta\sigma_x
- 2(\gamma_\parallel-\xi_\parallel)\partial_z\delta u_y,
\\
0\=(\Gam_\perp+\partial_0)\delta\sigma_z
+ 2\gamma_\perp\partial_x\delta  u_y .
    \end{eqnarray}
\end{subequations}
We assume a single-mode solution 
for the flow perturbation 
\begin{eqnarray}
    \delta \bu (t,x,z) =  \delta \tilde \bu(\omega,k_\perp,k_\para) e^{- i \omega t + i k_\perp x + i k_\para z}
    , \label{eq:mode-exp}
\end{eqnarray}
and similar forms for other perturbations. 
Then, these equations can be summarized in an algebraic matrix form
\begin{eqnarray}
    0=M_1 (\delta {\tilde u_y},\delta {\tilde B_y},\delta {\tilde \sigma _x},\delta {\tilde \sigma _z})^T, \label{quartic eq1}
\end{eqnarray}
where $M_1$ is the coefficient matrix. 
On the other hand, the latter group reads 
\begin{subequations}
\label{f1}
    \begin{eqnarray}
0 \=  \partial_0 \delta \epsilon
+ \frac{h}{B} v_A^2  \partial_0\delta B+h\partial_x \delta u_x
+ h (1-v_A^2) \partial_z \delta u_z,\nnb
\\
0\=h{\partial _0 }{\delta u_x } + c_s^2{\partial _x }{\delta \epsilon } - \frac{h}{B} v_A^2  
( {\partial_z }\delta {B_x}- \partial_x\delta B_z)
\nnb
&&
-[(\zeta_\perp+\eta_\perp)\partial_x^2+(\eta_\parallel+\gamma_\parallel-2\xi_\parallel)\partial_z^2]\delta u_x
\nnb
&&
-(\zeta_\times+\eta_\parallel-\gamma_\parallel)\partial_z\partial_x\delta u_z
- \frac 4\chi(\gamma_\parallel-\xi_\parallel)\partial_z\delta\sigma_{y},
\\
0\=c_s^2{\partial _z }\delta \epsilon
+ h (1-v_A^2){\partial _0 }{\delta u_z } 
-(\zeta_\times+\eta_\parallel-\gamma_\parallel)\partial_z \partial_x\delta u_x
\nnb
&&
-(\zeta_\parallel\partial_z^2
+(\eta_\parallel+\gamma_\parallel+2\xi_\parallel)\partial_x^2)\delta u_z
\nnb
&&
+\frac{4}{\chi}(\xi_\parallel+\gamma_\parallel)\partial_x\delta\sigma_{y},
\\
0\=B\partial_z \delta u_x-\partial_0 \delta B_x
\nnb
&&
- \rho'_\perp[\partial_z\partial_x \delta B_z-\frac{c_s^2 }{1-v_A^2} \frac{B}{h} \partial_z\partial_x \delta \epsilon-\partial_z^2 \delta B_x],
\\
0\=
( \Gam_\para +\partial_0)\delta\sigma_y
\nnb
&&
+ 2[(\gamma_\parallel - \xi_\parallel)\partial_z\delta u_x - (\gamma_\parallel+\xi_\parallel)\partial_x\delta u_z]
.
    \end{eqnarray}
\end{subequations}
Here, we defined the so-called Alfven-wave velocity 
\begin{eqnarray}
v_A = \frac{B}{\sqrt{  \mu_m h }} .
\end{eqnarray}
Without loss of generality, we have set the transverse coordinate system in such a way that the derivatives along the $y$ direction vanish. 
Then, one obtains a matrix equation 
\begin{eqnarray}
    0=M_2 (\delta {\tilde \epsilon},\delta {\tilde u_x},\delta {\tilde u_z},\delta {\tilde B_x},\delta{\tilde \sigma_y})^T, \label{quintic eq1}
\end{eqnarray}
where the $M_2$ is the coefficient matrix of these five equations. 
For nontrivial solutions, we require that 
\begin{eqnarray}
    \det(M_1)=0 \quad {\rm and} \quad  \det(M_2)=0.
\end{eqnarray}
{
Corresponding to the number of above equations, we will find nine modes. 
While six of them are modifications of the conventional gapless MHD modes, the remaining three are gapped spin modes in the totally antisymmetric pseudo-gauge. 
}

%We can first consider the solution in some special cases. For example, we can assume the variables are independent of spatial coordinates, we have ${k_x},{k_y},{k_z} = 0$. By using this method, we can simplify the equations, and if we choose a suitable assumption, we can diagnalize the coefficient matrix, and leave the equation solvable.

\section{Solutions for the linearized equations}\label{s4}

%\com{We should state this general property in the beginning.}

In this section, we seek solutions for the linearized equations obtained in Sec.~\ref{s3}. 
{
Because of the anisotropy induced by a magnetic field, the solutions depend on the angle $\theta$ between the momentum $\vec k$ and the magnetic field in equilibrium. Accordingly, we introduce $k_z=k \cos \theta$ and $k_x=k\sin\theta$. 
It is not difficult to find the solutions at specific angles $\theta=0, \ \pi/2$ as shown in Appendix~\ref{Appendix A}. 
However, it is challenging to find solutions at an arbitrary angle. 
We first show the first-order solutions in the small-$k$ expansion for an arbitrary angle. 
We use the method developed in Ref.~\cite{Fang:2024skm} and obtain the solutions up to the $k^2$ order as we are considering the first-order hydrodynamics.  
}

{Next, we compare the solutions in the small-$k$ expansion and the solutions at specific angles (without any expansion). 
We point out that the small-$k$ expansion breaks down when the angle approaches the perpendicular direction to the magnetic field, i.e., $\theta=\pi/2$, because the $n$-th order term blows up as $\sim 1/\cos^n \theta$. 
In this regime, we show an alternative series expression for a small value of $\cos\theta$. 
Furthermore, we confirm that these two series expressions agree with numerical solutions in appropriate angle regimes. 
Remarkably, there is a critical angle in between these regimes, where some propagating modes change to purely dissipative modes. 
}

\subsection{Leading-order solutions}\label{leading order solution}

To illustrate the crucial steps for the method developed in Ref.~\cite{Fang:2024skm}, we begin with the leading-order solutions. 
Let us take Eq.~(\ref{quartic eq1}) which is a quartic equation in $\omega$. 
Up to the $k^1$ order, the secular equation from $\det(M_1)=0$ should be factorized as 
\begin{eqnarray}
\det(M_1)
&\propto &
(\omega - \tilde v_1 k)(\omega-\tilde v_2 k)(\omega+ i\tilde \Gamma_1)(\omega+i \tilde\Gamma_2) . \quad \label{quantic LO eq}
\end{eqnarray} 
This is because one should have two gapless modes and two gapped modes. 
The latter stems from the non-conservation equation for spin (\ref{eq:spin-cons}).

By comparing the coefficients on the left- and right-hand sides of Eq.~(\ref{quantic LO eq}) on an order-by-order basis in $k$, one can determine $\tilde v_{1,2}$ and $\tilde\Gamma_{1,2}$. 
At the leading order, one can easily get  
\begin{subequations}
\label{quantic LO}
\begin{eqnarray}
&&
\tilde v_{1,2}=\pm v_A\cos\theta, \quad 
\\
&&
\tilde\Gamma_1=\Gamma_\parallel, 
\quad 
\tilde \Gamma_2=\Gamma_\perp.
\end{eqnarray} 
\end{subequations}
We find that $0\leq v_A \leq 1$ when the enthalpy is given as in Eq.~(\ref{eq:enthalpy}).

{
One can apply the same method to Eq.~(\ref{quintic eq1}). Since there is only one spin variable in Eq.~(\ref{quintic eq1}), one should have one gapped mode in the set of solutions. 
Therefore, the leading-order secular equation should be factorized as 
\begin{eqnarray}
    \det(M_2)&\propto&(\omega-\tilde v_1')(\omega-\tilde v_2')
    \nnb
    && \times
    (\omega-\tilde v_3')(\omega-\tilde v_4')(\omega+i\tilde\Gamma).
\end{eqnarray} 
We then obtain the leading-order solutions 
\begin{subequations}
    \begin{eqnarray}
    &&
        \tilde v_{1,2}' = \pm v_1 k , \quad 
        \tilde v_{3,4}' = \pm v_2 k ,
        \\
        &&
        \tilde \Gam = \Gam_\para ,
    \end{eqnarray}
\end{subequations}
where we defined 
\begin{subequations}
    \begin{eqnarray} 
    v_{1,2} \=\frac{1}{\sqrt{2}}\Big( \, V^2\pm\sqrt{V^4-4v_A^2 c_s^2 \cos^2\theta} \, \Big)^{\frac 12},
    \label{v_12}
    \\
     V \= \sqrt{v_A^2+ c_s^2(1-v_A^2\sin^2 \theta)} .
\end{eqnarray}
\end{subequations}
Here, we get four propagating modes, 
which are known as the magneto-sonic waves. When the sign in Eq.~(\ref{v_12}) takes plus, they are called the fast modes; 
In the other case, they are called the slow modes. We find that the leading-order solutions are exactly the same as the case without spin (see, e.g., Refs.~\cite{Grozdanov:2016tdf,Biswas:2020rps, Fang:2024skm} and references therein). 
It can be shown that $0\leq v_{1,2} \leq 1$ assuming that $0 \leq c_s \leq 1$ and  $0\leq v_A \leq 1$ (see Ref.~\cite{Fang:2024skm}). 
}

\subsection{Next-to-leading order solutions}\label{s4.3}

Now, we step forward to the $k^2$ order. As we are discussing solutions within the first-order hydrodynamics, the terms beyond the $k^2$ order are uncertain. Therefore even though we get the expansion to the $k^3$ or even higher orders, we cannot judge if it is a better solution than the expansion up to the $k^2$ order. In all, since higher-order terms are not under control, one should stop at the $k^2$ order.

It is straightforward to extend 
the discussion in Sec.~\ref{leading order solution} to the next-to-leading order. 
We can now assume the factorization for Eq.~(\ref{quartic eq1}) up to the $k^2$ order as 
\begin{eqnarray}
\det(M_1) &\propto& (\omega - v_A k_\parallel+i\tilde\omega_1 k^2)(\omega+ v_A k_\parallel+i\tilde\omega_2 k^2)
\nnb
&&\times
(\omega+ i \Gamma_\parallel+i\tilde\omega_3 k^2)(\omega+i \Gamma_\perp+i\tilde\omega_4 k^2) , \quad \label{quantic NLO eq}
\end{eqnarray} 
where $\tilde \omega_i$ should be independent of $k$. By comparing $\det(M_1)$ with the above factorized form order-by-order in $k$, 
one can determine $\tilde\omega_i$. 
Getting rid of the irrelevant higher-order terms allows us to efficiently identify the corrections 
\begin{subequations}
    \begin{eqnarray}
    \label{NLO11}
    \tilde\omega_1 \= \tilde\omega_2=\frac{1}{2h}\big[( \tilde \eta_\para +h\rho_\perp')\cos^2\theta
    \nnb
    &&
+(\eta_\perp+h\rho_\parallel')\sin^2\theta\big], 
    \\
    \label{NLO12}
    \tilde\omega_{3} \=\frac{(\gamma_\parallel-\xi_\parallel)^2}{h\gamma_\parallel}\cos^2\theta, \,\,
    \\
     \tilde\omega_{4} \= \frac{\gamma_\perp}{h} \sin^2\theta,
\end{eqnarray}\label{NLO}
\end{subequations}
where we defined 
\begin{eqnarray}
\label{eq:eta-tilde}
\tilde \eta_\parallel =    \eta_\parallel-\frac{\xi^2_\parallel}{\gamma_\parallel} .
\end{eqnarray}
Then, the first-order  solutions can be summarized as 
\begin{subequations}
\label{quantic NLO solution-set1}
\begin{eqnarray}
\label{quantic NLO solution1}
\omega \=\pm v_A k \cos\theta-i\tilde\omega_{1,2}k^2 ,
\\
\label{quantic NLO solution2}
\omega \= -i\Gamma_\parallel -i\tilde\omega_3 k^2 ,
\\
\label{quantic NLO solution3}
\omega \= -i\Gamma_\perp -i\tilde \omega_4 k^2 .
\end{eqnarray}\label{quantic NLO solution}
\end{subequations}

Next, let us discuss the $k^2$-order solutions for Eq.~(\ref{quintic eq1}). We assume the factorization of $\det(M_2)$ as
\begin{eqnarray}
\label{quartic NLO equation}
    \det(M_2) &\propto& (\omega-v_1 k+i\tilde\omega_1' k^2)(\omega+v_1 k+i\tilde\omega_2' k^2)
    \nnb
    && \times
    (\omega-v_2 k+i\tilde\omega_3' k^2)(\omega+v_2 k+i\tilde\omega_4' k^2)
    \nnb&& \times
(\omega+i\Gamma_\parallel+i\tilde\omega_5 k^2).
\end{eqnarray}
Again, by comparing this ansatz with the determinant on an order-by-order basis, one can obtain $\tilde \omega_i'$ and $\omega_5$. 
We find that $\tilde\omega_1 ' = \tilde\omega_2'$ and that $\tilde\omega_3'=\tilde\omega_4'$.
This indicates that a pair of waves propagating in opposite directions with the same speed acquire the same damping rate at the $k^2$ order.
%\gray{The form of $\tilde\omega_i$ is quite complex here, and we find it simpler to show the solution up to $k^2 $ order as} 
The solutions are summarized as 
\begin{subequations}
\label{quartic NLO solution}
\begin{eqnarray}
\label{quartic NLO solution1}
    \omega \= \pm v_1 k+i k^2 \frac{W_1-W_2 v_1^2}{2(v_1^2-v_2^2)},  
    \\ 
    \label{quartic NLO solution2}
    \omega \= \pm v_2 k-i k^2\frac{W_1-W_2 v_2^2}{2(v_1^2-v_2^2)}, 
    \\
    \label{quartic NLO solution3}
    \omega \= -i\Gamma_\parallel-{ ik^2 \left(  {W_3}-  {W_2}\right)}, 
\end{eqnarray}
\end{subequations}
where the explicit forms of $W_i$ are found to be 
\begin{subequations}
\label{eq:W-sol}
\begin{eqnarray} 
W_1  \= \frac{1}{h}\Big[
%(\eta_\para-\frac{\xi_\parallel^2}{\gamma_\parallel}) 
\tilde \eta_\para
\Big( c_s^2  \cos^2 (2\theta)
+  \frac{v_A^2  \sin^2 \theta}{1-v_A^2}
 \Big) 
\\
&& 
+ h \rho'_\perp c_s^2 \frac{ 1 - v_A^2 \cos^2 \theta }{1-v_A^2}  + 
\zeta_\para   \frac{v_A^2}{1-v_A^2}  \cos^2 \theta
\nnb
&& 
+ ( \zeta_\para +\zeta_\perp- 2 \zeta_\times 
+ \eta_\perp ) c_s^2  \sin^2 \theta \cos^2 \theta \Big]
,  
\nnb
W_2 \=  \frac{1}{h}\Big[
%(\eta_\para-\frac{\xi_\parallel^2}{\gamma_\parallel}) 
\tilde \eta_\para
\frac{1 - v_A^2  \cos^2 \theta}{1-v_A^2}
%\Big( 1 + \frac{v_A^2}{1-v_A^2} \sin^2 \theta\Big)  
+  \zeta_\para\frac{\cos^2 \theta }{1-v_A^2} 
\\
&& 
+ ( \zeta_\perp  + \eta_\perp ) \sin^2\theta
+ h\rho'_\perp \Big( 1 + \frac{v_A^2  c_s^2 }{1-v_A^2} \sin^2\theta  \Big)
\Big]
,
\nnb
W_3 \= \frac{1}{h}\Big[\cos^2\theta(\eta_\parallel+\gamma_\parallel-2\xi_\parallel)
\\
&&
+\frac{\sin^2\theta}{1-v_A^2}(\eta_\parallel+\gamma_\parallel+2\xi_\parallel)
+\zeta_\parallel\frac{\cos^2\theta}{1-v_A^2}
\nnb
&&
+(\zeta_\perp+\eta_\perp)\sin^2\theta 
+ h\rho_\perp^\prime(1+\frac{v_A^2 c_s^2}{1-v_A^2}\sin^2\theta)\Big]
\nn
.
\end{eqnarray}\label{quartic NLO W}
\end{subequations}
The above solutions in the small-$k$ expansion are one of the main results in this paper. 
However, we will further see that the above series expressions are not valid when $\cos \theta$ becomes small in the next subsection. 
Before going into this point, we give a couple of comments on the above solutions.

One can compare the above results with the first-order solutions without spin \cite{Fang:2024skm}. 
First, the propagating modes in the $k$ expansion, which are shown in Eqs.~(\ref{quantic NLO solution1}), (\ref{quartic NLO solution1}), and (\ref{quartic NLO solution2}), 
have the same forms as the damping Alfven and magneto-sonic waves up to the replacement of the shear viscosity $\eta_\para$ by $\tilde \eta_\para$ defined in Eq.~(\ref{eq:eta-tilde}). 
In other words, only when there is a nonzero cross rotational viscosity $\xi_\para$, the propagating modes are modified by dynamics of angular momentum. 
In Eq.~(\ref{con12}), the cross rotational viscosity converts the antisymmetric perturbation to the symmetric stress, and modifies the Alfven and sound modes. 
{The other three modes in Eqs.~(\ref{quantic NLO solution2}), (\ref{quantic NLO solution3}), and (\ref{quartic NLO solution3}) are damping spin modes. The damping factor remains in the zero momentum limit, and spin density does not persist in the strict equilibrium limit because spin angular momentum is not a conserved quantity by itself \cite{Hattori:2019lfp}.}

\cout{
This finding is consistent with the preceding linear-mode analyses (in the absence of a magnetic field) \cite{Hattori:2019lfp, Hongo:2021ona,Hongo:2021ona,Hu:2022azy,Daher:2022wzf} where the propagating modes are not modified by dynamics of angular momentum in the first-order solutions at the $k^2$ order. 
The cross rotational viscosity can exist only in the presence of an order-one vector such as a magnetic field and/or strong vorticity \cite{Cao:2022aku}. 
}

%%Xie:2023gbo,

{
One can also confirm that all the eigenmodes obtained in Eqs.~(\ref{quantic NLO solution}) and (\ref{quartic NLO solution}) are damping in time according to the inequalities on transport coefficients in Eq.~(\ref{con20}) from the entropy-current analysis. 
It is clear that all the $\tilde\omega_i$ in Eq.~(\ref{NLO}) are semi-positive definite for any $\theta$, indicating that all the modes in Eq.~(\ref{quantic NLO solution}) are damping in the $k^2$ order. 
It is less obvious that the other set of solutions (\ref{quartic NLO solution}) are also all damping in the $k^2$ order. 
In Ref.~\cite{Fang:2024skm} without spin, this has been shown for the first two modes for $\eta_\para \geq 0$. 
Replacing $\eta_\para$ by $\tilde \eta_\para$, we find that these two modes are still damping. 
One can show that the third mode is also damping. 
Its coefficient of the $k^2$ order reads
\begin{eqnarray} 
W_3-W_2 \=\frac{1}{h(1-v_A^2){\gam_\para}}
\big[\, (1-v_A^2\cos^2\theta)(\gamma_\parallel^2+\xi_\parallel^2)
\nnb
&&
+ 2\{  \sin^2\theta - { (1-v_A^2) \cos^2\theta } \} \xi_\parallel \gamma_\parallel \, \big] .
\end{eqnarray}
Since we have $4(1-v_A^2\cos^2\theta)^2 
- 2^2\{  \sin^2\theta -  (1-v_A^2) \cos^2\theta  \}^2
= 4(1-v_A^2)\sin^2(2\theta) \geq 0$, 
we find that   
\begin{eqnarray}
W_3-W_2 &\geq& \frac{1}{h(1-v_A^2){\gam_\para}}
(1-v_A^2\cos^2\theta)
[ (\gamma_\parallel^2+\xi_\parallel^2) -2\xi_\parallel \gamma_\parallel]  
\nnb
&\geq& 0 .
\end{eqnarray}
\cout{
\begin{eqnarray}
&&
W_3-W_2=
\nnb
&&
\frac{1}{h}\Big[\cos^2\theta(\eta_\parallel+\gamma_\parallel-2\xi_\parallel)+\frac{\sin^2\theta}{1-v_A^2}(\eta_\parallel+\gam_\parallel+2\xi_\parallel)
\nnb
&&
-\tilde\eta_\parallel \frac{1-v_A^2 \cos^2\theta}{1-v_A^2}\Big]
\nnb
&&
=\frac{1}{h(1-v_A^2)}[(1-v_A^2\cos^2\theta)(\gamma_\parallel^2+\xi_\parallel^2)+2 \sin^2\theta\xi_\parallel \gamma_\parallel]
\nnb
&&
\geq \frac{1}{h(a-v_A^2)}\sin^2\theta(\gamma_\parallel+\xi_\parallel)^2\geq 0 .
\end{eqnarray}
}
%\com{Is there a typo in the first line?}
Thus, we can conclude that all the solutions in Eq.~(\ref{quartic NLO solution}) are also damping. Then, all the nine modes in spin MHD are damping within the current linear-mode analysis in the rest frame.

%because $  (\eta_\para + \gam_\para )^2 - (2\xi_\para)^2 \geq (2\sqrt{ \eta_\para  \gam_\para})^2 - (2\xi_\para)^2 \geq 0$ according to the inequality $\gam_\para \eta_\para \geq \xi_\para^2$ in Eq.~(\ref{con20}). 

}

\subsection{Breakdown of the small-momentum expansion}

\label{sec:critical-mom}

{

We have obtained the solutions for an arbitrary angle $\theta$ up to the $k^2$ order in Sec.~\ref{s4.3}. 
Here, however, we point out that the small-$k$ expansion breaks down depending on the angle $\theta$, and thus provide an alternative method of expansion in such an angle regime. 
One can realize the breakdown of the small-$k$ expansion by comparing the solutions in Eqs.~(\ref{quantic NLO solution}) and (\ref{quartic NLO solution}) with the solutions at specific angles shown in Appendix.~\ref{Appendix A}.
The latter solutions can be obtained without any expansion. 

}

When $\theta=\pi/2$, 
the solution (\ref{quantic NLO solution1}) does not agree with any of the solutions in the limit $\theta=\pi/2$.\footnote{
Consistent cases can be confirmed as follows. 
For the set of solutions (\ref{quantic NLO solution}) containing the Alfven modes, the solutions (\ref{quantic NLO solution}) are consistent with those in Eq.~(\ref{solution 35}) and Eq.~(\ref{36}) when $\theta=0$. 
When $\theta=\pi/2$, the solutions (\ref{quantic NLO solution2}) and (\ref{quantic NLO solution3}) are still consistent with the one in Eq.~(\ref{o44}) and the first one in Eq.~(\ref{o22}). } 
{Therefore, one should investigate the behavior of the solution near $\theta=\pi/2$ more closely.} 
{As discussed below}, the exact solution (without the small-$k$ expansion) 
%correspond to Eq.~(\ref{quantic NLO solution1}) 
has a non-analytic boundary at $k=k_c$, where $k_c$ {depends on $\theta$}. The $k$ expansion gets invalid when $k$ goes beyond this boundary. The critical momentum $k_c$ approaches zero when $\cos\theta$ approaches zero. 
%\com{What is the condition?}
Accordingly, the Taylor expansion for $k$ breaks down when $\cos\theta$ approaches zero. 
{
This non-analytic property arises from that of a square root. 
Since waves propagating in opposite directions appear in pairs, the secular equation is an equation for $\omega^2$, and its solution for $\omega$ is in general given by a square root. 
}

This singular behavior was recently discussed in MHD without spin by the present authors \cite{Fang:2024skm}. 
Here, we recapitulate the discussions therein. 
The simplest mode, which nevertheless involves the singularity, is the Alfven waves 
\begin{eqnarray}
\label{eq:sol-Alfven}
\omega  \= \pm \sqrt{v_A^2  k^2 \cos^2  \theta
- \frac14  (\tilde \rho - \tilde \eta )^2k^4  }
- \frac{i}{2} (\tilde \rho + \tilde \eta) k^2
, 
\end{eqnarray}
where $\tilde \rho =  \rho'_\perp \cos^2 \theta
+\rho'_\para  \sin^2 \theta$ and $
\tilde \eta =\frac{1}{h}( \eta_\para \cos^2 \theta + \eta_\perp \sin^2 \theta ) $. 
This is the solution without any expansion. 
The square root in Eq.~(\ref{eq:sol-Alfven}) becomes a pure imaginary number when $k$ grows larger than $2v_A\left|\frac{\cos\theta}{\tilde \rho-\tilde\eta}\right|$. 
Namely, the dispersion relation loses the real part, turning to a purely diffusive mode. 
Clearly, this behavior is different from that of the solution in the small-$k$ expansion that always has the Alfven velocity as the group velocity. 
%However, finite expansion of $k$ near $k=0$ failed to predict this behavior, so the expansion get invalid after this point. 
In this regime where $k> 2v_A\left|\frac{\cos\theta}{\tilde \rho-\tilde\eta}\right|$, one can instead expand Eq.~(\ref{eq:sol-Alfven}) with $\cos\theta$ near $\cos\theta=0$. 
This alternative expansion is valid in the regime where the square root is imaginary.

\cout{To make this point more clear, we should illustrate the exact solution and show when the $k$ expansion is invalid.
However, the exact solution for spin-MHD is too complex to illustrate here. We choose to ignore the effect spin and focus on the main principle. 
When the spin vanishes, it is just the MHD. Then the equation for arbitrary angle reads
\begin{eqnarray}\label{eq:matrix-Alfven}
\det
\begin{pmatrix}
\omega+\rho_\perp' k_\para^2 + \rho_\para '  k_\perp^2   &  B k_\para 
\\
 h \frac{v_A^2}{B} k_\para &  h \omega + \eta_\para  k_\para^2 + \eta_\perp  k_\perp^2
 % \mu_m^{-1}  B k_\para &   
\end{pmatrix} 
=0
,  
\end{eqnarray} 
whose solutions read 
\begin{eqnarray}
\label{eq:sol-Alfven}
\omega  \= \pm \sqrt{v_A^2  k^2 \cos^2  \theta
- \frac14  (\tilde \rho - \tilde \eta )^2k^4  }
- \frac{i}{2} (\tilde \rho + \tilde \eta) k^2
.
\end{eqnarray}
where $\tilde \rho =  \rho'_\perp \cos^2 \theta
+\rho'_\para  \sin^2 \theta$ and $
\tilde \eta =\frac{1}{h}( \eta_\para \cos^2 \theta + \eta_\perp \sin^2 \theta ) $. It's obvious that the terms in the radical sign equal 0 when we take $k = 0,\,\,\cos\theta = 0$, so it is a non-analytical point in the complex plane. 
Next, to compare $k$ with $\cos\theta$, we need to introduce a value to eliminate the dimension of $k$. 
As hydrodynamics is a low energy effective theory, we assume there's an cutoff of $k$, beyond which the hydrodynamics will fail. 
We take $k_{cut}$ as an arbitrarily chosen value in this article, then we can denote $\tilde k=\frac{k}{k_{cut}}$ as a dimensionless value. In MHD, the solution in Eq.~(\ref{eq:sol-Alfven}) can be expanded as
\begin{subequations}
\label{eq:Alfven-various-expansion}
    \begin{eqnarray}
\omega \= \pm v_A k_\para 
- \frac{i}{2} (\tilde \rho + \tilde \eta) k^2 + \order \big(\frac{k}{|\cos\theta|} k^2 \big) 
,
\nnb
%\lambda < 1
&&\hat{k} < |\cos\theta| ,
%\quad \hat{k} < |\cos\theta| ,
\\
\omega  \= - i k^2 \rho'_\parallel
\nnb
&-& i \Big( ( \rho'_\perp - \rho'_\para) k^2
+ \frac{h v_A^2}{\eta_\perp - h\rho'_\para} \Big)
\cos^2\theta +\order(\cos^4\theta) , 
\nnb
&-& \frac {i k^2}{h} \eta_\perp- i \Big( (\eta_\para - \eta_\perp) \frac{k^2}{h}
-\frac{h v_A^2}{\eta_\perp- h\rho'_\para}\Big)\cos^2\theta\nnb
&+&\order(\cos^4\theta)
%,\quad |\cos\theta| < \hat{k} 
,\quad 
%\lambda > 1
|\cos\theta| < \hat{k} 
.
    \end{eqnarray}
\end{subequations}
This illustrate that when $\tilde k << \cos\theta$, the $k$ expansion is effective. While when $\tilde k >> \cos\theta$, a $\cos\theta$ expansion is more close to accurate solution.

Note that taking $\theta=\frac{\pi}{2}$ before $k$ expansion as we did in \ref{s4.2}, we always have $\tilde k >> \cos\theta$, so we should use $\cos\theta$ expansion as
\begin{eqnarray}
    \omega= a(k)+b(k) \cos^2\theta+\order ({\cos^4\theta}).
\end{eqnarray}
But since we have take $\cos\theta=0$, the exact solution is just $\omega=a(k)$. Then we can write $a(k)$ as expansion of $k$; In this case, we get a result which is close to accurate result. 
However, if we take $\theta$ as an arbitrary angle and expand solution by order of $k$ first, we will suffer from an issue.  
Simply taking $\theta\rightarrow \frac{\pi}{2}$ limit in the expansion of $k$ makes no sense, because there must be a point where $\cos\theta$ grow too small that the $k$ expansion get invalid.
}

\begin{figure}
    \centering
    \includegraphics[width=\hsize]{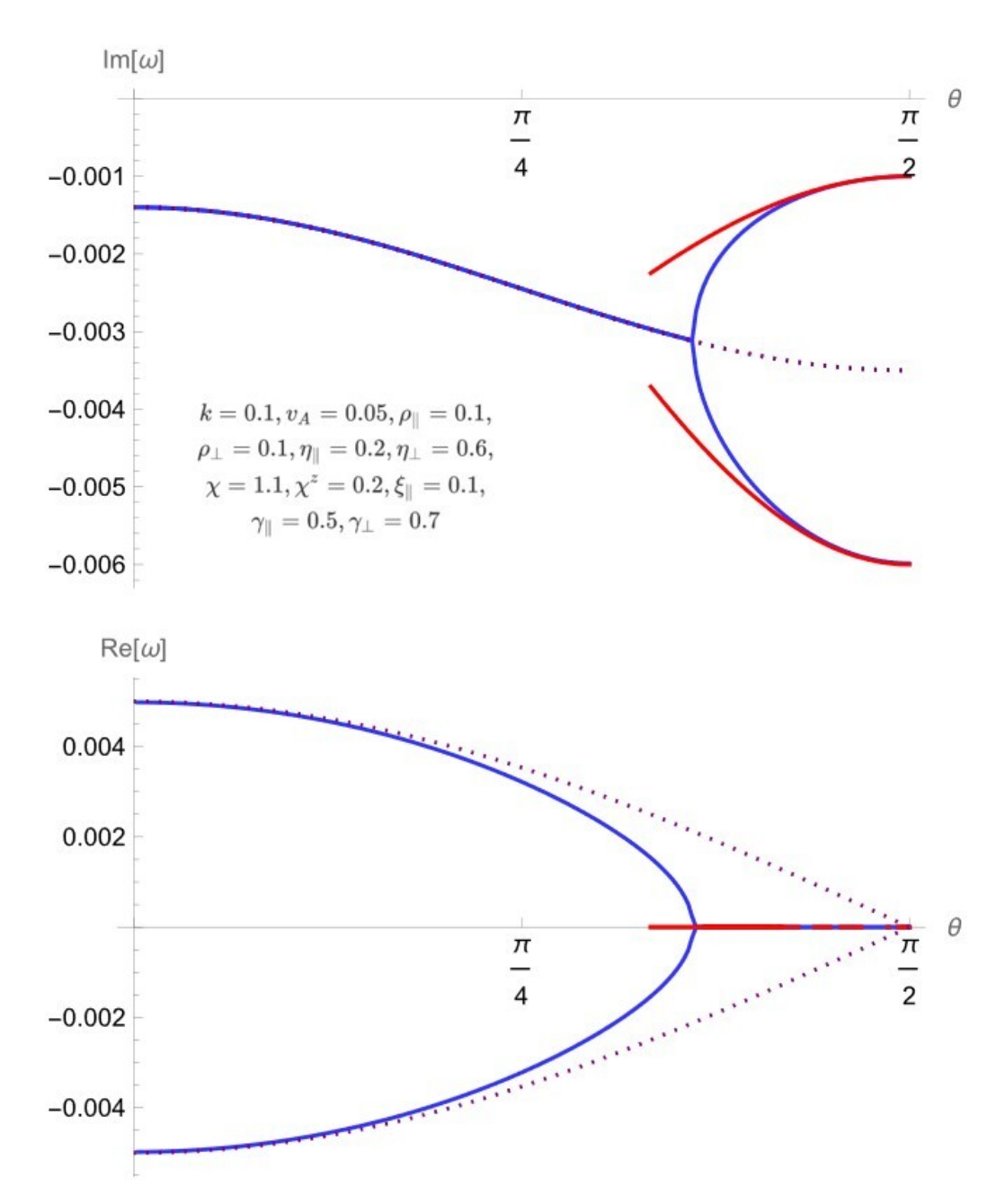}
    \caption{The dispersion relations for the Alfven modes. Blue curves show the ``exact solution'' without any expansion. 
    Dotted curves show the small-$k$ expansion in Eq.~(\ref{NLO11}). 
    Red curves show the small-cosine expansion in Eq.~(\ref{quartic solution simple}). 
    } 
     \label{fig:spin-MHD-4}
\end{figure}

We find the same issue in spin MHD. 
The two Alfven modes shown in Eq.~(\ref{NLO11}) are effective only when $\cos\theta$ is large. When $\cos\theta$ gets smaller than a certain value, the real parts of the dispersion relations merge to zero and the imaginary parts split into two distinct ones. 
We confirm these merging and splitting behaviors with numerical plots in Fig.~\ref{fig:spin-MHD-4}. 
The dotted line shows the small-$k$ expansion, while the red solid line shows the cosine expansion. 
We find a good agreement with the blue solid line that shows the solution without any expansion. 
In Appendix.~\ref{Appendix B}, we obtain the cosine expansion as 
\begin{eqnarray}
{ \omega } =-\frac{i {\eta_\perp} k^2}{h}+\Omega_1\cos ^2\theta, \quad
-i\rho_\parallel' k^2+\Omega_2\cos^2 \theta.
\label{quartic solution simple}
\end{eqnarray}
The explicit forms of $\Omega_i$ are somewhat involved, and are shown in Appendix.~\ref{Appendix B} together with their derivations.

\begin{figure}
    \centering
    \includegraphics[width=\hsize]{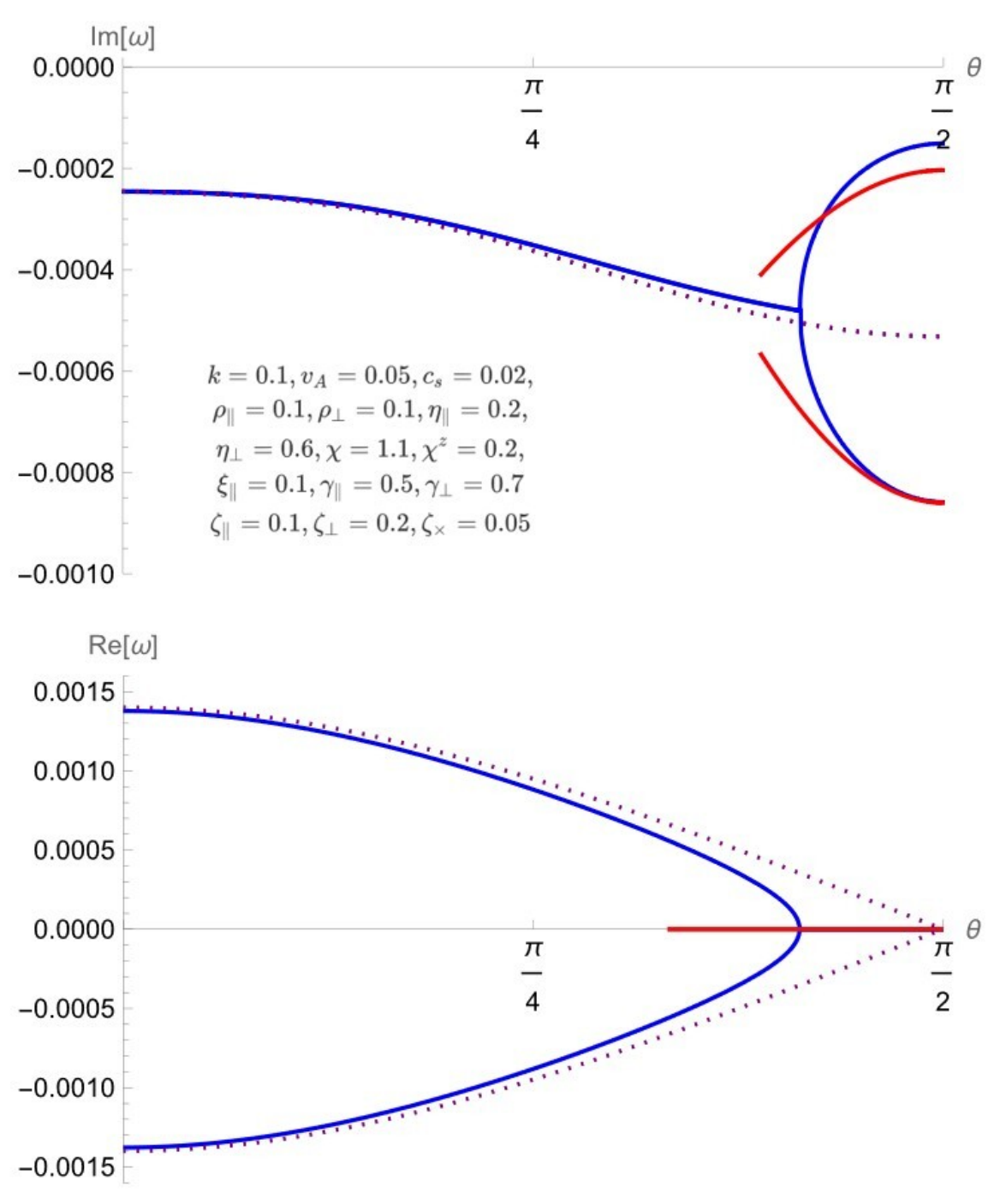}
    \caption{The dispersion relations for the slow magneto-sonic modes. The blue curves represent the numerical solutions of the quartic equation. 
   Dotted curves show the small-$k$ expansion in Eq.~(\ref{quartic NLO solution2}). 
    Red curves show the small-cosine expansion in Eq.~(\ref{quantic solution simple-sonic}). 
    %The deviation at $\theta = \pi/2$ is only due to the terms beyond the $k^2$ order. 
    } 
    %\label{fig:Sonic2}
     \label{fig:spin-MHD-5}
\end{figure}

The same issue also exists in the quintic equation containing the magneto-sonic modes. The limit $\theta\rightarrow \pi/2$ for Eq.~(\ref{quartic NLO solution}) can be readily obtained as
\begin{subequations}
\begin{eqnarray}
\label{final 3}   
\omega_{1,\pm} &\to& \pm v_f k \nnb
&&-\frac{i k^2}{2} \Big( \frac{ v_A^2}{v_f^2} (1 - c_s^2)^2   \rho'_\perp
+ \frac {\eta_\perp + \zeta_\perp} { h}  \Big),
\\
\label{final 4}
\omega_{2,\pm} &\to& -\frac{i k^2 }{2h (1 - v_A^2) }( \tilde \eta_\parallel  + \frac{c_s^2}{v_f^2}  h \rho'_\perp) ,
\\
\label{final damping 2}
\omega &\to& -i\Gamma_\parallel-\frac{i k^2 }{(1-v_A^2) h  } \frac{(\gamma_\parallel +\xi_\parallel )^2}{\gamma_\parallel} ,
\end{eqnarray}
\end{subequations}
where $v_f^2=  c_s^2 + v_A^2 (1 - c_s^2)$. 
However, we notice that the limiting behavior of the two slow magneto-sonic modes (\ref{final 4}) disagrees with Eq.~(\ref{e11}) and the second solution in Eq.~(\ref{o11}) that are obtained without any expansion. 
{
This disagreement also stems from the existence of a non-analytic boundary in the $k$-$\theta$ plane as in the case of the Alfven wave. 
}
%in between $\theta=\frac {\pi}{2},\,\,k = 0$. 
%\com{What is the condition? between sth and sth?}
%That is, the $k$ expansion gets invalid near the non-analytic point. 
The correct result at this limit is obtained in Eq.~(\ref{s4.2}) by taking $\theta=\pi/2$ without the $k$ expansion.

In Fig.~\ref{fig:spin-MHD-5}, we compare the $k$ expansion (\ref{quartic NLO solution2}) to the numerical solutions. 
%obtained by Mathematica without expansion (of which the explicit form is not shown for its complicated form). 
The former and latter are shown by the dotted and blue solid lines, respectively. 
While we find a good agreement between them when $\cos\theta$ is large, they do not at all agree with each other near $\theta=\pi/2$. 
Instead, one can organize the cosine expansion 
\begin{eqnarray}
\omega \= -\frac{i k^2 \tilde \eta_\parallel }{h \left(1- {v_A}^2\right)}+\Omega_1' \cos^2 \theta,\nnb
&&-\frac{i  {c_s}^2 k^2  { \rho'_\perp}}{\left(1-{v_A}^2\right) v_f^2}+\Omega_2 '\cos^2 \theta.
\label{quantic solution simple-sonic}
\end{eqnarray}
The explicit forms of $\Omega_i'$ are shown in Appendix.~\ref{Appendix B}.
These series expressions, shown by the red solid lines, well reproduce the solution in blue. 
{ 
These solutions for $\Omega_i'$ explicitly depend on the rotational viscosities even when the cross rotational viscosity $\xi_\para$ is vanishing. 
Whereas the magneto-sonic modes (\ref{quartic NLO solution2}) depend on the rotational viscosity only via $\tilde \eta_\para$, they get explicit dependences on the rotational viscosities after turning into the purely diffusive modes (\ref{quantic solution simple-sonic}) near $\theta = \pi/2$. 

}

{
These solutions in Eqs.~(\ref{quartic solution simple}) and (\ref{quantic solution simple-sonic}) are expected to be stable as seen in Figs.~\ref{fig:spin-MHD-4} and \ref{fig:spin-MHD-5}. 
According to the solutions in Appendix~\ref{s4.2}, one can easily confirm that all the modes are damping at $\theta=\pi/2$. 
As we decrease $\theta$, the imaginary parts of the solutions approach those of the Alfven and magneto-sonic modes that have the definite signs as shown in Sec.~\ref{s4.3}. 
Therefore, the solutions in Eqs.~(\ref{quartic solution simple}) and (\ref{quantic solution simple-sonic}) are stable (unless there arise non-monotonic behaviors with respect to $\theta$). 

}

\cout{
\subsection{Comparison to MHD and stability in spin MHD}\label{s4.4}
\com{This section may be removed.}
\com{It would be enough to mention the correspondence between $\eta_\para$ and $\tilde \eta_\para$ in the k expansion. 
Is there anything to add more?
Rather, we should discuss effects of spin in the cosine expansion.}

\begin{figure}
    \centering
    \includegraphics[width=\hsize]{IS.pdf}
    \caption{
    The red line is the dispersion relation of IS hydro without external field, while the blue line is the dispersion relation when take relaxation time to $0$, that is, the normal hydro.
    } 
    \label{fig:Normal-IS}
\end{figure}
}

{

\subsection{Extension to the Israel-Stewart equations}

%The causality issue in spin-MHD}

%As shown in Fig. \ref{fig:spin-MHD-4} and Fig. \ref{fig:spin-MHD-5}, non-propagating modes emerge near the vertical limit (i.e., $\theta\rightarrow \frac{\pi}{2}$).In first-order hydrodynamics, such modes typically  lead to violation of causality, as they imply instantaneous signal propagation. This issue is addressed in Israel-Stewart (IS) equation by introducing a finite relaxation time $\tau$. However, even within the IS construction, causality problems persist at the vertical limit.  

In the first-order hydrodynamics, diffusive modes typically lead to violation of causality as diffusion equations imply instantaneous signal propagation. Since a prescription for this issue was provided by Israel and Stewart \cite{Israel:1976tn, Israel:1976efz, Israel:1979wp}, causality has been one of the important topics in relativistic hydrodynamics and is still a hot topic for discussion (see, e.g., Refs.~\cite{Bemfica:2017wps, Bemfica:2019knx, Bemfica:2020zjp, Gavassino:2021owo, Heller:2022ejw, Gavassino:2023myj, Heller:2023jtd} for recent developments). 
%\com{Please add refs if some more.} 

Our first-order solutions, shown in Figs.~\ref{fig:spin-MHD-4} and \ref{fig:spin-MHD-5}, contain non-propagating regimes near the vertical limit, i.e., $\theta\rightarrow \pi/2$. 
%our solutions indicate that the Alfven and sound modes turn into non-propagating modes near the vertical limit (i.e., $\theta\rightarrow \frac{\pi}{2}$). 
We investigate these diffusive modes based on the IS equations by introducing a finite relaxation time $\tau$. 
We will find that, near the vertical limit, there remain diffusive modes in the IS equations in a low-momentum region, while they are modified to become propagating modes in the higher momentum region. Therefore, the IS-modified dispersion relation exhibits a diffusive window in the intermediate momentum region as shown below.

%\com{Some highlights may be added later.}

%However, even within the IS construction, causality problems persist at the vertical limit.  
%We restrict our analysis to the Alfven modes of MHD framework, which proves sufficient for our purposes. Furthermore, we employ the isotropic approximation — assuming that hydrodynamic variables are orientation-independent — since this simplification does not affect the qualitative conclusions regarding causality. 

We focus on the Alfven modes to demonstrate how the diffusive modes, emerging near the vertical limit, are modified in the IS equations. For simplicity, we assume that the transport coefficients take isotropic values, i.e., $\eta := \eta_\parallel=\eta_\perp$ and $ \rho: = \rho_\parallel=\rho_\perp$, 
since this simplification does not affect a conclusion regarding the presence of residual diffusive modes in the IS equations at the qualitative level; We also suppress the spin sector for the same reason. 
%employ the isotropic approximation — assuming that hydrodynamic variables are orientation-independent — since this simplification does not affect the qualitative conclusions regarding causality. 
Then, the IS equations for the Alfven modes read  
\begin{subequations}
    \begin{eqnarray}
    &&
           \partial_0\delta B_y = B \partial_z\delta u_y+\partial_x \delta \P_{xy}-\partial_z \delta \P_{yz}, \\
    &&
		\tau\partial_0\delta \P_{xy}+\delta \P_{xy}= \frac{1}{\mu_m} {\rho }\partial_{x}\delta B_{y}, \\
    &&
		\tau\partial_0\delta \P_{yz}+\delta \P_{yz}= -\frac{1}{\mu_m} {\rho }\partial_{z}\delta B_{y}, \\
    &&
        h{\partial _0 }{\delta u^y } = 
        B \partial_z \delta B_y +\partial_x\delta\pi_{xy} + \partial_z \delta \pi_{yz},\\
    &&
    \tau\partial_0\delta \pi_{xy}+\delta \pi_{xy}= {\eta }\partial_{x} \delta u_{y}, \\
    &&	
    \tau\partial_0\delta \pi_{yz}+\delta \pi_{yz}= {\eta }\partial_{z} \delta u_{y}.
    \end{eqnarray}
\label{eq:IS-eqs}
\end{subequations}
%\com{Are these full equations including $\pd_z^2$?}
%\com{We may choose a better notation for $P_{xy}, \ P_{yz}$.}
where we denote the dynamical variables in the IS equations as $\pi_{ij}$ 
%$, \Pi$, 
and $\P_{ij}$ and the corresponding relaxation times as $\tau_R$ and $\tau_B$, respectively. 
We take both relaxation times to be the same value $\tau$. 
%And we define $\eta_\parallel=\eta_\perp=\eta,\xi_\parallel=\xi_\perp=\eta$ since we considered the isotropic limit. 
When $\tau\rightarrow 0$, these equations go back the first-order equations (\ref{f2}) discussed in the previous sections (without spin). 

%to Alfven equations Eq.~\ref{f1} without spin. }
%\com{Why do you mention an equation in the ref? Alfven modes should be seen in the previous sections in this manuscript. If not any reason, be self-contained.}

We first provide analytic solutions at the vertical limit, which are obtained as 
\begin{align}
\label{eq:IS-solutions}
    \omega = -\frac{ih\pm\sqrt{4 \eta  h k^2 \tau -h^2}}{2 h \tau }
    \, , \quad 
    -\frac{i\mu_m \pm \sqrt{4 \mu_m \rho k^2 \tau - \mu_m^2}}{2 \mu_m \tau } \, .
\end{align}
%\gray{\begin{align}
%\nn%\label{eq:IS-solutions}
 %   \omega = -\frac{ih\pm\sqrt{4 \eta  h k^2 \tau -h^2}}{2 h \tau }
%    \, , \quad 
%    -\frac{ih\pm\sqrt{4 \rho  h k^2 \tau -h^2}}{2 h \tau } \, .
%\end{align}
%}
The other two equations for $\delta \P_{yz}$ and $\delta \pi_{yz}$ are decoupled in this limit. 
Expanding these solutions for a small $k$, one finds two diffusive modes and two gapped modes 
\begin{subequations}
\begin{eqnarray}
    \omega \= -i \frac{\eta}{  h}k^2  \, , \quad 
    -\frac{i}{\tau} + i\frac{\eta}{ h} k^2 \, ,  \\
    \omega \= -i \frac{\rho}{\mu_m}k^2  \, ,  \quad 
    -\frac{i}{\tau} + i\frac{\rho}{\mu_m} k^2 \, . \label{eq:IS-small-k}
\end{eqnarray}
\end{subequations}
Two of these modes, which depend on $\tau$, are gapped modes introduced by the IS relaxation equations. 
The other two gapless diffusive modes are independent of the relaxation time at the leading order in $k$ and reduce to the Alfven modes in the first-order solutions in Fig.~\ref{fig:spin-MHD-4} when $\tau\to0$; Thus, these modes are identified as the ``IS-modified Alfven modes.'' 
In Panel 1 of Fig.~\ref{fig:IS-dispersions}, blue curves show the IS-modified Alfven modes, while orange curves show the the Alfven modes in the first-order solutions at $\tau=0$. 
%, which are characterized by $\eta$ and $\rho$ separately. 
We take $\eta h^{-3/4} =0.2$, $\rho h^{1/4} =0.2$, $\tau h^{1/4} = 1$, and $v_A = 0.7$, where dimensionful quantities are rescaled by $h$ in Eq.~(\ref{eq:enthalpy}).  

The small-$k$ expansion (\ref{eq:IS-small-k}) shows that the IS-modified Alfven modes remain diffusive modes at the vertical limit even after the IS relaxation equations are introduced. 
This behavior is in clear contrast to the typical behavior of dispersion relations from the IS equations where diffusive modes in the first-order equations are modified to become propagating modes in a small momentum region down to vanishing $k$ (see, e.g., Ref.~\cite{Denicol:2008ha}). 
As we increase $k$, the square-root arguments in the full solution~(\ref{eq:IS-solutions}) turn into positive values at the critical momenta 
\begin{eqnarray}
    k_c^{\rm IS} 
    = \sqrt{ \frac{h}{ 4 \eta \tau} }
    \, , \quad 
    \sqrt{ \frac{\mu_m}{ 4 \rho \tau} } 
    \, .
    \label{eq:IS-critical-mom}
\end{eqnarray}
At each of these critical momenta, a diffusive mode merges with one of the gapped relaxation modes to form a pair of propagating modes as shown with the blue curves in the same figure (Panel 1 of Fig.~\ref{fig:IS-dispersions}).

\begin{widetext}

\vspace{1cm} %%%%%%%%%%%%%

\begin{figure}
%\vspace{-1cm}
\begin{center}
   \includegraphics[width=0.9\hsize]{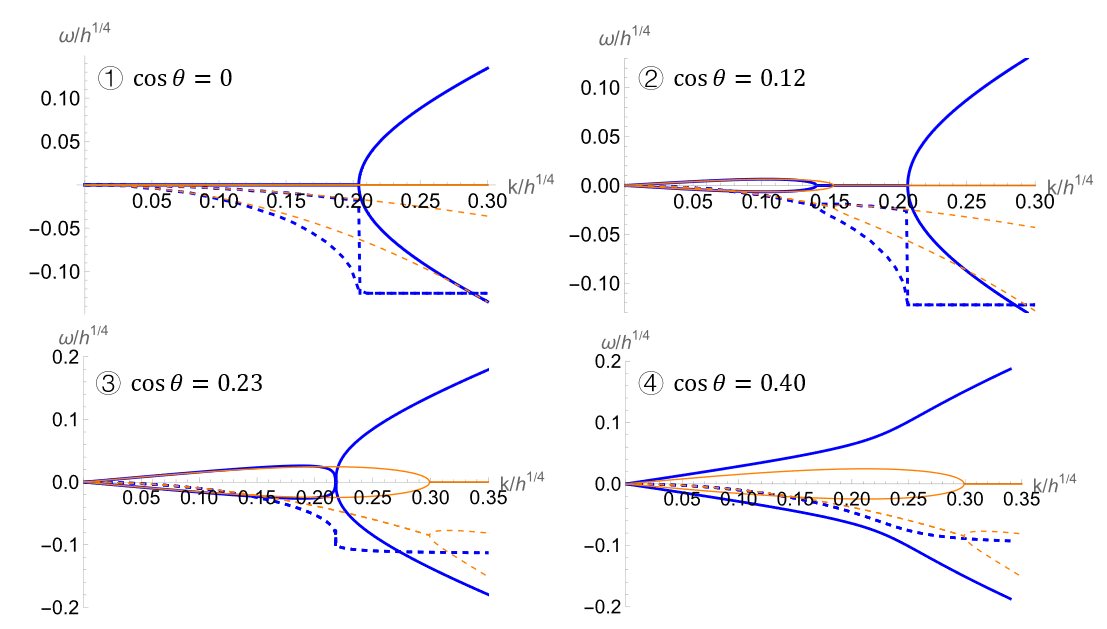}
\end{center}
\vspace{-0.5cm}
\caption{
{
Dispersion relations of the IS-modified Alfven modes (blue curves) and the first-order Alfven modes (orange curves) at fixed angles. 
Solid and dotted curves show the real and imaginary parts of the dispersion relations. 
See also Fig.~\ref{fig:IS-phase} for general angles. 
%Real parts of the dispersion relations from the IS equations at the angles indicated in the panels. 
%\com{What are other parameter values such as the transport coefficients?}
}
}
  \label{fig:IS-dispersions}
\end{figure}

\end{widetext}

\begin{figure}[t]
%\vspace{-1cm}
\begin{center}
   \includegraphics[width=\hsize]{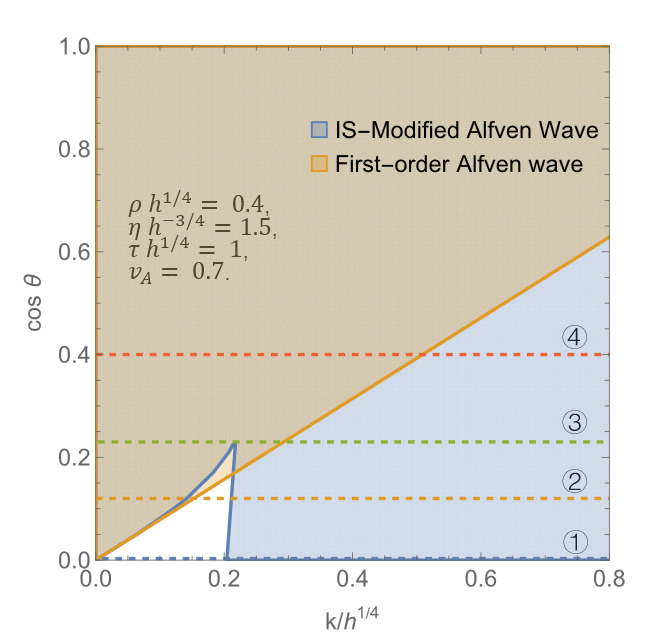}
\end{center}
\vspace{-0.5cm}
\caption{
{
Unshaded and shaded regions show diffusive and propagating regimes, respectively, in the IS-modified Alfven modes (blue shade) and the first-order Alfven modes (orange shade). 
%The vertical limit corresponds to $\theta=\pi/2$ at the bottom. 
The IS-modified Alfven modes retain a diffusive window in the unshaded ``shark-fin'' region. 
}
}
  \label{fig:IS-phase}
\end{figure}

On top of the above observations at the vertical limit, we examine the angle dependence of the dispersion relations. In Fig.~\ref{fig:IS-phase}, diffusive and propagating regimes are shown by the unshaded and shaded regions, respectively, as a phase diagram with respect to $\theta$ and $k$. The numbered cross-sections at different angles correspond to the dispersion relations shown in Fig.~\ref{fig:IS-dispersions}. 
We take the same parameters in Figs.~\ref{fig:IS-dispersions} and \ref{fig:IS-phase}.

First, let us examine the low-momentum regions. In case of the first-order MHD discussed in the previous sections, we have the propagating Alfven waves as we decrease the angle from the vertical limit, which corresponds to the region shaded in dark orange. 
However, the small-momentum expansion breaks down at the critical momentum $k_c (\theta)$ discussed in Sec.~\ref{sec:critical-mom}; The Alfven modes in the first-order MHD become diffusive above $k_c (\theta)$. 
As shown in Fig.~\ref{fig:IS-phase}, 
the IS-modified Alfven modes basically succeed these properties, though 
%the Alfven modes remain gapless propagating modes (as long as $\theta \not=\pi/2$) and become diffusive modes above the critical momentum $k_c (\theta)$. 
the critical momentum is slightly modified as we find a slight deviation between the orange and blue boundaries. These propagating modes are also confirmed in the low-momentum regions in Panels 2, 3, and 4 of Fig.~\ref{fig:IS-dispersions}.

In the high-momentum region, diffusive modes are modified by the relaxation dynamics introduced in the IS equations in general. Indeed, as already discussed above at the vertical limit, the diffusive regime in the first-order solutions turns into the propagating regime. This modification is found in all the angles in Fig.~\ref{fig:IS-dispersions} in the high-momentum regions. Recall, however, that this modification is effective only above the other critical momentum $ k_c^{\rm IS}$ defined in Eq.~(\ref{eq:IS-critical-mom}) that is controlled by the relaxation time $\tau$ and depends on $\theta$ in general.

It is, therefore, interesting to focus on a momentum window defined by the two critical momenta, $k_c (\theta)\leq k \leq k_c^{\rm IS}(\theta)$. 
Since the critical momenta depend on the angle $\theta$, the window is closed when $k_c (\theta) = k_c^{\rm IS}(\theta)$. 
At this angle, the low- and high-momentum propagating regimes touch each other (see Panel 3 in Fig.~\ref{fig:IS-dispersions}). 
Above this angle (closer to the vertical limit), a diffusive window persists even after a finite relaxation time is introduced with the IS equations. This regime corresponds to the unshaded ``shark-fin'' region in Fig.~\ref{fig:IS-phase}. 
%On the other hand, below this angle (closer to the parallel direction to a magnetic field), there is a pair of gapless propagating modes in the whole momentum region as shown in Panel 4 of Fig.~\ref{fig:IS-dispersions}, which are the modified Alfven waves. 
The presence of the diffusive window in an intermediate momentum region deserves further study in future works. 

}

\section{conclusion}\label{s5}
In this paper, we formulated the first-order spin MHD. 
The magnetic field breaks the rotational symmetry, and gives rise to anisotropic spin damping rates in terms of the rotational viscosities in the antisymmetric part of the energy-momentum tensor (\ref{con12}). 
The energy-momentum tensor also acquires a novel cross term that converts a vorticity into the symmetric stress and a shear into the antisymmetric torque. 
This effect arises only in the presence of a zeroth-order vector in derivative, that is, a magnetic field in case of spin MHD and a vorticity in case of gryohydrodynamics \cite{Cao:2022aku}.

We then performed the linear-mode analysis to get the nine dispersion relations in Sec.~\ref{s3}. 
In the absence of spin degrees of freedom, 
three pairs among them are known as the Alfven waves and the fast and slow magneto-sonic waves. 
We found that the presence of spin damping rates modifies the shear viscosity in a particular combination (\ref{eq:eta-tilde}). 
There are three remaining modes in the totally antisymmetric pseudogauge, which are damping spin modes. 
They are quasihydrodynamic modes that have nonvanishing damping rates even at a vanishing momentum. 
That is, spin is not a conserved quantity by itself and is converted into the orbital angular momentum until the spin potential and the thermal vorticity match each other. 
We also proved the linear stability of these modes in the rest frame by extending our previous analysis for MHD \cite{Fang:2024skm}.

We stress that the above analytic solutions are obtained for an arbitrary angle $\theta$ in the small-momentum expansion. 
However, these solutions indicate that the small-momentum expansion is spoiled due to the blow up of the higher-order terms when the angle $\theta$ approaches the right angle where the linear waves propagate nearly perpendicular to an equilibrium magnetic field. 
We showed that there is a critical angle where the two competing expansion parameters, the momentum $k$ and $\cos \theta$, have similar magnitudes. 
Since the critical angle depends on the momentum, the solution has a non-analytic boundary in the $k$-$\theta$ plane. 
{This boundary defines the critical momentum $k_c(\theta)$ when we consider the dispersion relations as functions of $k$ at a fixed $\theta$.}
On the other side of the boundary, we provided the alternative solution in the small-cosine expansion. 
We confirmed that the momentum and cosine expansions well reproduce the numerical solutions in the appropriate domains separated by the boundary. 
We point out that this non-analytic structure is a prevalent issue in anisotropic systems.

{
We also included the relaxation dynamics for the electric resistivity and shear viscosity with the IS framework. 
These higher-order corrections modify the dispersion relations in a high momentum region. 
In particular, the purely diffusive regime, which was found in the first-order solutions above the critical momentum $k_c(\theta)$, turns into a propagating regime due to the relaxation dynamics. 
The boundary between the diffusive and propagating regimes defines another critical momentum $k_c^{\rm IS} (\theta)$. 
We found that there remains a diffusive window between the lower critical momentum $k_c$ and the upper critical momentum $k_c^{\rm IS}$ as shown in Fig.~\ref{fig:IS-phase}. 
%As we reduce the angle $\theta$, the diffusive window is closed. 
%with $k_c = k_c^{\rm IS}$ at an angle $\theta$. 
}

We would like to give a future outlook. 
To address the splitting between $\Lambda$ and $\bar \Lambda$ spin polarization in the framework of spin MHD, it is important to include a finite chemical potential since a net spin polarization is not induced in a charge neutral fluid. 
Then, one may need to include a finite electric current and an electric field accordingly. 
They are quasihydrodynamic modes, and require an extended framework. 
Also, it is interesting to investigate the spin relaxation to a nonzero thermal vorticity. 
One needs to include a zeroth-order vorticity into the constitutive equations (cf. Ref.~\cite{Cao:2022aku}) and perform the linear-mode analysis near a vortical equilibrium configuration. 
Last but not least, extensions to the Israel-Stewart type frameworks will be useful for numerical implementation toward applications to relativistic-heavy ion collisions.

\begin{acknowledgements}
The authors thank Yi-hui Tu, Xu-Guang Huang, and Yi Yin for useful comments. 
This work is  supported by the NSFC under grant No.12505149, and the JSPS  KAKENHI under grant No.20K03948 and No.22H01216. 
\end{acknowledgements}

%\bibliographystyle{plain}
%\bibliography{ref}

%\newpage

\appendix

\section{Solutions at two specific angles}\label{Appendix A}

In this appendix, we focus on the solutions at two specific angles at $k_\perp=0$ and $k_\para=0$. 

\subsection{Eigenmodes at $k_\perp = 0$}\label{s4.1}

When $k_\perp = 0$, the Gauss law constraint in Eq.~(\ref{don26}) indicates that $\delta B_z = - k_\perp \delta B_x/k_\para = 0$, so that 
one can get rid of the last equation in Eq.~(\ref{don26}). 
Notice also that the linearized equations are further factorized into block-diagonal forms: 
\begin{widetext}
\begin{subequations}
    \begin{eqnarray}
0 \=  
\left( 
  \begin{array}{ccc|ccc}
   B   k_\para &   \omega + i  \rho_\perp'  k_\para^2  & 0 &  &  
  %&0 
\\
  h \omega + i (\eta_\para  -2\xi_\para+\gam_\para ) k_\para^2& 
\frac{  h}{  B}  v_A^2 k_\para   &  \frac{4}{\chi}(\gam_\para - \xi_\para) k_\para  &  & 
%& 0 
\\
 -2 ( \gam_\para-\xi_\para)k_\para &   0 & 
\omega + i \Gam_\para &  &  
%&  0
\\ \hline
 &   &  &\omega & & -   h  (1- v_A^2) k_\para 
%&  \frac{  h}{  B} v_A^2 \omega 
\\
  &    &   & - c_s^2 k_\para & &
 h  (1- v_A^2) \omega  + i \zeta_\para k_\para^2 
%& \gr{ 0 }
%\\
% 0  & 0& 0& 0  & 0  & \omega  
\end{array}
\right)
\begin{pmatrix}
\delta u_x \\ \delta B_x  \\ \delta \sigma_y \\
\delta \epsilon \\  \delta u_z 
%\\  \delta B_\para 
\end{pmatrix} 
,
\\
0 \= 
\left(
  \begin{array}{ccc|c}
  B k_\para & \omega + i   \rho_\perp' k_\para^2   &0 & 
\\
  h \omega +i( \eta_\para - 2 \xi_\para +\gam_\para ) k_\para^2   
& \frac{  h}{  B} v_A^2 k_\para  & 
- \frac{4}{\chi} (\gam_\para - \xi_\para)  k_\para  & 
\\
2  (\gam_\para - \xi_\para) k_\para  & 0 & 
\omega +  i \Gam_\para & 
\\ \hline
 &  &  & \omega +  i \Gam_\perp
\end{array}
\right)
\begin{pmatrix}
\delta u_y \\   \delta B_y \\ \delta \sigma_x \\ \delta \sigma_z
\end{pmatrix} 
\, .
\end{eqnarray}
\end{subequations}
\end{widetext} 
%Then, the secular equations are obtaiend as 
%\begin{subequations}
%    \begin{eqnarray}
%&& \big[ \big]^2 = 0
%\\
%&& \omega^2 + i \zeta'_\para k_\para^2  %\omega - c_s^2  k_\para^2 =0 ,
%\\
%&& \omega + 8 i \Gam_\perp = 0
%.
%    \end{eqnarray}
%\end{subequations}
By solving the secular equations, 
one can straightforwardly obtain the dispersion relations of the eigenmodes. 
One will find degenerate eigenmodes for $\delta u_x,\delta B_x,\delta\sigma_y$ and $\delta u_y,\delta B_y,\delta\sigma_x$,
because of the rotational symmetry around the direction of a magnetic field. 
The degeneracy manifests itself in the determinants for the upper block-diagonal parts of the two matrices that result in the same cubic equations.\footnote{Whereas there are sign differences between the coefficients in front of $\delta \sigma_x$ and $\delta \sigma_y$, they cancel out in the determinants as immediately confirmed with the Laplace expansion.} 
Up to the order of $k_\para^2$, the three dispersion relations are obtained  as  
  \begin{eqnarray}
\omega \=  -  i \Gam_\para 
- i \frac{(\gam_\para - \xi_\para)^2}{  h \gam_\para} k_\para^2  , \quad 
\nnb
&&
\pm v_A k_\para 
-  \frac{i}{2} \Big(  \rho'_\perp  
+ \frac{ \eta_\para \gam_\para - \xi_\para^2 }{  h \gam_\para}  \Big) k_\para^2 .\label{solution 35}
\end{eqnarray} 
%These are six modes including the degeneracy. 
Notice that these modes are always damped out due to the first-order transport coefficients that satisfy the thermodynamic constraints (\ref{con20}). 
The spin mode is a gapped and intrinsically damping mode at the hydrodynamic limit $k_\para \to 0$, because of $\Gamma_\para$ from the non-vanishing right-hand side in Eq.~(\ref{eq:spin-cons}). The other two modes propagate in opposite directions with the Alfven-wave velocity $v_A$, while they are damped out due to the dissipative transport coefficients. 
All those modes have transverse polarizations with respect to the momentum. 
%{When we ignore the new transport coefficient $\xi_\para$, the Alfven modes reduce to the results without spin (see, e.g., Refs.~\cite{Grozdanov:2016tdf}), regardless of the value of $\gamma_\parallel$. }

The remaining blocks in the above two matrices yield three dispersion relations 
\begin{subequations}  \label{36}
  \begin{eqnarray} 
  \label{36a}
\omega \= -  i \Gam_\perp ,
\\
%\= - \frac{i}{2} \zeta'_\para k_\para^2\pm \frac12 \sqrt{ 4 c_s^2  k_\para^2 - ( \zeta'_\para k_\para^2 )^2  }
%\\
\label{36b}
\omega \= \pm c_s k_\para 
- \frac{i}{2} \zeta'_\para k_\para^2 
,
\end{eqnarray}  
\end{subequations} 
where $\zeta'_\para = \frac{  \zeta_\para}{  h (1-v_A^2) }$. 
Again, the spin mode is damped out in the hydrodynamic limit with $\Gam_\perp$ from the non-vanishing right-hand side in Eq.~(\ref{eq:spin-cons}). 
The other two modes are sound waves that are damped out due to one of the bulk viscosities. 
All these modes have longitudinal polarizations with respect to the momentum.

\cout{

As we focus on Alfven mode, we apply the perturbation $\delta {u_ \bot },\delta {B_ \bot },\delta {\sigma _ \bot }$, which rely on $t,z$. And the equations can be separated into two parts:

\begin{equation}
    \begin{array}{l}
0=-ih{\omega}{\delta u^x } -i\frac{B}{\mu_m} {k_z }\delta {B^x} + (\eta_\parallel+\gamma_\parallel-2\xi_\parallel)k_z^2\delta u_x\\+\frac {4i}{\chi}(-\gamma_\parallel+\xi_\parallel)k_z\delta\sigma^{y}\\
0 = i\omega\delta B_x+ik_z   B\delta u_x-\frac{\rho_\perp}{\mu_m}k_z^2 \delta B_x\\
0=2i(\xi_\parallel-\gamma_\parallel)k_z\delta u_x{-}(\frac{8\gamma_\parallel}{\chi}-i\omega)\delta\sigma_y.
\end{array}\label{e01}
\end{equation}

\begin{equation}
    \begin{array}{l}
0=-ih{\omega}{\delta u^y } -i\frac{B}{\mu_m} {k_z }\delta {B^y}+(\eta_\parallel+\gamma_\parallel-2\xi_\parallel)k_z^2\delta u_y\\
-\frac{4i}{\chi}(-\gamma_\parallel+\xi_\parallel)k_z\delta\sigma^{x}\\
0 = i\omega\delta B_y+ik_z   B\delta u_y-\frac{\rho_\perp}{\mu_m}k_z^2 \delta B_y\\
0=2i(\xi_\parallel-\gamma_\parallel)k_z\delta u_y{+}(\frac{8\gamma_\parallel}{\chi}-i\omega)\delta\sigma^x
\end{array}\label{e02}
\end{equation}

One can notice that equations Eq.~(\ref{e02}) and Eq.~(\ref{e01}) are symmetry, i.e. the solution in x and y direction is degenerate. In \ref{e01}, we can get solution of $\omega$:
\begin{equation}
    \begin{array}{l}
\omega_\pm=\pm \frac{B}{\sqrt{h\mu_m}}k_z-\frac{i}{2}(\frac{\gamma_\parallel\eta_\parallel-\xi_\parallel^2}{h\gamma_\parallel}+\frac{\rho_\perp}{\mu_m}) k_z^2\\
\omega_0= -i\frac{8\gamma_\parallel}{\chi}+\frac{i}{h}\frac{{(\xi _\parallel  - \gamma _\parallel )^2}}{{{\gamma _\parallel }}}k_z^2.
\end{array}\label{e03}
\end{equation}	  

}

\subsection{Eigenmodes at $k_\para=0$}\label{s4.2}

Next, we discuss the case at $k_\para=0$. 
In this case, we have $\delta B_x = - k_\para \delta B_z/k_\perp =0$. 
The remaining non-vanishing elements of the matrices are block-diagonalized as 
\begin{widetext}
    \begin{subequations}
        \begin{eqnarray}
0\= 
\left( 
  \begin{array}{ccc|ccc}
-  i \frac{ c_s^2 \rho_\perp' }{  1 -v_A^2 } 
\frac{  B}{h}  k_\perp^2 & - k_\perp   B   & \omega + i\rho'_\perp k_\perp^2   & &  
\\
\omega & -  h k_\perp  & \frac{ h}{  B} v_A^2 \omega  &  & 
\\
- c_s^2 k_\perp &   h \omega 
+i  (\zeta_\perp + \eta_\perp  ) k_\perp^2 
 & - \frac{  h}{  B}  v_A^2  k_\perp &    
\\ \hline
 &  & &  h  ( 1- v_A^2 ) \omega 
+ i(\eta_\para + 2 \xi_\para + \gam_\para ) k_\perp^2 & &
- \frac{4}{\chi}(\xi_\para + \gam_\para) k_\perp 
\\
 &  &   &   2( \gam_\para + \xi_\para ) k_\perp & & \omega +  i \Gam_\para   
%\\
% &  &  &   &   &  \omega 
\end{array}
\right)
\begin{pmatrix}
\delta \epsilon \\ \delta u_x \\ \delta B_z  \\  
\delta u_z  \\ \delta \sigma_y %\\ \delta B_x
\end{pmatrix}
,
\nnb\\
0\= 
\left(  
\begin{array}{cc|cc}
  h \omega +i (\eta_\perp  + \gam_\perp) k_\perp^2 &  \frac{4}{\chi} \gam_\perp  k_\perp  &  & 
\\
-  2 \gam_\perp k_\perp & \omega + i  \Gam_\perp &  & 
\\ \hline
 &  & \omega + i \rho_\para '  k_\perp^2 & 
\\
 &  &  & \omega +i  \Gam_\para
\end{array}
\right)
\begin{pmatrix}
\delta u_y  \\ \delta \sigma_z \\   \delta B_y \\ \delta \sigma_x
\end{pmatrix} 
.\label{quartic half pi}
        \end{eqnarray}
    \end{subequations}
\end{widetext}
The $3\times3$ block-diagonal part yields the three eigenmodes 
\begin{eqnarray}
 \label{e11}
\omega \= 
 -i\frac {c_s^2} {v_f^2}\frac { \rho'_\perp} {1-v_A^2}k_\perp^2
% -i\frac {c_s^2} {v_f^2}\frac {h \rho_\perp} {\mu_m h - B^2}k_\perp^2
,
\\
   && 
\label{e22}
\pm v_f k_\perp - 
\frac{i}{2} \Big( \frac{ v_A^2}{v_f^2} (1 - c_s^2)^2   \rho'_\perp
+ \frac {\eta_\perp + \zeta_\perp} { h}  \Big)k_\perp^2
%   \pm v_A k_\perp - 
%\frac{i}{2} \Big(\frac {B^2} 
%{\mu_m}\frac {(1 - c_s^2)^2} { h v_f^2}  \rho'_\perp
%+ \frac {\eta_\perp + \zeta_\perp} { h}  \Big)k_\perp^2 
, 
\end{eqnarray} 
where $v_f^2=  c_s^2 + v_A^2 (1 - c_s^2)$. 
The two propagating modes are the fast magneto-sonic waves with the velocity $v_f$. 
While there are two fast and two slow magneto-sonic waves in a general momentum direction, 
the two slow waves vanish when $k_\para=0$ (see, e.g., Refs.~\cite{biskamp1997nonlinear, 
Hattori:2017usa, Biswas:2020rps}). 
The remaining blocks yield six modes 
\begin{subequations}\label{eq:sol-pi/2}
\begin{eqnarray}  
\label{o11}
\omega 
\= - i \Gam_\para
- i \frac{(\gam_\para + \xi_\para)^2}{  h (1 -v_A^2) \gam_\para } k_\perp^2  
 ,  \nnb
 && 
- i \frac{ \tilde \eta_\para  }
{ h (1 -v_A^2) } k_\perp^2 ,
\\
\label{o22}
\omega 
\= - i \Gam_\perp 
- i \frac{ \gam_\perp}{   h } k_\perp^2 
,  \, \, 
- i\frac{ \eta_\perp}{  h } k_\perp^2 ,
\\ 
\label{o33}
 \omega\= - i \rho_\para '  k_\perp^2 ,
\\
\label{o44}
 \omega \= -  i  \Gam_\para
 ,
\end{eqnarray}
\end{subequations}
where $\tilde \eta_\para$ is defined in Eq.~(\ref{eq:eta-tilde}). 
The first four modes stem from the mixing between the spin and shear modes. 
According to the thermodynamic constraints (\ref{con20}), all the imaginary parts are negative, indicating that all those modes are purely diffusive modes.

%\gray{In the general solutions, the counter-propagating modes reduce to a pair of purely diffusive modes when $k_\para \to 0$. In this limit, the diffusive modes should have the same damping rates because the counter-propagating modes at an infinitesimal $k_\para $ had the same imaginary parts. However, this degeneracy in the imaginary part is not correctly reproduced in the above solutions because of the non-commutative orders in the diagonalization and the limit. }

\cout{

As we have discussed in \ref{s3.2}, the equations of magnetic field should be:
\begin{equation}
\begin{array}{l}
0 = i(\omega\delta B_y+k_z   B\delta u_y) - \frac{2{  \rho_ \parallel }  T}{\mu_m}k_x^2 \delta (\beta {B_y}) \\
0 = i(\omega\delta B_z-k_x   B\delta u^x) + \frac{2{  \rho_ \perp }  T}{\mu_m}k_x^2\delta (\beta {B_z}),\label{e04}
\end{array}
\end{equation}	

where we have take $k_y=0$ without lose of generality. As we focus on magnetic-sonic wave, we focus on the perturbations $\delta \epsilon,\delta {u^x},\delta B^z$: 

\begin{equation}
\begin{array}{l}
0 =  - \omega \delta \epsilon  - \omega \frac{1}{{{\mu _m}}}  B\delta B_z + (  e +  {  p_ \perp })({k_x}\delta {u^x} ), \\
        0=-ih{\omega }{\delta u^x } + ic_s^2{k_x }{\delta \epsilon } +i\frac{B}{\mu_m}k_x\delta B^z\\
        +(\zeta_\perp+\eta_\perp)k_x^2\delta u_x\\
0 =\frac{2  B c_s^2   \rho_\perp}{(  e+  p_\parallel)\mu_m}k_x^2\delta \epsilon-i k_x B\delta u^x\\
+ [i\omega-\frac{2  \rho_\perp}{\mu_m}k_x^2]\delta B_z. \\
\end{array}\label{e07}
\end{equation}

The solution of these equations are:
\begin{equation}
    \begin{array}{l}
      \omega_ {\pm} = \pm v_f k_x - 
  i (\frac {\eta_\perp + \zeta_\perp} {2 h} + \frac {B^2} 
{\mu_m^2}\frac {(1 - c_s^2)^2 \rho_\perp} {2 h v_f^2})k_x^2
   \\
   \omega_0=-i\frac {c_s^2} {v_f^2}\frac {h 
\rho_\perp} {\mu_m h - B^2}k_x^2
    \end{array}\label{e11}
\end{equation}

Where $v_f^2=\frac {1} {h} (hc_s^2 + (1 - c_s^2)\frac {B^2} {\mu_m})$. We can see there’re two magnetic-sonic wave in the solution. In general case, there should be 4, but the slow wave vanishes when $k_z=0$ [\cite{Biswas:2020rps}]. 

}

\section{Explicit forms of the cosine expansion}\label{Appendix B}

{
Here, starting from the solutions in the limit of $\theta=\pi/ 2$, we provide the explicit forms of the cosine expansion. 
For the quartic equation (\ref{quartic half pi}), we have already got the solution in Eq.~(\ref{eq:sol-pi/2}) as 
}
\begin{eqnarray}
{\tilde\omega_1} \=-\frac{i {\eta_\perp} k^2}{h}, 
\nnb
{\tilde\omega_2} \=-i\rho_\parallel' k^2  , 
\nnb
{\tilde\omega_3} \=-i\Gamma_\perp-\frac{i {\gamma_\perp} k^2}{h} , 
\nnb
{\tilde\omega_4} \=-i\Gamma_\parallel.  
\end{eqnarray}
The determinant of coefficient matrix $M_1$ should be factorized as
\begin{eqnarray}
\det&&(M_1)\propto \\
&&(\omega-\tilde\omega_1-\omega_1\cos^2\theta)(\omega-\tilde\omega_2-\omega_2\cos^2\theta)
\nnb
&&\times(\omega-\tilde\omega_3-\omega_3\cos^2\theta)(\omega-\tilde\omega_4-\omega_4\cos^2\theta)
 . \nn
\end{eqnarray}
%Denote the coefficient of order $\omega^i \cos^2\theta$ as $X_{i}$. 
By contracting the terms order-by-order in the cosine factor, we conclude that
\begin{eqnarray}
{ \omega _i}= \tilde\omega_i-\frac{ \left( \tilde\omega_i^3  {X_{3}}+ \tilde\omega_i^2  {X_{2}}+ \tilde\omega_i  {X_{1}}+ {X_{0}}\right)}{\prod_{j\neq i}( {\tilde\omega_i}- {\tilde\omega_j}) }\cos ^2 \theta. \label{cos quartic}
\end{eqnarray}
The coefficients $X_{i}$ are obtained as
\begin{subequations}
\begin{eqnarray}
X_{3} \=\frac{k^2}{h}  { \Gamma _ \perp} \Gamma_\parallel 
\Big(  h  {v_A}^2+k^2 \big(  \tilde \eta_\parallel  \rho_\parallel' +   { \eta   _ \perp} ( { \rho _ \perp'}-2 \rho_\parallel' ) \big)\Big),  
\\
X_{2} \=-\frac{ik^2}{h} [(\Gamma _\parallel +\Gamma _ \perp) h {v_A}^2
-\Gamma_\parallel   { \Gamma _ \perp} \big(-\tilde \eta_\parallel + { \eta _ \perp}+h \rho_\parallel' -h  { \rho _ \perp'}\big)]
\nnb
&-&\frac{ik^4}{h} \Big[ 
  \Gamma_\parallel\Big(  \rho_\parallel'  \tilde \eta_\parallel+  ({ \rho   _ \perp'}-2\rho_\para') ( { \gamma   _ \perp}+ { \eta   _ \perp})\Big) 
\nnb
&+&\Gamma _ \perp  \big(\rho_\parallel'  (\gamma_\parallel +\eta_\parallel -2  { \eta _ \perp}-2\xi_\parallel )+ { \eta _ \perp}  { \rho _ \perp'}\big) \Big],  
\\
X_{1} \= \frac{k^2}{h}\Big[-h  {v_A}^2
+\Gamma _\parallel \big(  { \gamma _ \perp}-\tilde \eta_\parallel + { \eta _ \perp}+h \rho_\parallel' -h  { \rho _ \perp'}\big)
\nnb
&-&{ \Gamma _ \perp} (\gamma_\parallel +\eta_\parallel - { \eta _ \perp}-h \rho_\parallel' +h  { \rho _ \perp'}-2 \xi_\parallel )\Big]
\\
&-& \frac{k^4}{h} \Big[\Big(\rho_\parallel'  (\gamma_\parallel +\eta_\parallel -2 \xi_\parallel )+ ({ \rho _ \perp'}-2\rho_\para') ( { \gamma _ \perp}+ { \eta  _ \perp})\Big)
\Big],  
\nnb
X_{0}\= -\frac{i k^2}{h}\big({ \gamma _ \perp}-\gamma_\parallel -\eta_\parallel + { \eta _ \perp}+2 \xi_\parallel
+h (\rho_\parallel' -  { \rho _ \perp'})  \big).\nnb
\end{eqnarray}
\end{subequations}
As already discussed, the $\order (k^2)$ terms are outside of our interest, as they are beyond the first-order spin MHD. Expanding Eq.~(\ref{cos quartic}) up to the $k^2$ order, one gets the parameters $\Omega_i$ for Eq.~(\ref{quartic solution simple}) as
\begin{eqnarray}
\Omega_1 \= \frac{i h {v_A^2}}{{\eta_\perp}-h \rho_\parallel' }\nnb
&+&i k^2 \left(\frac 1 h({{\eta _\perp}-\tilde \eta_\parallel )}+\frac{{v_A^2} {\chi}}{8 h \rho_\parallel' -8 {\eta_\perp}}\right),
\\
\Omega_2 \= -\frac{i h v_A^2}{{\eta_\perp}-h \rho_\parallel' }\nnb
&+&  i k^2 \left(\rho_\parallel' - {\rho_\perp'}-\frac{{v_A^2} {\chi}}{8 h \rho_\parallel' -8 {\eta_\perp}}\right).
\end{eqnarray}

In the same way, we obtain $\Omega_i$ for Eq.~(\ref{quantic solution simple-sonic}): 
\begin{eqnarray}
\Omega_1' \= \frac{i    {c_s}^2 h {v_A}^2 \left(1-{v_A}^2\right)}{  {c_s}^2 h {\rho _\perp}-{v_f}^2  \tilde  \eta_\parallel  }\nnb
&+&\frac{i k^2}{8 \gamma_\parallel^2  h^3 \left(1-{v_A}^2\right) {v_f}^2 ({v_f}^2 \tilde  \eta_\parallel - {c_s}^2 h {\rho_\perp} )}
\nnb
&&
\Big[{c_s}^2 {v_A}^2 \Big(2 \Big(\frac{{\rho _\perp} }{{v_f}^2}  \gamma_\parallel  \left(1-{c_s}^2\right)^2 h^2 {v_A}^2 \left(1-{v_A}^2 \right) \nnb
&-& 2 h \gamma_\parallel  \tilde  \eta_\parallel +\gamma_\parallel  h \left(1-{v_A}^2\right) ({\zeta _\perp}+{\eta _\perp})\Big)^2
\nnb
&+&
h^3 \left(1-{v_A}^2\right) {v_f}^2 \chi  \left( (\gamma_\parallel + \xi_\para)^2 -\gamma_\parallel  \tilde \eta_\parallel \right)\Big)
\nnb
&-&h^2 {v_f}^2 \Big(8 \gamma_\parallel ^2 {c_s}^2 h {\rho _\perp} \Big(\zeta_\parallel -\left(4-3 {v_A}^2\right) \tilde  \eta_\parallel\nnb
&+&\left(1-{v_A}^2\right)^2 ({\zeta _\perp}+{\eta _\perp})-2 {\zeta _\times} \left(1-{v_A}^2\right)\Big)
\nnb
&-& \gamma_\parallel \tilde  \eta_\parallel\Big( {c_s}^2 \Big(-h {v_A}^2 \left(1-{v_A}^2\right) \chi
\nnb
&+&8 \gamma_\parallel \left(1-{v_A}^2\right) (\zeta_\parallel +{\zeta _\perp}-2 {\zeta _\times}+{\eta _\perp})\Big)\nnb
&-&8 \left({c_s}^2 \left(4-3 {v_A}^2\right)+{v_A}^2\right) \gamma_\parallel\tilde  \eta_\parallel 
+ 8 \gamma_\parallel  \zeta_\parallel  {v_A}^2\Big)\Big] ,  
\end{eqnarray}
and 
\begin{eqnarray}
\Omega_2' \= -\frac{i \gamma  {c_s}^2 h {v_A}^2 \left(1-{v_A}^2\right)}{\gamma  {c_s}^2 h {\rho _\perp}-{v_f}^2 \gamma_\parallel\tilde \eta_\parallel }\nnb
&-&\frac{i {c_s}^2 k^2}{8 h \left(1-{v_A}^2\right) {v_f}^2\gamma_\parallel \left({v_f}^2 \tilde \eta_\parallel -{c_s}^2 h {\rho_ \perp}\right)}
\nnb
&&
\times \Big[{v_A}^2 \Big(2 \gamma_\parallel \Big(\left(1-{v_A}^2\right) \frac{\left(1-{c_s}^2\right)^2 h {\rho_\perp} {v_A}^2}{{v_f}^2}\nnb
&&+(1-{v_A}^2)({\zeta _\perp}+{\eta _\perp})-\frac{2 {c_s}^2 h {\rho _\perp}}{{v_f}^2}\Big)^2
\nnb
&+&
h \left(1-{v_A}^2\right) \chi  \big({c_s}^2 (-h {\rho _\perp})+\frac{{v_f}^2 (\gamma_\parallel +\xi_\parallel )^2}{\gamma_\parallel }\big)\Big)
\nnb
&-&
h {\rho _\perp} \Big(\frac{8 \gamma_\parallel  {c_s}^4 h {\rho _\perp} {v_A}^2}{{v_f}^2}-8 {v_A}^2 \gamma_\parallel  (\zeta_\parallel -\tilde\eta_\parallel ) \nnb
&+&{c_s}^2 \big[h {v_A}^2 \left(8 \gamma_\parallel  {\rho _\perp}+\left({v_A^2}-1\right) \chi \right)
\nnb
&+&
8 \left({v_A^2}-1\right) \big(\gamma_\parallel  (\zeta_\parallel +{\zeta _\perp}-2 {\zeta _\times}-4 \tilde \eta_\parallel +{\eta _\perp}) \big)\big]
\nnb
&+&
8 {v_f}^2 \gamma_\parallel  \big(\zeta_\parallel +({\eta _\perp}+{\zeta _\perp} )\left({v_A^2}-1\right)^2+2 {\zeta_ \times} \left({v_A^2}-1\right)\nnb
&+&2 \tilde \eta_\parallel \left({v_A}^2-2\right)  \big)\Big)\Big]. 
\end{eqnarray}

\bibliography{ref}

%apsrev4-2.bst 2019-01-14 (MD) hand-edited version of apsrev4-1.bst
%Control: key (0)
%Control: author (8) initials jnrlst
%Control: editor formatted (1) identically to author
%Control: production of article title (0) allowed
%Control: page (0) single
%Control: year (1) truncated
%Control: production of eprint (0) enabled
\begin{thebibliography}{117}%
\makeatletter
\providecommand \@ifxundefined [1]{%
 \@ifx{#1\undefined}
}%
\providecommand \@ifnum [1]{%
 \ifnum #1\expandafter \@firstoftwo
 \else \expandafter \@secondoftwo
 \fi
}%
\providecommand \@ifx [1]{%
 \ifx #1\expandafter \@firstoftwo
 \else \expandafter \@secondoftwo
 \fi
}%
\providecommand \natexlab [1]{#1}%
\providecommand \enquote  [1]{``#1''}%
\providecommand \bibnamefont  [1]{#1}%
\providecommand \bibfnamefont [1]{#1}%
\providecommand \citenamefont [1]{#1}%
\providecommand \href@noop [0]{\@secondoftwo}%
\providecommand \href [0]{\begingroup \@sanitize@url \@href}%
\providecommand \@href[1]{\@@startlink{#1}\@@href}%
\providecommand \@@href[1]{\endgroup#1\@@endlink}%
\providecommand \@sanitize@url [0]{\catcode `\\12\catcode `\$12\catcode
  `\&12\catcode `\#12\catcode `\^12\catcode `\_12\catcode `\%12\relax}%
\providecommand \@@startlink[1]{}%
\providecommand \@@endlink[0]{}%
\providecommand \url  [0]{\begingroup\@sanitize@url \@url }%
\providecommand \@url [1]{\endgroup\@href {#1}{\urlprefix }}%
\providecommand \urlprefix  [0]{URL }%
\providecommand \Eprint [0]{\href }%
\providecommand \doibase [0]{https://doi.org/}%
\providecommand \selectlanguage [0]{\@gobble}%
\providecommand \bibinfo  [0]{\@secondoftwo}%
\providecommand \bibfield  [0]{\@secondoftwo}%
\providecommand \translation [1]{[#1]}%
\providecommand \BibitemOpen [0]{}%
\providecommand \bibitemStop [0]{}%
\providecommand \bibitemNoStop [0]{.\EOS\space}%
\providecommand \EOS [0]{\spacefactor3000\relax}%
\providecommand \BibitemShut  [1]{\csname bibitem#1\endcsname}%
\let\auto@bib@innerbib\@empty
%</preamble>
\bibitem [{\citenamefont {Liang}\ and\ \citenamefont
  {Wang}(2005{\natexlab{a}})}]{Liang:2004ph}%
  \BibitemOpen
  \bibfield  {author} {\bibinfo {author} {\bibfnamefont {Z.-T.}\ \bibnamefont
  {Liang}}\ and\ \bibinfo {author} {\bibfnamefont {X.-N.}\ \bibnamefont
  {Wang}},\ }\bibfield  {title} {\bibinfo {title} {{Globally polarized
  quark-gluon plasma in non-central A+A collisions}},\ }\href
  {https://doi.org/10.1103/PhysRevLett.94.102301} {\bibfield  {journal}
  {\bibinfo  {journal} {Phys. Rev. Lett.}\ }\textbf {\bibinfo {volume} {94}},\
  \bibinfo {pages} {102301} (\bibinfo {year} {2005}{\natexlab{a}})},\ \bibinfo
  {note} {[Erratum: Phys.Rev.Lett. 96, 039901 (2006)]},\ \Eprint
  {https://arxiv.org/abs/nucl-th/0410079} {arXiv:nucl-th/0410079} \BibitemShut
  {NoStop}%
\bibitem [{\citenamefont {Liang}\ and\ \citenamefont
  {Wang}(2005{\natexlab{b}})}]{Liang:2004xn}%
  \BibitemOpen
  \bibfield  {author} {\bibinfo {author} {\bibfnamefont {Z.-T.}\ \bibnamefont
  {Liang}}\ and\ \bibinfo {author} {\bibfnamefont {X.-N.}\ \bibnamefont
  {Wang}},\ }\bibfield  {title} {\bibinfo {title} {{Spin alignment of vector
  mesons in non-central A+A collisions}},\ }\href
  {https://doi.org/10.1016/j.physletb.2005.09.060} {\bibfield  {journal}
  {\bibinfo  {journal} {Phys. Lett. B}\ }\textbf {\bibinfo {volume} {629}},\
  \bibinfo {pages} {20} (\bibinfo {year} {2005}{\natexlab{b}})},\ \Eprint
  {https://arxiv.org/abs/nucl-th/0411101} {arXiv:nucl-th/0411101} \BibitemShut
  {NoStop}%
\bibitem [{\citenamefont {Betz}\ \emph {et~al.}(2007)\citenamefont {Betz},
  \citenamefont {Gyulassy},\ and\ \citenamefont {Torrieri}}]{Betz:2007kg}%
  \BibitemOpen
  \bibfield  {author} {\bibinfo {author} {\bibfnamefont {B.}~\bibnamefont
  {Betz}}, \bibinfo {author} {\bibfnamefont {M.}~\bibnamefont {Gyulassy}},\
  and\ \bibinfo {author} {\bibfnamefont {G.}~\bibnamefont {Torrieri}},\
  }\bibfield  {title} {\bibinfo {title} {{Polarization probes of vorticity in
  heavy ion collisions}},\ }\href {https://doi.org/10.1103/PhysRevC.76.044901}
  {\bibfield  {journal} {\bibinfo  {journal} {Phys. Rev. C}\ }\textbf {\bibinfo
  {volume} {76}},\ \bibinfo {pages} {044901} (\bibinfo {year} {2007})},\
  \Eprint {https://arxiv.org/abs/0708.0035} {arXiv:0708.0035 [nucl-th]}
  \BibitemShut {NoStop}%
\bibitem [{\citenamefont {Becattini}\ \emph {et~al.}(2008)\citenamefont
  {Becattini}, \citenamefont {Piccinini},\ and\ \citenamefont
  {Rizzo}}]{Becattini:2007sr}%
  \BibitemOpen
  \bibfield  {author} {\bibinfo {author} {\bibfnamefont {F.}~\bibnamefont
  {Becattini}}, \bibinfo {author} {\bibfnamefont {F.}~\bibnamefont
  {Piccinini}},\ and\ \bibinfo {author} {\bibfnamefont {J.}~\bibnamefont
  {Rizzo}},\ }\bibfield  {title} {\bibinfo {title} {{Angular momentum
  conservation in heavy ion collisions at very high energy}},\ }\href
  {https://doi.org/10.1103/PhysRevC.77.024906} {\bibfield  {journal} {\bibinfo
  {journal} {Phys. Rev. C}\ }\textbf {\bibinfo {volume} {77}},\ \bibinfo
  {pages} {024906} (\bibinfo {year} {2008})},\ \Eprint
  {https://arxiv.org/abs/0711.1253} {arXiv:0711.1253 [nucl-th]} \BibitemShut
  {NoStop}%
\bibitem [{\citenamefont {Gao}\ \emph {et~al.}(2008)\citenamefont {Gao},
  \citenamefont {Chen}, \citenamefont {Deng}, \citenamefont {Liang},
  \citenamefont {Wang},\ and\ \citenamefont {Wang}}]{Gao:2007bc}%
  \BibitemOpen
  \bibfield  {author} {\bibinfo {author} {\bibfnamefont {J.-H.}\ \bibnamefont
  {Gao}}, \bibinfo {author} {\bibfnamefont {S.-W.}\ \bibnamefont {Chen}},
  \bibinfo {author} {\bibfnamefont {W.-t.}\ \bibnamefont {Deng}}, \bibinfo
  {author} {\bibfnamefont {Z.-T.}\ \bibnamefont {Liang}}, \bibinfo {author}
  {\bibfnamefont {Q.}~\bibnamefont {Wang}},\ and\ \bibinfo {author}
  {\bibfnamefont {X.-N.}\ \bibnamefont {Wang}},\ }\bibfield  {title} {\bibinfo
  {title} {{Global quark polarization in non-central A+A collisions}},\ }\href
  {https://doi.org/10.1103/PhysRevC.77.044902} {\bibfield  {journal} {\bibinfo
  {journal} {Phys. Rev. C}\ }\textbf {\bibinfo {volume} {77}},\ \bibinfo
  {pages} {044902} (\bibinfo {year} {2008})},\ \Eprint
  {https://arxiv.org/abs/0710.2943} {arXiv:0710.2943 [nucl-th]} \BibitemShut
  {NoStop}%
\bibitem [{\citenamefont {Becattini}\ \emph {et~al.}(2017)\citenamefont
  {Becattini}, \citenamefont {Karpenko}, \citenamefont {Lisa}, \citenamefont
  {Upsal},\ and\ \citenamefont {Voloshin}}]{Becattini:2016gvu}%
  \BibitemOpen
  \bibfield  {author} {\bibinfo {author} {\bibfnamefont {F.}~\bibnamefont
  {Becattini}}, \bibinfo {author} {\bibfnamefont {I.}~\bibnamefont {Karpenko}},
  \bibinfo {author} {\bibfnamefont {M.}~\bibnamefont {Lisa}}, \bibinfo {author}
  {\bibfnamefont {I.}~\bibnamefont {Upsal}},\ and\ \bibinfo {author}
  {\bibfnamefont {S.}~\bibnamefont {Voloshin}},\ }\bibfield  {title} {\bibinfo
  {title} {{Global hyperon polarization at local thermodynamic equilibrium with
  vorticity, magnetic field and feed-down}},\ }\href
  {https://doi.org/10.1103/PhysRevC.95.054902} {\bibfield  {journal} {\bibinfo
  {journal} {Phys. Rev. C}\ }\textbf {\bibinfo {volume} {95}},\ \bibinfo
  {pages} {054902} (\bibinfo {year} {2017})},\ \Eprint
  {https://arxiv.org/abs/1610.02506} {arXiv:1610.02506 [nucl-th]} \BibitemShut
  {NoStop}%
\bibitem [{\citenamefont {Adamczyk}\ \emph {et~al.}(2017)\citenamefont
  {Adamczyk} \emph {et~al.}}]{STAR:2017ckg}%
  \BibitemOpen
  \bibfield  {author} {\bibinfo {author} {\bibfnamefont {L.}~\bibnamefont
  {Adamczyk}} \emph {et~al.} (\bibinfo {collaboration} {STAR}),\ }\bibfield
  {title} {\bibinfo {title} {{Global $\Lambda$ hyperon polarization in nuclear
  collisions: evidence for the most vortical fluid}},\ }\href
  {https://doi.org/10.1038/nature23004} {\bibfield  {journal} {\bibinfo
  {journal} {Nature}\ }\textbf {\bibinfo {volume} {548}},\ \bibinfo {pages}
  {62} (\bibinfo {year} {2017})},\ \Eprint {https://arxiv.org/abs/1701.06657}
  {arXiv:1701.06657 [nucl-ex]} \BibitemShut {NoStop}%
\bibitem [{\citenamefont {Adam}\ \emph {et~al.}(2018)\citenamefont {Adam} \emph
  {et~al.}}]{STAR:2018gyt}%
  \BibitemOpen
  \bibfield  {author} {\bibinfo {author} {\bibfnamefont {J.}~\bibnamefont
  {Adam}} \emph {et~al.} (\bibinfo {collaboration} {STAR}),\ }\bibfield
  {title} {\bibinfo {title} {{Global polarization of $\Lambda$ hyperons in
  Au+Au collisions at $\sqrt{s_{_{NN}}}$ = 200 GeV}},\ }\href
  {https://doi.org/10.1103/PhysRevC.98.014910} {\bibfield  {journal} {\bibinfo
  {journal} {Phys. Rev. C}\ }\textbf {\bibinfo {volume} {98}},\ \bibinfo
  {pages} {014910} (\bibinfo {year} {2018})},\ \Eprint
  {https://arxiv.org/abs/1805.04400} {arXiv:1805.04400 [nucl-ex]} \BibitemShut
  {NoStop}%
\bibitem [{\citenamefont {Mohanty}\ \emph {et~al.}(2021)\citenamefont
  {Mohanty}, \citenamefont {Kundu}, \citenamefont {Singha},\ and\ \citenamefont
  {Singh}}]{mohanty_spin_2021}%
  \BibitemOpen
  \bibfield  {author} {\bibinfo {author} {\bibfnamefont {B.}~\bibnamefont
  {Mohanty}}, \bibinfo {author} {\bibfnamefont {S.}~\bibnamefont {Kundu}},
  \bibinfo {author} {\bibfnamefont {S.}~\bibnamefont {Singha}},\ and\ \bibinfo
  {author} {\bibfnamefont {R.}~\bibnamefont {Singh}},\ }\bibfield  {title}
  {\bibinfo {title} {{Spin alignment measurement of vector mesons produced in
  high energy collisions}},\ }\href {https://doi.org/10.1142/S0217732321300263}
  {\bibfield  {journal} {\bibinfo  {journal} {Mod. Phys. Lett. A}\ }\textbf
  {\bibinfo {volume} {36}},\ \bibinfo {pages} {2130026} (\bibinfo {year}
  {2021})},\ \Eprint {https://arxiv.org/abs/2112.04816} {arXiv:2112.04816
  [nucl-ex]} \BibitemShut {NoStop}%
\bibitem [{\citenamefont {Adam}\ \emph {et~al.}(2019)\citenamefont {Adam} \emph
  {et~al.}}]{STAR:2019erd}%
  \BibitemOpen
  \bibfield  {author} {\bibinfo {author} {\bibfnamefont {J.}~\bibnamefont
  {Adam}} \emph {et~al.} (\bibinfo {collaboration} {STAR}),\ }\bibfield
  {title} {\bibinfo {title} {{Polarization of $\Lambda$ ($\bar{\Lambda}$)
  hyperons along the beam direction in Au+Au collisions at $\sqrt{s_{_{NN}}}$ =
  200 GeV}},\ }\href {https://doi.org/10.1103/PhysRevLett.123.132301}
  {\bibfield  {journal} {\bibinfo  {journal} {Phys. Rev. Lett.}\ }\textbf
  {\bibinfo {volume} {123}},\ \bibinfo {pages} {132301} (\bibinfo {year}
  {2019})},\ \Eprint {https://arxiv.org/abs/1905.11917} {arXiv:1905.11917
  [nucl-ex]} \BibitemShut {NoStop}%
\bibitem [{\citenamefont {Adam}\ \emph {et~al.}(2021)\citenamefont {Adam} \emph
  {et~al.}}]{STAR:2020xbm}%
  \BibitemOpen
  \bibfield  {author} {\bibinfo {author} {\bibfnamefont {J.}~\bibnamefont
  {Adam}} \emph {et~al.} (\bibinfo {collaboration} {STAR}),\ }\bibfield
  {title} {\bibinfo {title} {{Global Polarization of $\Xi$ and $\Omega$
  Hyperons in Au+Au Collisions at $\sqrt {s_{NN}}$ = 200 GeV}},\ }\href
  {https://doi.org/10.1103/PhysRevLett.126.162301} {\bibfield  {journal}
  {\bibinfo  {journal} {Phys. Rev. Lett.}\ }\textbf {\bibinfo {volume} {126}},\
  \bibinfo {pages} {162301} (\bibinfo {year} {2021})},\ \Eprint
  {https://arxiv.org/abs/2012.13601} {arXiv:2012.13601 [nucl-ex]} \BibitemShut
  {NoStop}%
\bibitem [{\citenamefont {Acharya}\ \emph {et~al.}(2023)\citenamefont {Acharya}
  \emph {et~al.}}]{ALICE:2022dyy}%
  \BibitemOpen
  \bibfield  {author} {\bibinfo {author} {\bibfnamefont {S.}~\bibnamefont
  {Acharya}} \emph {et~al.} (\bibinfo {collaboration} {ALICE}),\ }\bibfield
  {title} {\bibinfo {title} {{Measurement of the J/\ensuremath{\psi}
  Polarization with Respect to the Event Plane in Pb-Pb Collisions at the
  LHC}},\ }\href {https://doi.org/10.1103/PhysRevLett.131.042303} {\bibfield
  {journal} {\bibinfo  {journal} {Phys. Rev. Lett.}\ }\textbf {\bibinfo
  {volume} {131}},\ \bibinfo {pages} {042303} (\bibinfo {year} {2023})},\
  \Eprint {https://arxiv.org/abs/2204.10171} {arXiv:2204.10171 [nucl-ex]}
  \BibitemShut {NoStop}%
\bibitem [{\citenamefont {Micheletti}(2023)}]{Micheletti:2023qlh}%
  \BibitemOpen
  \bibfield  {author} {\bibinfo {author} {\bibfnamefont {L.}~\bibnamefont
  {Micheletti}} (\bibinfo {collaboration} {ALICE}),\ }\bibfield  {title}
  {\bibinfo {title} {{Vector Mesons Polarization in Pb\textendash{}Pb and $pp$
  Collisions with ALICE}},\ }\href
  {https://doi.org/10.5506/APhysPolBSupp.16.1-A35} {\bibfield  {journal}
  {\bibinfo  {journal} {Acta Phys. Polon. Supp.}\ }\textbf {\bibinfo {volume}
  {16}},\ \bibinfo {pages} {1} (\bibinfo {year} {2023})}\BibitemShut {NoStop}%
\bibitem [{\citenamefont {Fang}\ \emph {et~al.}(2016)\citenamefont {Fang},
  \citenamefont {Pang}, \citenamefont {Wang},\ and\ \citenamefont
  {Wang}}]{Fang:2016vpj}%
  \BibitemOpen
  \bibfield  {author} {\bibinfo {author} {\bibfnamefont {R.-h.}\ \bibnamefont
  {Fang}}, \bibinfo {author} {\bibfnamefont {L.-g.}\ \bibnamefont {Pang}},
  \bibinfo {author} {\bibfnamefont {Q.}~\bibnamefont {Wang}},\ and\ \bibinfo
  {author} {\bibfnamefont {X.-n.}\ \bibnamefont {Wang}},\ }\bibfield  {title}
  {\bibinfo {title} {{Polarization of massive fermions in a vortical fluid}},\
  }\href {https://doi.org/10.1103/PhysRevC.94.024904} {\bibfield  {journal}
  {\bibinfo  {journal} {Phys. Rev. C}\ }\textbf {\bibinfo {volume} {94}},\
  \bibinfo {pages} {024904} (\bibinfo {year} {2016})},\ \Eprint
  {https://arxiv.org/abs/1604.04036} {arXiv:1604.04036 [nucl-th]} \BibitemShut
  {NoStop}%
\bibitem [{\citenamefont {Karpenko}\ and\ \citenamefont
  {Becattini}(2017)}]{Karpenko:2016jyx}%
  \BibitemOpen
  \bibfield  {author} {\bibinfo {author} {\bibfnamefont {I.}~\bibnamefont
  {Karpenko}}\ and\ \bibinfo {author} {\bibfnamefont {F.}~\bibnamefont
  {Becattini}},\ }\bibfield  {title} {\bibinfo {title} {{Study of $\Lambda $
  polarization in relativistic nuclear collisions at $\sqrt{s_\mathrm
  {NN}}=7.7$ \textendash{}200 GeV}},\ }\href
  {https://doi.org/10.1140/epjc/s10052-017-4765-1} {\bibfield  {journal}
  {\bibinfo  {journal} {Eur. Phys. J. C}\ }\textbf {\bibinfo {volume} {77}},\
  \bibinfo {pages} {213} (\bibinfo {year} {2017})},\ \Eprint
  {https://arxiv.org/abs/1610.04717} {arXiv:1610.04717 [nucl-th]} \BibitemShut
  {NoStop}%
\bibitem [{\citenamefont {Shi}\ \emph {et~al.}(2019)\citenamefont {Shi},
  \citenamefont {Li},\ and\ \citenamefont {Liao}}]{Shi:2017wpk}%
  \BibitemOpen
  \bibfield  {author} {\bibinfo {author} {\bibfnamefont {S.}~\bibnamefont
  {Shi}}, \bibinfo {author} {\bibfnamefont {K.}~\bibnamefont {Li}},\ and\
  \bibinfo {author} {\bibfnamefont {J.}~\bibnamefont {Liao}},\ }\bibfield
  {title} {\bibinfo {title} {{Searching for the Subatomic Swirls in the CuCu
  and CuAu Collisions}},\ }\href
  {https://doi.org/10.1016/j.physletb.2018.09.066} {\bibfield  {journal}
  {\bibinfo  {journal} {Phys. Lett. B}\ }\textbf {\bibinfo {volume} {788}},\
  \bibinfo {pages} {409} (\bibinfo {year} {2019})},\ \Eprint
  {https://arxiv.org/abs/1712.00878} {arXiv:1712.00878 [nucl-th]} \BibitemShut
  {NoStop}%
\bibitem [{\citenamefont {Fu}\ \emph {et~al.}(2021{\natexlab{a}})\citenamefont
  {Fu}, \citenamefont {Xu}, \citenamefont {Huang},\ and\ \citenamefont
  {Song}}]{Fu:2020oxj}%
  \BibitemOpen
  \bibfield  {author} {\bibinfo {author} {\bibfnamefont {B.}~\bibnamefont
  {Fu}}, \bibinfo {author} {\bibfnamefont {K.}~\bibnamefont {Xu}}, \bibinfo
  {author} {\bibfnamefont {X.-G.}\ \bibnamefont {Huang}},\ and\ \bibinfo
  {author} {\bibfnamefont {H.}~\bibnamefont {Song}},\ }\bibfield  {title}
  {\bibinfo {title} {{Hydrodynamic study of hyperon spin polarization in
  relativistic heavy ion collisions}},\ }\href
  {https://doi.org/10.1103/PhysRevC.103.024903} {\bibfield  {journal} {\bibinfo
   {journal} {Phys. Rev. C}\ }\textbf {\bibinfo {volume} {103}},\ \bibinfo
  {pages} {024903} (\bibinfo {year} {2021}{\natexlab{a}})},\ \Eprint
  {https://arxiv.org/abs/2011.03740} {arXiv:2011.03740 [nucl-th]} \BibitemShut
  {NoStop}%
\bibitem [{\citenamefont {Liu}\ and\ \citenamefont
  {Huang}(2022)}]{Liu:2021nyg}%
  \BibitemOpen
  \bibfield  {author} {\bibinfo {author} {\bibfnamefont {Y.-C.}\ \bibnamefont
  {Liu}}\ and\ \bibinfo {author} {\bibfnamefont {X.-G.}\ \bibnamefont
  {Huang}},\ }\bibfield  {title} {\bibinfo {title} {{Spin polarization formula
  for Dirac fermions at local equilibrium}},\ }\href
  {https://doi.org/10.1007/s11433-022-1903-8} {\bibfield  {journal} {\bibinfo
  {journal} {Sci. China Phys. Mech. Astron.}\ }\textbf {\bibinfo {volume}
  {65}},\ \bibinfo {pages} {272011} (\bibinfo {year} {2022})},\ \Eprint
  {https://arxiv.org/abs/2109.15301} {arXiv:2109.15301 [nucl-th]} \BibitemShut
  {NoStop}%
\bibitem [{\citenamefont {Becattini}\ \emph
  {et~al.}(2021{\natexlab{a}})\citenamefont {Becattini}, \citenamefont
  {Buzzegoli}, \citenamefont {Inghirami}, \citenamefont {Karpenko},\ and\
  \citenamefont {Palermo}}]{Becattini:2021iol}%
  \BibitemOpen
  \bibfield  {author} {\bibinfo {author} {\bibfnamefont {F.}~\bibnamefont
  {Becattini}}, \bibinfo {author} {\bibfnamefont {M.}~\bibnamefont
  {Buzzegoli}}, \bibinfo {author} {\bibfnamefont {G.}~\bibnamefont
  {Inghirami}}, \bibinfo {author} {\bibfnamefont {I.}~\bibnamefont
  {Karpenko}},\ and\ \bibinfo {author} {\bibfnamefont {A.}~\bibnamefont
  {Palermo}},\ }\bibfield  {title} {\bibinfo {title} {{Local Polarization and
  Isothermal Local Equilibrium in Relativistic Heavy Ion Collisions}},\ }\href
  {https://doi.org/10.1103/PhysRevLett.127.272302} {\bibfield  {journal}
  {\bibinfo  {journal} {Phys. Rev. Lett.}\ }\textbf {\bibinfo {volume} {127}},\
  \bibinfo {pages} {272302} (\bibinfo {year} {2021}{\natexlab{a}})},\ \Eprint
  {https://arxiv.org/abs/2103.14621} {arXiv:2103.14621 [nucl-th]} \BibitemShut
  {NoStop}%
\bibitem [{\citenamefont {Wu}\ \emph {et~al.}(2019)\citenamefont {Wu},
  \citenamefont {Pang}, \citenamefont {Huang},\ and\ \citenamefont
  {Wang}}]{Wu:2019eyi}%
  \BibitemOpen
  \bibfield  {author} {\bibinfo {author} {\bibfnamefont {H.-Z.}\ \bibnamefont
  {Wu}}, \bibinfo {author} {\bibfnamefont {L.-G.}\ \bibnamefont {Pang}},
  \bibinfo {author} {\bibfnamefont {X.-G.}\ \bibnamefont {Huang}},\ and\
  \bibinfo {author} {\bibfnamefont {Q.}~\bibnamefont {Wang}},\ }\bibfield
  {title} {\bibinfo {title} {{Local spin polarization in high energy heavy ion
  collisions}},\ }\href {https://doi.org/10.1103/PhysRevResearch.1.033058}
  {\bibfield  {journal} {\bibinfo  {journal} {Phys. Rev. Research.}\ }\textbf
  {\bibinfo {volume} {1}},\ \bibinfo {pages} {033058} (\bibinfo {year}
  {2019})},\ \Eprint {https://arxiv.org/abs/1906.09385} {arXiv:1906.09385
  [nucl-th]} \BibitemShut {NoStop}%
\bibitem [{\citenamefont {Becattini}\ \emph {et~al.}(2024)\citenamefont
  {Becattini}, \citenamefont {Buzzegoli}, \citenamefont {Niida}, \citenamefont
  {Pu}, \citenamefont {Tang},\ and\ \citenamefont {Wang}}]{Becattini:2024uha}%
  \BibitemOpen
  \bibfield  {author} {\bibinfo {author} {\bibfnamefont {F.}~\bibnamefont
  {Becattini}}, \bibinfo {author} {\bibfnamefont {M.}~\bibnamefont
  {Buzzegoli}}, \bibinfo {author} {\bibfnamefont {T.}~\bibnamefont {Niida}},
  \bibinfo {author} {\bibfnamefont {S.}~\bibnamefont {Pu}}, \bibinfo {author}
  {\bibfnamefont {A.-H.}\ \bibnamefont {Tang}},\ and\ \bibinfo {author}
  {\bibfnamefont {Q.}~\bibnamefont {Wang}},\ }\bibfield  {title} {\bibinfo
  {title} {{Spin polarization in relativistic heavy-ion collisions}},\
  }\href@noop {} {\  (\bibinfo {year} {2024})},\ \Eprint
  {https://arxiv.org/abs/2402.04540} {arXiv:2402.04540 [nucl-th]} \BibitemShut
  {NoStop}%
\bibitem [{\citenamefont {Hattori}\ \emph
  {et~al.}(2019{\natexlab{a}})\citenamefont {Hattori}, \citenamefont {Hongo},
  \citenamefont {Huang}, \citenamefont {Matsuo},\ and\ \citenamefont
  {Taya}}]{Hattori:2019lfp}%
  \BibitemOpen
  \bibfield  {author} {\bibinfo {author} {\bibfnamefont {K.}~\bibnamefont
  {Hattori}}, \bibinfo {author} {\bibfnamefont {M.}~\bibnamefont {Hongo}},
  \bibinfo {author} {\bibfnamefont {X.-G.}\ \bibnamefont {Huang}}, \bibinfo
  {author} {\bibfnamefont {M.}~\bibnamefont {Matsuo}},\ and\ \bibinfo {author}
  {\bibfnamefont {H.}~\bibnamefont {Taya}},\ }\bibfield  {title} {\bibinfo
  {title} {{Fate of spin polarization in a relativistic fluid: An
  entropy-current analysis}},\ }\href
  {https://doi.org/10.1016/j.physletb.2019.05.040} {\bibfield  {journal}
  {\bibinfo  {journal} {Phys. Lett. B}\ }\textbf {\bibinfo {volume} {795}},\
  \bibinfo {pages} {100} (\bibinfo {year} {2019}{\natexlab{a}})},\ \Eprint
  {https://arxiv.org/abs/1901.06615} {arXiv:1901.06615 [hep-th]} \BibitemShut
  {NoStop}%
\bibitem [{\citenamefont {Fukushima}\ and\ \citenamefont
  {Pu}(2021)}]{Fukushima:2020ucl}%
  \BibitemOpen
  \bibfield  {author} {\bibinfo {author} {\bibfnamefont {K.}~\bibnamefont
  {Fukushima}}\ and\ \bibinfo {author} {\bibfnamefont {S.}~\bibnamefont {Pu}},\
  }\bibfield  {title} {\bibinfo {title} {{Spin hydrodynamics and symmetric
  energy-momentum tensors \textendash{} A current induced by the spin vorticity
  \textendash{}}},\ }\href {https://doi.org/10.1016/j.physletb.2021.136346}
  {\bibfield  {journal} {\bibinfo  {journal} {Phys. Lett. B}\ }\textbf
  {\bibinfo {volume} {817}},\ \bibinfo {pages} {136346} (\bibinfo {year}
  {2021})},\ \Eprint {https://arxiv.org/abs/2010.01608} {arXiv:2010.01608
  [hep-th]} \BibitemShut {NoStop}%
\bibitem [{\citenamefont {Gallegos}\ \emph {et~al.}(2021)\citenamefont
  {Gallegos}, \citenamefont {G\"ursoy},\ and\ \citenamefont
  {Yarom}}]{Gallegos:2021bzp}%
  \BibitemOpen
  \bibfield  {author} {\bibinfo {author} {\bibfnamefont {A.~D.}\ \bibnamefont
  {Gallegos}}, \bibinfo {author} {\bibfnamefont {U.}~\bibnamefont {G\"ursoy}},\
  and\ \bibinfo {author} {\bibfnamefont {A.}~\bibnamefont {Yarom}},\ }\bibfield
   {title} {\bibinfo {title} {{Hydrodynamics of spin currents}},\ }\href
  {https://doi.org/10.21468/SciPostPhys.11.2.041} {\bibfield  {journal}
  {\bibinfo  {journal} {SciPost Phys.}\ }\textbf {\bibinfo {volume} {11}},\
  \bibinfo {pages} {041} (\bibinfo {year} {2021})},\ \Eprint
  {https://arxiv.org/abs/2101.04759} {arXiv:2101.04759 [hep-th]} \BibitemShut
  {NoStop}%
\bibitem [{\citenamefont {Li}\ \emph {et~al.}(2021)\citenamefont {Li},
  \citenamefont {Stephanov},\ and\ \citenamefont {Yee}}]{Li:2020eon}%
  \BibitemOpen
  \bibfield  {author} {\bibinfo {author} {\bibfnamefont {S.}~\bibnamefont
  {Li}}, \bibinfo {author} {\bibfnamefont {M.~A.}\ \bibnamefont {Stephanov}},\
  and\ \bibinfo {author} {\bibfnamefont {H.-U.}\ \bibnamefont {Yee}},\
  }\bibfield  {title} {\bibinfo {title} {{Nondissipative Second-Order
  Transport, Spin, and Pseudogauge Transformations in Hydrodynamics}},\ }\href
  {https://doi.org/10.1103/PhysRevLett.127.082302} {\bibfield  {journal}
  {\bibinfo  {journal} {Phys. Rev. Lett.}\ }\textbf {\bibinfo {volume} {127}},\
  \bibinfo {pages} {082302} (\bibinfo {year} {2021})},\ \Eprint
  {https://arxiv.org/abs/2011.12318} {arXiv:2011.12318 [hep-th]} \BibitemShut
  {NoStop}%
\bibitem [{\citenamefont {Hu}(2021)}]{Hu:2021lnx}%
  \BibitemOpen
  \bibfield  {author} {\bibinfo {author} {\bibfnamefont {J.}~\bibnamefont
  {Hu}},\ }\bibfield  {title} {\bibinfo {title} {{Kubo formulae for first-order
  spin hydrodynamics}},\ }\href {https://doi.org/10.1103/PhysRevD.103.116015}
  {\bibfield  {journal} {\bibinfo  {journal} {Phys. Rev. D}\ }\textbf {\bibinfo
  {volume} {103}},\ \bibinfo {pages} {116015} (\bibinfo {year} {2021})},\
  \Eprint {https://arxiv.org/abs/2101.08440} {arXiv:2101.08440 [hep-ph]}
  \BibitemShut {NoStop}%
\bibitem [{\citenamefont {Hu}(2023)}]{Hu:2022azy}%
  \BibitemOpen
  \bibfield  {author} {\bibinfo {author} {\bibfnamefont {J.}~\bibnamefont
  {Hu}},\ }\bibfield  {title} {\bibinfo {title} {{Cross effects in spin
  hydrodynamics: Entropy analysis and statistical operator}},\ }\href
  {https://doi.org/10.1103/PhysRevC.107.024915} {\bibfield  {journal} {\bibinfo
   {journal} {Phys. Rev. C}\ }\textbf {\bibinfo {volume} {107}},\ \bibinfo
  {pages} {024915} (\bibinfo {year} {2023})},\ \Eprint
  {https://arxiv.org/abs/2209.10979} {arXiv:2209.10979 [nucl-th]} \BibitemShut
  {NoStop}%
\bibitem [{\citenamefont {Singh}\ \emph {et~al.}(2023)\citenamefont {Singh},
  \citenamefont {Shokri},\ and\ \citenamefont {Mehr}}]{Singh:2022ltu}%
  \BibitemOpen
  \bibfield  {author} {\bibinfo {author} {\bibfnamefont {R.}~\bibnamefont
  {Singh}}, \bibinfo {author} {\bibfnamefont {M.}~\bibnamefont {Shokri}},\ and\
  \bibinfo {author} {\bibfnamefont {S.~M. A.~T.}\ \bibnamefont {Mehr}},\
  }\bibfield  {title} {\bibinfo {title} {{Relativistic hydrodynamics with spin
  in the presence of electromagnetic fields}},\ }\href
  {https://doi.org/10.1016/j.nuclphysa.2023.122656} {\bibfield  {journal}
  {\bibinfo  {journal} {Nucl. Phys. A}\ }\textbf {\bibinfo {volume} {1035}},\
  \bibinfo {pages} {122656} (\bibinfo {year} {2023})},\ \Eprint
  {https://arxiv.org/abs/2202.11504} {arXiv:2202.11504 [hep-ph]} \BibitemShut
  {NoStop}%
\bibitem [{\citenamefont {Cao}\ \emph {et~al.}(2022)\citenamefont {Cao},
  \citenamefont {Hattori}, \citenamefont {Hongo}, \citenamefont {Huang},\ and\
  \citenamefont {Taya}}]{Cao:2022aku}%
  \BibitemOpen
  \bibfield  {author} {\bibinfo {author} {\bibfnamefont {Z.}~\bibnamefont
  {Cao}}, \bibinfo {author} {\bibfnamefont {K.}~\bibnamefont {Hattori}},
  \bibinfo {author} {\bibfnamefont {M.}~\bibnamefont {Hongo}}, \bibinfo
  {author} {\bibfnamefont {X.-G.}\ \bibnamefont {Huang}},\ and\ \bibinfo
  {author} {\bibfnamefont {H.}~\bibnamefont {Taya}},\ }\bibfield  {title}
  {\bibinfo {title} {{Gyrohydrodynamics: Relativistic spinful fluid with strong
  vorticity}},\ }\href {https://doi.org/10.1093/ptep/ptac091} {\bibfield
  {journal} {\bibinfo  {journal} {PTEP}\ }\textbf {\bibinfo {volume} {2022}},\
  \bibinfo {pages} {071D01} (\bibinfo {year} {2022})},\ \Eprint
  {https://arxiv.org/abs/2205.08051} {arXiv:2205.08051 [hep-th]} \BibitemShut
  {NoStop}%
\bibitem [{\citenamefont {Daher}\ \emph {et~al.}(2023)\citenamefont {Daher},
  \citenamefont {Das},\ and\ \citenamefont {Ryblewski}}]{Daher:2022wzf}%
  \BibitemOpen
  \bibfield  {author} {\bibinfo {author} {\bibfnamefont {A.}~\bibnamefont
  {Daher}}, \bibinfo {author} {\bibfnamefont {A.}~\bibnamefont {Das}},\ and\
  \bibinfo {author} {\bibfnamefont {R.}~\bibnamefont {Ryblewski}},\ }\bibfield
  {title} {\bibinfo {title} {{Stability studies of first-order
  spin-hydrodynamic frameworks}},\ }\href
  {https://doi.org/10.1103/PhysRevD.107.054043} {\bibfield  {journal} {\bibinfo
   {journal} {Phys. Rev. D}\ }\textbf {\bibinfo {volume} {107}},\ \bibinfo
  {pages} {054043} (\bibinfo {year} {2023})},\ \Eprint
  {https://arxiv.org/abs/2209.10460} {arXiv:2209.10460 [nucl-th]} \BibitemShut
  {NoStop}%
\bibitem [{\citenamefont {Sarwar}\ \emph {et~al.}(2023)\citenamefont {Sarwar},
  \citenamefont {Hasanujjaman}, \citenamefont {Bhatt}, \citenamefont {Mishra},\
  and\ \citenamefont {Alam}}]{Sarwar:2022yzs}%
  \BibitemOpen
  \bibfield  {author} {\bibinfo {author} {\bibfnamefont {G.}~\bibnamefont
  {Sarwar}}, \bibinfo {author} {\bibfnamefont {M.}~\bibnamefont
  {Hasanujjaman}}, \bibinfo {author} {\bibfnamefont {J.~R.}\ \bibnamefont
  {Bhatt}}, \bibinfo {author} {\bibfnamefont {H.}~\bibnamefont {Mishra}},\ and\
  \bibinfo {author} {\bibfnamefont {J.-e.}\ \bibnamefont {Alam}},\ }\bibfield
  {title} {\bibinfo {title} {{Causality and stability of relativistic spin
  hydrodynamics}},\ }\href {https://doi.org/10.1103/PhysRevD.107.054031}
  {\bibfield  {journal} {\bibinfo  {journal} {Phys. Rev. D}\ }\textbf {\bibinfo
  {volume} {107}},\ \bibinfo {pages} {054031} (\bibinfo {year} {2023})},\
  \Eprint {https://arxiv.org/abs/2209.08652} {arXiv:2209.08652 [nucl-th]}
  \BibitemShut {NoStop}%
\bibitem [{\citenamefont {Kiamari}\ \emph {et~al.}(2024)\citenamefont
  {Kiamari}, \citenamefont {Sadooghi},\ and\ \citenamefont
  {Jafari}}]{Kiamari:2023fbe}%
  \BibitemOpen
  \bibfield  {author} {\bibinfo {author} {\bibfnamefont {M.}~\bibnamefont
  {Kiamari}}, \bibinfo {author} {\bibfnamefont {N.}~\bibnamefont {Sadooghi}},\
  and\ \bibinfo {author} {\bibfnamefont {M.~S.}\ \bibnamefont {Jafari}},\
  }\bibfield  {title} {\bibinfo {title} {{Relativistic magnetohydrodynamics of
  a spinful and vortical fluid: Entropy current analysis}},\ }\href
  {https://doi.org/10.1103/PhysRevD.109.036024} {\bibfield  {journal} {\bibinfo
   {journal} {Phys. Rev. D}\ }\textbf {\bibinfo {volume} {109}},\ \bibinfo
  {pages} {036024} (\bibinfo {year} {2024})},\ \Eprint
  {https://arxiv.org/abs/2310.01874} {arXiv:2310.01874 [nucl-th]} \BibitemShut
  {NoStop}%
\bibitem [{\citenamefont {Xie}\ \emph {et~al.}(2023)\citenamefont {Xie},
  \citenamefont {Wang}, \citenamefont {Yang},\ and\ \citenamefont
  {Pu}}]{Xie:2023gbo}%
  \BibitemOpen
  \bibfield  {author} {\bibinfo {author} {\bibfnamefont {X.-Q.}\ \bibnamefont
  {Xie}}, \bibinfo {author} {\bibfnamefont {D.-L.}\ \bibnamefont {Wang}},
  \bibinfo {author} {\bibfnamefont {C.}~\bibnamefont {Yang}},\ and\ \bibinfo
  {author} {\bibfnamefont {S.}~\bibnamefont {Pu}},\ }\bibfield  {title}
  {\bibinfo {title} {{Causality and stability analysis for the minimal causal
  spin hydrodynamics}},\ }\href {https://doi.org/10.1103/PhysRevD.108.094031}
  {\bibfield  {journal} {\bibinfo  {journal} {Phys. Rev. D}\ }\textbf {\bibinfo
  {volume} {108}},\ \bibinfo {pages} {094031} (\bibinfo {year} {2023})},\
  \Eprint {https://arxiv.org/abs/2306.13880} {arXiv:2306.13880 [hep-ph]}
  \BibitemShut {NoStop}%
\bibitem [{\citenamefont {Ren}\ \emph {et~al.}(2024)\citenamefont {Ren},
  \citenamefont {Yang}, \citenamefont {Wang},\ and\ \citenamefont
  {Pu}}]{Ren:2024pur}%
  \BibitemOpen
  \bibfield  {author} {\bibinfo {author} {\bibfnamefont {X.}~\bibnamefont
  {Ren}}, \bibinfo {author} {\bibfnamefont {C.}~\bibnamefont {Yang}}, \bibinfo
  {author} {\bibfnamefont {D.-L.}\ \bibnamefont {Wang}},\ and\ \bibinfo
  {author} {\bibfnamefont {S.}~\bibnamefont {Pu}},\ }\bibfield  {title}
  {\bibinfo {title} {{Thermodynamic stability in relativistic viscous and spin
  hydrodynamics}},\ }\href {https://doi.org/10.1103/PhysRevD.110.034010}
  {\bibfield  {journal} {\bibinfo  {journal} {Phys. Rev. D}\ }\textbf {\bibinfo
  {volume} {110}},\ \bibinfo {pages} {034010} (\bibinfo {year} {2024})},\
  \Eprint {https://arxiv.org/abs/2405.03105} {arXiv:2405.03105 [nucl-th]}
  \BibitemShut {NoStop}%
\bibitem [{\citenamefont {Florkowski}\ \emph {et~al.}(2018)\citenamefont
  {Florkowski}, \citenamefont {Friman}, \citenamefont {Jaiswal},\ and\
  \citenamefont {Speranza}}]{Florkowski:2017ruc}%
  \BibitemOpen
  \bibfield  {author} {\bibinfo {author} {\bibfnamefont {W.}~\bibnamefont
  {Florkowski}}, \bibinfo {author} {\bibfnamefont {B.}~\bibnamefont {Friman}},
  \bibinfo {author} {\bibfnamefont {A.}~\bibnamefont {Jaiswal}},\ and\ \bibinfo
  {author} {\bibfnamefont {E.}~\bibnamefont {Speranza}},\ }\bibfield  {title}
  {\bibinfo {title} {{Relativistic fluid dynamics with spin}},\ }\href
  {https://doi.org/10.1103/PhysRevC.97.041901} {\bibfield  {journal} {\bibinfo
  {journal} {Phys. Rev. C}\ }\textbf {\bibinfo {volume} {97}},\ \bibinfo
  {pages} {041901} (\bibinfo {year} {2018})},\ \Eprint
  {https://arxiv.org/abs/1705.00587} {arXiv:1705.00587 [nucl-th]} \BibitemShut
  {NoStop}%
\bibitem [{\citenamefont {Weickgenannt}\ \emph {et~al.}(2021)\citenamefont
  {Weickgenannt}, \citenamefont {Speranza}, \citenamefont {Sheng},
  \citenamefont {Wang},\ and\ \citenamefont {Rischke}}]{Weickgenannt:2020aaf}%
  \BibitemOpen
  \bibfield  {author} {\bibinfo {author} {\bibfnamefont {N.}~\bibnamefont
  {Weickgenannt}}, \bibinfo {author} {\bibfnamefont {E.}~\bibnamefont
  {Speranza}}, \bibinfo {author} {\bibfnamefont {X.-l.}\ \bibnamefont {Sheng}},
  \bibinfo {author} {\bibfnamefont {Q.}~\bibnamefont {Wang}},\ and\ \bibinfo
  {author} {\bibfnamefont {D.~H.}\ \bibnamefont {Rischke}},\ }\bibfield
  {title} {\bibinfo {title} {{Generating Spin Polarization from Vorticity
  through Nonlocal Collisions}},\ }\href
  {https://doi.org/10.1103/PhysRevLett.127.052301} {\bibfield  {journal}
  {\bibinfo  {journal} {Phys. Rev. Lett.}\ }\textbf {\bibinfo {volume} {127}},\
  \bibinfo {pages} {052301} (\bibinfo {year} {2021})},\ \Eprint
  {https://arxiv.org/abs/2005.01506} {arXiv:2005.01506 [hep-ph]} \BibitemShut
  {NoStop}%
\bibitem [{\citenamefont {Yang}\ \emph {et~al.}(2020)\citenamefont {Yang},
  \citenamefont {Hattori},\ and\ \citenamefont {Hidaka}}]{Yang:2020hri}%
  \BibitemOpen
  \bibfield  {author} {\bibinfo {author} {\bibfnamefont {D.-L.}\ \bibnamefont
  {Yang}}, \bibinfo {author} {\bibfnamefont {K.}~\bibnamefont {Hattori}},\ and\
  \bibinfo {author} {\bibfnamefont {Y.}~\bibnamefont {Hidaka}},\ }\bibfield
  {title} {\bibinfo {title} {{Effective quantum kinetic theory for spin
  transport of fermions with collsional effects}},\ }\href
  {https://doi.org/10.1007/JHEP07(2020)070} {\bibfield  {journal} {\bibinfo
  {journal} {JHEP}\ }\textbf {\bibinfo {volume} {07}},\ \bibinfo {pages}
  {070}},\ \Eprint {https://arxiv.org/abs/2002.02612} {arXiv:2002.02612
  [hep-ph]} \BibitemShut {NoStop}%
\bibitem [{\citenamefont {Bhadury}\ \emph {et~al.}(2021)\citenamefont
  {Bhadury}, \citenamefont {Florkowski}, \citenamefont {Jaiswal}, \citenamefont
  {Kumar},\ and\ \citenamefont {Ryblewski}}]{Bhadury:2020cop}%
  \BibitemOpen
  \bibfield  {author} {\bibinfo {author} {\bibfnamefont {S.}~\bibnamefont
  {Bhadury}}, \bibinfo {author} {\bibfnamefont {W.}~\bibnamefont {Florkowski}},
  \bibinfo {author} {\bibfnamefont {A.}~\bibnamefont {Jaiswal}}, \bibinfo
  {author} {\bibfnamefont {A.}~\bibnamefont {Kumar}},\ and\ \bibinfo {author}
  {\bibfnamefont {R.}~\bibnamefont {Ryblewski}},\ }\bibfield  {title} {\bibinfo
  {title} {{Dissipative Spin Dynamics in Relativistic Matter}},\ }\href
  {https://doi.org/10.1103/PhysRevD.103.014030} {\bibfield  {journal} {\bibinfo
   {journal} {Phys. Rev. D}\ }\textbf {\bibinfo {volume} {103}},\ \bibinfo
  {pages} {014030} (\bibinfo {year} {2021})},\ \Eprint
  {https://arxiv.org/abs/2008.10976} {arXiv:2008.10976 [nucl-th]} \BibitemShut
  {NoStop}%
\bibitem [{\citenamefont {Wang}\ \emph {et~al.}(2021)\citenamefont {Wang},
  \citenamefont {Guo},\ and\ \citenamefont {Zhuang}}]{Wang:2020pej}%
  \BibitemOpen
  \bibfield  {author} {\bibinfo {author} {\bibfnamefont {Z.}~\bibnamefont
  {Wang}}, \bibinfo {author} {\bibfnamefont {X.}~\bibnamefont {Guo}},\ and\
  \bibinfo {author} {\bibfnamefont {P.}~\bibnamefont {Zhuang}},\ }\bibfield
  {title} {\bibinfo {title} {{Equilibrium Spin Distribution From Detailed
  Balance}},\ }\href {https://doi.org/10.1140/epjc/s10052-021-09586-8}
  {\bibfield  {journal} {\bibinfo  {journal} {Eur. Phys. J. C}\ }\textbf
  {\bibinfo {volume} {81}},\ \bibinfo {pages} {799} (\bibinfo {year} {2021})},\
  \Eprint {https://arxiv.org/abs/2009.10930} {arXiv:2009.10930 [hep-th]}
  \BibitemShut {NoStop}%
\bibitem [{\citenamefont {Sheng}\ \emph {et~al.}(2021)\citenamefont {Sheng},
  \citenamefont {Weickgenannt}, \citenamefont {Speranza}, \citenamefont
  {Rischke},\ and\ \citenamefont {Wang}}]{Sheng:2021kfc}%
  \BibitemOpen
  \bibfield  {author} {\bibinfo {author} {\bibfnamefont {X.-L.}\ \bibnamefont
  {Sheng}}, \bibinfo {author} {\bibfnamefont {N.}~\bibnamefont {Weickgenannt}},
  \bibinfo {author} {\bibfnamefont {E.}~\bibnamefont {Speranza}}, \bibinfo
  {author} {\bibfnamefont {D.~H.}\ \bibnamefont {Rischke}},\ and\ \bibinfo
  {author} {\bibfnamefont {Q.}~\bibnamefont {Wang}},\ }\bibfield  {title}
  {\bibinfo {title} {{From Kadanoff-Baym to Boltzmann equations for massive
  spin-1/2 fermions}},\ }\href {https://doi.org/10.1103/PhysRevD.104.016029}
  {\bibfield  {journal} {\bibinfo  {journal} {Phys. Rev. D}\ }\textbf {\bibinfo
  {volume} {104}},\ \bibinfo {pages} {016029} (\bibinfo {year} {2021})},\
  \Eprint {https://arxiv.org/abs/2103.10636} {arXiv:2103.10636 [nucl-th]}
  \BibitemShut {NoStop}%
\bibitem [{\citenamefont {M\"uller}\ and\ \citenamefont
  {Yang}(2022)}]{Muller:2021hpe}%
  \BibitemOpen
  \bibfield  {author} {\bibinfo {author} {\bibfnamefont {B.}~\bibnamefont
  {M\"uller}}\ and\ \bibinfo {author} {\bibfnamefont {D.-L.}\ \bibnamefont
  {Yang}},\ }\bibfield  {title} {\bibinfo {title} {{Anomalous spin polarization
  from turbulent color fields}},\ }\href
  {https://doi.org/10.1103/PhysRevD.105.L011901} {\bibfield  {journal}
  {\bibinfo  {journal} {Phys. Rev. D}\ }\textbf {\bibinfo {volume} {105}},\
  \bibinfo {pages} {L011901} (\bibinfo {year} {2022})},\ \bibinfo {note}
  {[Erratum: Phys.Rev.D 106, 039904 (2022)]},\ \Eprint
  {https://arxiv.org/abs/2110.15630} {arXiv:2110.15630 [nucl-th]} \BibitemShut
  {NoStop}%
\bibitem [{\citenamefont {Kumar}\ \emph
  {et~al.}(2023{\natexlab{a}})\citenamefont {Kumar}, \citenamefont {M\"uller},\
  and\ \citenamefont {Yang}}]{Kumar:2022ylt}%
  \BibitemOpen
  \bibfield  {author} {\bibinfo {author} {\bibfnamefont {A.}~\bibnamefont
  {Kumar}}, \bibinfo {author} {\bibfnamefont {B.}~\bibnamefont {M\"uller}},\
  and\ \bibinfo {author} {\bibfnamefont {D.-L.}\ \bibnamefont {Yang}},\
  }\bibfield  {title} {\bibinfo {title} {{Spin polarization and correlation of
  quarks from the glasma}},\ }\href
  {https://doi.org/10.1103/PhysRevD.107.076025} {\bibfield  {journal} {\bibinfo
   {journal} {Phys. Rev. D}\ }\textbf {\bibinfo {volume} {107}},\ \bibinfo
  {pages} {076025} (\bibinfo {year} {2023}{\natexlab{a}})},\ \Eprint
  {https://arxiv.org/abs/2212.13354} {arXiv:2212.13354 [nucl-th]} \BibitemShut
  {NoStop}%
\bibitem [{\citenamefont {Kumar}\ \emph
  {et~al.}(2023{\natexlab{b}})\citenamefont {Kumar}, \citenamefont {M\"uller},\
  and\ \citenamefont {Yang}}]{Kumar:2023ghs}%
  \BibitemOpen
  \bibfield  {author} {\bibinfo {author} {\bibfnamefont {A.}~\bibnamefont
  {Kumar}}, \bibinfo {author} {\bibfnamefont {B.}~\bibnamefont {M\"uller}},\
  and\ \bibinfo {author} {\bibfnamefont {D.-L.}\ \bibnamefont {Yang}},\
  }\bibfield  {title} {\bibinfo {title} {{Spin alignment of vector mesons by
  glasma fields}},\ }\href {https://doi.org/10.1103/PhysRevD.108.016020}
  {\bibfield  {journal} {\bibinfo  {journal} {Phys. Rev. D}\ }\textbf {\bibinfo
  {volume} {108}},\ \bibinfo {pages} {016020} (\bibinfo {year}
  {2023}{\natexlab{b}})},\ \Eprint {https://arxiv.org/abs/2304.04181}
  {arXiv:2304.04181 [nucl-th]} \BibitemShut {NoStop}%
\bibitem [{\citenamefont {Peng}\ \emph {et~al.}(2021)\citenamefont {Peng},
  \citenamefont {Zhang}, \citenamefont {Sheng},\ and\ \citenamefont
  {Wang}}]{Peng:2021ago}%
  \BibitemOpen
  \bibfield  {author} {\bibinfo {author} {\bibfnamefont {H.-H.}\ \bibnamefont
  {Peng}}, \bibinfo {author} {\bibfnamefont {J.-J.}\ \bibnamefont {Zhang}},
  \bibinfo {author} {\bibfnamefont {X.-L.}\ \bibnamefont {Sheng}},\ and\
  \bibinfo {author} {\bibfnamefont {Q.}~\bibnamefont {Wang}},\ }\bibfield
  {title} {\bibinfo {title} {{Ideal Spin Hydrodynamics from the Wigner Function
  Approach}},\ }\href {https://doi.org/10.1088/0256-307X/38/11/116701}
  {\bibfield  {journal} {\bibinfo  {journal} {Chin. Phys. Lett.}\ }\textbf
  {\bibinfo {volume} {38}},\ \bibinfo {pages} {116701} (\bibinfo {year}
  {2021})},\ \Eprint {https://arxiv.org/abs/2107.00448} {arXiv:2107.00448
  [hep-th]} \BibitemShut {NoStop}%
\bibitem [{\citenamefont {Hu}(2022{\natexlab{a}})}]{Hu:2021pwh}%
  \BibitemOpen
  \bibfield  {author} {\bibinfo {author} {\bibfnamefont {J.}~\bibnamefont
  {Hu}},\ }\bibfield  {title} {\bibinfo {title} {{Relativistic first-order spin
  hydrodynamics via the Chapman-Enskog expansion}},\ }\href
  {https://doi.org/10.1103/PhysRevD.105.076009} {\bibfield  {journal} {\bibinfo
   {journal} {Phys. Rev. D}\ }\textbf {\bibinfo {volume} {105}},\ \bibinfo
  {pages} {076009} (\bibinfo {year} {2022}{\natexlab{a}})},\ \Eprint
  {https://arxiv.org/abs/2111.03571} {arXiv:2111.03571 [hep-ph]} \BibitemShut
  {NoStop}%
\bibitem [{\citenamefont {Hu}(2022{\natexlab{b}})}]{Hu:2022lpi}%
  \BibitemOpen
  \bibfield  {author} {\bibinfo {author} {\bibfnamefont {J.}~\bibnamefont
  {Hu}},\ }\bibfield  {title} {\bibinfo {title} {{Linear mode analysis from
  spin transport equation}},\ }\href
  {https://doi.org/10.1103/PhysRevD.106.036004} {\bibfield  {journal} {\bibinfo
   {journal} {Phys. Rev. D}\ }\textbf {\bibinfo {volume} {106}},\ \bibinfo
  {pages} {036004} (\bibinfo {year} {2022}{\natexlab{b}})},\ \Eprint
  {https://arxiv.org/abs/2202.07373} {arXiv:2202.07373 [hep-ph]} \BibitemShut
  {NoStop}%
\bibitem [{\citenamefont {Hu}(2022{\natexlab{c}})}]{Hu:2022xjn}%
  \BibitemOpen
  \bibfield  {author} {\bibinfo {author} {\bibfnamefont {J.}~\bibnamefont
  {Hu}},\ }\bibfield  {title} {\bibinfo {title} {{Linear mode analysis and spin
  relaxation}},\ }\href {https://doi.org/10.1103/PhysRevD.105.096021}
  {\bibfield  {journal} {\bibinfo  {journal} {Phys. Rev. D}\ }\textbf {\bibinfo
  {volume} {105}},\ \bibinfo {pages} {096021} (\bibinfo {year}
  {2022}{\natexlab{c}})},\ \Eprint {https://arxiv.org/abs/2204.12946}
  {arXiv:2204.12946 [hep-ph]} \BibitemShut {NoStop}%
\bibitem [{\citenamefont {Hongo}\ \emph {et~al.}(2022)\citenamefont {Hongo},
  \citenamefont {Huang}, \citenamefont {Kaminski}, \citenamefont {Stephanov},\
  and\ \citenamefont {Yee}}]{Hongo:2022izs}%
  \BibitemOpen
  \bibfield  {author} {\bibinfo {author} {\bibfnamefont {M.}~\bibnamefont
  {Hongo}}, \bibinfo {author} {\bibfnamefont {X.-G.}\ \bibnamefont {Huang}},
  \bibinfo {author} {\bibfnamefont {M.}~\bibnamefont {Kaminski}}, \bibinfo
  {author} {\bibfnamefont {M.}~\bibnamefont {Stephanov}},\ and\ \bibinfo
  {author} {\bibfnamefont {H.-U.}\ \bibnamefont {Yee}},\ }\bibfield  {title}
  {\bibinfo {title} {{Spin relaxation rate for heavy quarks in weakly coupled
  QCD plasma}},\ }\href {https://doi.org/10.1007/JHEP08(2022)263} {\bibfield
  {journal} {\bibinfo  {journal} {JHEP}\ }\textbf {\bibinfo {volume} {08}},\
  \bibinfo {pages} {263}},\ \Eprint {https://arxiv.org/abs/2201.12390}
  {arXiv:2201.12390 [hep-th]} \BibitemShut {NoStop}%
\bibitem [{\citenamefont {Hidaka}\ \emph {et~al.}(2024)\citenamefont {Hidaka},
  \citenamefont {Hongo}, \citenamefont {Stephanov},\ and\ \citenamefont
  {Yee}}]{Hidaka:2023oze}%
  \BibitemOpen
  \bibfield  {author} {\bibinfo {author} {\bibfnamefont {Y.}~\bibnamefont
  {Hidaka}}, \bibinfo {author} {\bibfnamefont {M.}~\bibnamefont {Hongo}},
  \bibinfo {author} {\bibfnamefont {M.~A.}\ \bibnamefont {Stephanov}},\ and\
  \bibinfo {author} {\bibfnamefont {H.-U.}\ \bibnamefont {Yee}},\ }\bibfield
  {title} {\bibinfo {title} {{Spin relaxation rate for baryons in a thermal
  pion gas}},\ }\href {https://doi.org/10.1103/PhysRevC.109.054909} {\bibfield
  {journal} {\bibinfo  {journal} {Phys. Rev. C}\ }\textbf {\bibinfo {volume}
  {109}},\ \bibinfo {pages} {054909} (\bibinfo {year} {2024})},\ \Eprint
  {https://arxiv.org/abs/2312.08266} {arXiv:2312.08266 [hep-ph]} \BibitemShut
  {NoStop}%
\bibitem [{\citenamefont {Weickgenannt}\ \emph {et~al.}(2022)\citenamefont
  {Weickgenannt}, \citenamefont {Wagner}, \citenamefont {Speranza},\ and\
  \citenamefont {Rischke}}]{Weickgenannt:2022zxs}%
  \BibitemOpen
  \bibfield  {author} {\bibinfo {author} {\bibfnamefont {N.}~\bibnamefont
  {Weickgenannt}}, \bibinfo {author} {\bibfnamefont {D.}~\bibnamefont
  {Wagner}}, \bibinfo {author} {\bibfnamefont {E.}~\bibnamefont {Speranza}},\
  and\ \bibinfo {author} {\bibfnamefont {D.~H.}\ \bibnamefont {Rischke}},\
  }\bibfield  {title} {\bibinfo {title} {{Relativistic second-order dissipative
  spin hydrodynamics from the method of moments}},\ }\href
  {https://doi.org/10.1103/PhysRevD.106.096014} {\bibfield  {journal} {\bibinfo
   {journal} {Phys. Rev. D}\ }\textbf {\bibinfo {volume} {106}},\ \bibinfo
  {pages} {096014} (\bibinfo {year} {2022})},\ \Eprint
  {https://arxiv.org/abs/2203.04766} {arXiv:2203.04766 [nucl-th]} \BibitemShut
  {NoStop}%
\bibitem [{\citenamefont {Weickgenannt}(2023)}]{Weickgenannt:2023btk}%
  \BibitemOpen
  \bibfield  {author} {\bibinfo {author} {\bibfnamefont {N.}~\bibnamefont
  {Weickgenannt}},\ }\bibfield  {title} {\bibinfo {title} {{Linearly stable and
  causal relativistic first-order spin hydrodynamics}},\ }\href
  {https://doi.org/10.1103/PhysRevD.108.076011} {\bibfield  {journal} {\bibinfo
   {journal} {Phys. Rev. D}\ }\textbf {\bibinfo {volume} {108}},\ \bibinfo
  {pages} {076011} (\bibinfo {year} {2023})},\ \Eprint
  {https://arxiv.org/abs/2307.13561} {arXiv:2307.13561 [nucl-th]} \BibitemShut
  {NoStop}%
\bibitem [{\citenamefont {Bhadury}(2024)}]{Bhadury:2024ckc}%
  \BibitemOpen
  \bibfield  {author} {\bibinfo {author} {\bibfnamefont {S.}~\bibnamefont
  {Bhadury}},\ }\bibfield  {title} {\bibinfo {title} {{Relativistic spin
  hydrodynamics with momentum and spin-dependent relaxation time}},\
  }\href@noop {} {\  (\bibinfo {year} {2024})},\ \Eprint
  {https://arxiv.org/abs/2408.14462} {arXiv:2408.14462 [hep-ph]} \BibitemShut
  {NoStop}%
\bibitem [{\citenamefont {Fang}\ and\ \citenamefont {Pu}(2024)}]{Fang:2024vds}%
  \BibitemOpen
  \bibfield  {author} {\bibinfo {author} {\bibfnamefont {S.}~\bibnamefont
  {Fang}}\ and\ \bibinfo {author} {\bibfnamefont {S.}~\bibnamefont {Pu}},\
  }\bibfield  {title} {\bibinfo {title} {{Collisional corrections to spin
  polarization from quantum kinetic theory using Chapman-Enskog expansion}},\
  }\href@noop {} {\  (\bibinfo {year} {2024})},\ \Eprint
  {https://arxiv.org/abs/2408.09877} {arXiv:2408.09877 [hep-ph]} \BibitemShut
  {NoStop}%
\bibitem [{\citenamefont {Lin}\ and\ \citenamefont {Tang}(2024)}]{Lin:2024cxo}%
  \BibitemOpen
  \bibfield  {author} {\bibinfo {author} {\bibfnamefont {S.}~\bibnamefont
  {Lin}}\ and\ \bibinfo {author} {\bibfnamefont {H.}~\bibnamefont {Tang}},\
  }\bibfield  {title} {\bibinfo {title} {{Transient spin modes from
  relaxational axial kinetic theory}},\ }\href@noop {} {\  (\bibinfo {year}
  {2024})},\ \Eprint {https://arxiv.org/abs/2406.17632} {arXiv:2406.17632
  [nucl-th]} \BibitemShut {NoStop}%
\bibitem [{\citenamefont {Garbiso}\ and\ \citenamefont
  {Kaminski}(2020)}]{Garbiso:2020puw}%
  \BibitemOpen
  \bibfield  {author} {\bibinfo {author} {\bibfnamefont {M.}~\bibnamefont
  {Garbiso}}\ and\ \bibinfo {author} {\bibfnamefont {M.}~\bibnamefont
  {Kaminski}},\ }\bibfield  {title} {\bibinfo {title} {{Hydrodynamics of simply
  spinning black holes \& hydrodynamics for spinning quantum fluids}},\ }\href
  {https://doi.org/10.1007/JHEP12(2020)112} {\bibfield  {journal} {\bibinfo
  {journal} {JHEP}\ }\textbf {\bibinfo {volume} {12}},\ \bibinfo {pages}
  {112}},\ \Eprint {https://arxiv.org/abs/2007.04345} {arXiv:2007.04345
  [hep-th]} \BibitemShut {NoStop}%
\bibitem [{\citenamefont {Gallegos}\ and\ \citenamefont
  {G\"ursoy}(2020)}]{Gallegos:2020otk}%
  \BibitemOpen
  \bibfield  {author} {\bibinfo {author} {\bibfnamefont {A.~D.}\ \bibnamefont
  {Gallegos}}\ and\ \bibinfo {author} {\bibfnamefont {U.}~\bibnamefont
  {G\"ursoy}},\ }\bibfield  {title} {\bibinfo {title} {{Holographic spin
  liquids and Lovelock Chern-Simons gravity}},\ }\href
  {https://doi.org/10.1007/JHEP11(2020)151} {\bibfield  {journal} {\bibinfo
  {journal} {JHEP}\ }\textbf {\bibinfo {volume} {11}},\ \bibinfo {pages}
  {151}},\ \Eprint {https://arxiv.org/abs/2004.05148} {arXiv:2004.05148
  [hep-th]} \BibitemShut {NoStop}%
\bibitem [{\citenamefont {Hashimoto}\ \emph {et~al.}(2015)\citenamefont
  {Hashimoto}, \citenamefont {Iizuka},\ and\ \citenamefont
  {Kimura}}]{Hashimoto:2013bna}%
  \BibitemOpen
  \bibfield  {author} {\bibinfo {author} {\bibfnamefont {K.}~\bibnamefont
  {Hashimoto}}, \bibinfo {author} {\bibfnamefont {N.}~\bibnamefont {Iizuka}},\
  and\ \bibinfo {author} {\bibfnamefont {T.}~\bibnamefont {Kimura}},\
  }\bibfield  {title} {\bibinfo {title} {{Towards Holographic Spintronics}},\
  }\href {https://doi.org/10.1103/PhysRevD.91.086003} {\bibfield  {journal}
  {\bibinfo  {journal} {Phys. Rev. D}\ }\textbf {\bibinfo {volume} {91}},\
  \bibinfo {pages} {086003} (\bibinfo {year} {2015})},\ \Eprint
  {https://arxiv.org/abs/1304.3126} {arXiv:1304.3126 [hep-th]} \BibitemShut
  {NoStop}%
\bibitem [{\citenamefont {Montenegro}\ \emph {et~al.}(2017)\citenamefont
  {Montenegro}, \citenamefont {Tinti},\ and\ \citenamefont
  {Torrieri}}]{Montenegro:2017rbu}%
  \BibitemOpen
  \bibfield  {author} {\bibinfo {author} {\bibfnamefont {D.}~\bibnamefont
  {Montenegro}}, \bibinfo {author} {\bibfnamefont {L.}~\bibnamefont {Tinti}},\
  and\ \bibinfo {author} {\bibfnamefont {G.}~\bibnamefont {Torrieri}},\
  }\bibfield  {title} {\bibinfo {title} {{Ideal relativistic fluid limit for a
  medium with polarization}},\ }\href
  {https://doi.org/10.1103/PhysRevD.96.056012} {\bibfield  {journal} {\bibinfo
  {journal} {Phys. Rev. D}\ }\textbf {\bibinfo {volume} {96}},\ \bibinfo
  {pages} {056012} (\bibinfo {year} {2017})},\ \bibinfo {note} {[Addendum:
  Phys.Rev.D 96, 079901 (2017)]},\ \Eprint {https://arxiv.org/abs/1701.08263}
  {arXiv:1701.08263 [hep-th]} \BibitemShut {NoStop}%
\bibitem [{\citenamefont {Montenegro}\ and\ \citenamefont
  {Torrieri}(2020)}]{Montenegro:2020paq}%
  \BibitemOpen
  \bibfield  {author} {\bibinfo {author} {\bibfnamefont {D.}~\bibnamefont
  {Montenegro}}\ and\ \bibinfo {author} {\bibfnamefont {G.}~\bibnamefont
  {Torrieri}},\ }\bibfield  {title} {\bibinfo {title} {{Linear response theory
  and effective action of relativistic hydrodynamics with spin}},\ }\href
  {https://doi.org/10.1103/PhysRevD.102.036007} {\bibfield  {journal} {\bibinfo
   {journal} {Phys. Rev. D}\ }\textbf {\bibinfo {volume} {102}},\ \bibinfo
  {pages} {036007} (\bibinfo {year} {2020})},\ \Eprint
  {https://arxiv.org/abs/2004.10195} {arXiv:2004.10195 [hep-th]} \BibitemShut
  {NoStop}%
\bibitem [{\citenamefont {Becattini}\ and\ \citenamefont
  {Tinti}(2010)}]{Becattini:2009wh}%
  \BibitemOpen
  \bibfield  {author} {\bibinfo {author} {\bibfnamefont {F.}~\bibnamefont
  {Becattini}}\ and\ \bibinfo {author} {\bibfnamefont {L.}~\bibnamefont
  {Tinti}},\ }\bibfield  {title} {\bibinfo {title} {{The Ideal relativistic
  rotating gas as a perfect fluid with spin}},\ }\href
  {https://doi.org/10.1016/j.aop.2010.03.007} {\bibfield  {journal} {\bibinfo
  {journal} {Annals Phys.}\ }\textbf {\bibinfo {volume} {325}},\ \bibinfo
  {pages} {1566} (\bibinfo {year} {2010})},\ \Eprint
  {https://arxiv.org/abs/0911.0864} {arXiv:0911.0864 [gr-qc]} \BibitemShut
  {NoStop}%
\bibitem [{\citenamefont {Becattini}\ and\ \citenamefont
  {Tinti}(2013)}]{Becattini:2012pp}%
  \BibitemOpen
  \bibfield  {author} {\bibinfo {author} {\bibfnamefont {F.}~\bibnamefont
  {Becattini}}\ and\ \bibinfo {author} {\bibfnamefont {L.}~\bibnamefont
  {Tinti}},\ }\bibfield  {title} {\bibinfo {title} {{Nonequilibrium
  Thermodynamical Inequivalence of Quantum Stress-energy and Spin Tensors}},\
  }\href {https://doi.org/10.1103/PhysRevD.87.025029} {\bibfield  {journal}
  {\bibinfo  {journal} {Phys. Rev. D}\ }\textbf {\bibinfo {volume} {87}},\
  \bibinfo {pages} {025029} (\bibinfo {year} {2013})},\ \Eprint
  {https://arxiv.org/abs/1209.6212} {arXiv:1209.6212 [hep-th]} \BibitemShut
  {NoStop}%
\bibitem [{\citenamefont {Becattini}\ \emph {et~al.}(2019)\citenamefont
  {Becattini}, \citenamefont {Florkowski},\ and\ \citenamefont
  {Speranza}}]{Becattini:2018duy}%
  \BibitemOpen
  \bibfield  {author} {\bibinfo {author} {\bibfnamefont {F.}~\bibnamefont
  {Becattini}}, \bibinfo {author} {\bibfnamefont {W.}~\bibnamefont
  {Florkowski}},\ and\ \bibinfo {author} {\bibfnamefont {E.}~\bibnamefont
  {Speranza}},\ }\bibfield  {title} {\bibinfo {title} {{Spin tensor and its
  role in non-equilibrium thermodynamics}},\ }\href
  {https://doi.org/10.1016/j.physletb.2018.12.016} {\bibfield  {journal}
  {\bibinfo  {journal} {Phys. Lett. B}\ }\textbf {\bibinfo {volume} {789}},\
  \bibinfo {pages} {419} (\bibinfo {year} {2019})},\ \Eprint
  {https://arxiv.org/abs/1807.10994} {arXiv:1807.10994 [hep-th]} \BibitemShut
  {NoStop}%
\bibitem [{\citenamefont {Hongo}\ \emph {et~al.}(2021)\citenamefont {Hongo},
  \citenamefont {Huang}, \citenamefont {Kaminski}, \citenamefont {Stephanov},\
  and\ \citenamefont {Yee}}]{Hongo:2021ona}%
  \BibitemOpen
  \bibfield  {author} {\bibinfo {author} {\bibfnamefont {M.}~\bibnamefont
  {Hongo}}, \bibinfo {author} {\bibfnamefont {X.-G.}\ \bibnamefont {Huang}},
  \bibinfo {author} {\bibfnamefont {M.}~\bibnamefont {Kaminski}}, \bibinfo
  {author} {\bibfnamefont {M.}~\bibnamefont {Stephanov}},\ and\ \bibinfo
  {author} {\bibfnamefont {H.-U.}\ \bibnamefont {Yee}},\ }\bibfield  {title}
  {\bibinfo {title} {{Relativistic spin hydrodynamics with torsion and linear
  response theory for spin relaxation}},\ }\href
  {https://doi.org/10.1007/JHEP11(2021)150} {\bibfield  {journal} {\bibinfo
  {journal} {JHEP}\ }\textbf {\bibinfo {volume} {11}},\ \bibinfo {pages}
  {150}},\ \Eprint {https://arxiv.org/abs/2107.14231} {arXiv:2107.14231
  [hep-th]} \BibitemShut {NoStop}%
\bibitem [{\citenamefont {Tiwari}\ and\ \citenamefont
  {Patra}(2024)}]{Tiwari:2024trl}%
  \BibitemOpen
  \bibfield  {author} {\bibinfo {author} {\bibfnamefont {A.}~\bibnamefont
  {Tiwari}}\ and\ \bibinfo {author} {\bibfnamefont {B.~K.}\ \bibnamefont
  {Patra}},\ }\bibfield  {title} {\bibinfo {title} {{Second-order spin
  hydrodynamics from Zubarev's nonequilibrium statistical operator
  formalism}},\ }\href@noop {} {\  (\bibinfo {year} {2024})},\ \Eprint
  {https://arxiv.org/abs/2408.11514} {arXiv:2408.11514 [hep-th]} \BibitemShut
  {NoStop}%
\bibitem [{\citenamefont {Wagner}(2024)}]{Wagner:2024fry}%
  \BibitemOpen
  \bibfield  {author} {\bibinfo {author} {\bibfnamefont {D.}~\bibnamefont
  {Wagner}},\ }\bibfield  {title} {\bibinfo {title} {{Resummed spin
  hydrodynamics from quantum kinetic theory}},\ }\href@noop {} {\  (\bibinfo
  {year} {2024})},\ \Eprint {https://arxiv.org/abs/2409.07143}
  {arXiv:2409.07143 [nucl-th]} \BibitemShut {NoStop}%
\bibitem [{\citenamefont {Wagner}\ \emph {et~al.}(2024)\citenamefont {Wagner},
  \citenamefont {Shokri},\ and\ \citenamefont {Rischke}}]{Wagner:2024fhf}%
  \BibitemOpen
  \bibfield  {author} {\bibinfo {author} {\bibfnamefont {D.}~\bibnamefont
  {Wagner}}, \bibinfo {author} {\bibfnamefont {M.}~\bibnamefont {Shokri}},\
  and\ \bibinfo {author} {\bibfnamefont {D.~H.}\ \bibnamefont {Rischke}},\
  }\bibfield  {title} {\bibinfo {title} {{On the damping of spin waves}},\
  }\href@noop {} {\  (\bibinfo {year} {2024})},\ \Eprint
  {https://arxiv.org/abs/2405.00533} {arXiv:2405.00533 [nucl-th]} \BibitemShut
  {NoStop}%
\bibitem [{\citenamefont {Liu}\ and\ \citenamefont {Yin}(2021)}]{Liu:2021uhn}%
  \BibitemOpen
  \bibfield  {author} {\bibinfo {author} {\bibfnamefont {S.~Y.~F.}\
  \bibnamefont {Liu}}\ and\ \bibinfo {author} {\bibfnamefont {Y.}~\bibnamefont
  {Yin}},\ }\bibfield  {title} {\bibinfo {title} {{Spin polarization induced by
  the hydrodynamic gradients}},\ }\href
  {https://doi.org/10.1007/JHEP07(2021)188} {\bibfield  {journal} {\bibinfo
  {journal} {JHEP}\ }\textbf {\bibinfo {volume} {07}},\ \bibinfo {pages}
  {188}},\ \Eprint {https://arxiv.org/abs/2103.09200} {arXiv:2103.09200
  [hep-ph]} \BibitemShut {NoStop}%
\bibitem [{\citenamefont {Becattini}\ \emph
  {et~al.}(2021{\natexlab{b}})\citenamefont {Becattini}, \citenamefont
  {Buzzegoli},\ and\ \citenamefont {Palermo}}]{Becattini:2021suc}%
  \BibitemOpen
  \bibfield  {author} {\bibinfo {author} {\bibfnamefont {F.}~\bibnamefont
  {Becattini}}, \bibinfo {author} {\bibfnamefont {M.}~\bibnamefont
  {Buzzegoli}},\ and\ \bibinfo {author} {\bibfnamefont {A.}~\bibnamefont
  {Palermo}},\ }\bibfield  {title} {\bibinfo {title} {{Spin-thermal shear
  coupling in a relativistic fluid}},\ }\href
  {https://doi.org/10.1016/j.physletb.2021.136519} {\bibfield  {journal}
  {\bibinfo  {journal} {Phys. Lett. B}\ }\textbf {\bibinfo {volume} {820}},\
  \bibinfo {pages} {136519} (\bibinfo {year} {2021}{\natexlab{b}})},\ \Eprint
  {https://arxiv.org/abs/2103.10917} {arXiv:2103.10917 [nucl-th]} \BibitemShut
  {NoStop}%
\bibitem [{\citenamefont {Fu}\ \emph {et~al.}(2021{\natexlab{b}})\citenamefont
  {Fu}, \citenamefont {Liu}, \citenamefont {Pang}, \citenamefont {Song},\ and\
  \citenamefont {Yin}}]{Fu:2021pok}%
  \BibitemOpen
  \bibfield  {author} {\bibinfo {author} {\bibfnamefont {B.}~\bibnamefont
  {Fu}}, \bibinfo {author} {\bibfnamefont {S.~Y.~F.}\ \bibnamefont {Liu}},
  \bibinfo {author} {\bibfnamefont {L.}~\bibnamefont {Pang}}, \bibinfo {author}
  {\bibfnamefont {H.}~\bibnamefont {Song}},\ and\ \bibinfo {author}
  {\bibfnamefont {Y.}~\bibnamefont {Yin}},\ }\bibfield  {title} {\bibinfo
  {title} {{Shear-Induced Spin Polarization in Heavy-Ion Collisions}},\ }\href
  {https://doi.org/10.1103/PhysRevLett.127.142301} {\bibfield  {journal}
  {\bibinfo  {journal} {Phys. Rev. Lett.}\ }\textbf {\bibinfo {volume} {127}},\
  \bibinfo {pages} {142301} (\bibinfo {year} {2021}{\natexlab{b}})},\ \Eprint
  {https://arxiv.org/abs/2103.10403} {arXiv:2103.10403 [hep-ph]} \BibitemShut
  {NoStop}%
\bibitem [{\citenamefont {Lin}\ and\ \citenamefont {Wang}(2022)}]{Lin:2022tma}%
  \BibitemOpen
  \bibfield  {author} {\bibinfo {author} {\bibfnamefont {S.}~\bibnamefont
  {Lin}}\ and\ \bibinfo {author} {\bibfnamefont {Z.}~\bibnamefont {Wang}},\
  }\bibfield  {title} {\bibinfo {title} {{Shear induced polarization:
  collisional contributions}},\ }\href
  {https://doi.org/10.1007/JHEP12(2022)030} {\bibfield  {journal} {\bibinfo
  {journal} {JHEP}\ }\textbf {\bibinfo {volume} {12}},\ \bibinfo {pages}
  {030}},\ \Eprint {https://arxiv.org/abs/2206.12573} {arXiv:2206.12573
  [hep-ph]} \BibitemShut {NoStop}%
\bibitem [{\citenamefont {Lin}\ and\ \citenamefont {Wang}(2024)}]{Lin:2024zik}%
  \BibitemOpen
  \bibfield  {author} {\bibinfo {author} {\bibfnamefont {S.}~\bibnamefont
  {Lin}}\ and\ \bibinfo {author} {\bibfnamefont {Z.}~\bibnamefont {Wang}},\
  }\bibfield  {title} {\bibinfo {title} {{Steady state, displacement current
  and spin polarization for massless fermion in a shear flow}},\ }\href@noop {}
  {\  (\bibinfo {year} {2024})},\ \Eprint {https://arxiv.org/abs/2406.10003}
  {arXiv:2406.10003 [hep-ph]} \BibitemShut {NoStop}%
\bibitem [{\citenamefont {Kharzeev}\ \emph {et~al.}(2016)\citenamefont
  {Kharzeev}, \citenamefont {Liao}, \citenamefont {Voloshin},\ and\
  \citenamefont {Wang}}]{Kharzeev:2015znc}%
  \BibitemOpen
  \bibfield  {author} {\bibinfo {author} {\bibfnamefont {D.~E.}\ \bibnamefont
  {Kharzeev}}, \bibinfo {author} {\bibfnamefont {J.}~\bibnamefont {Liao}},
  \bibinfo {author} {\bibfnamefont {S.~A.}\ \bibnamefont {Voloshin}},\ and\
  \bibinfo {author} {\bibfnamefont {G.}~\bibnamefont {Wang}},\ }\bibfield
  {title} {\bibinfo {title} {{Chiral magnetic and vortical effects in
  high-energy nuclear collisions: A status report}},\ }\href
  {https://doi.org/10.1016/j.ppnp.2016.01.001} {\bibfield  {journal} {\bibinfo
  {journal} {Prog. Part. Nucl. Phys.}\ }\textbf {\bibinfo {volume} {88}},\
  \bibinfo {pages} {1} (\bibinfo {year} {2016})},\ \Eprint
  {https://arxiv.org/abs/1511.04050} {arXiv:1511.04050 [hep-ph]} \BibitemShut
  {NoStop}%
%%CITATION = ARXIV:1511.04050;%%
\bibitem [{\citenamefont {Skokov}\ \emph {et~al.}(2017)\citenamefont {Skokov},
  \citenamefont {Sorensen}, \citenamefont {Koch}, \citenamefont {Schlichting},
  \citenamefont {Thomas}, \citenamefont {Voloshin}, \citenamefont {Wang},\ and\
  \citenamefont {Yee}}]{Skokov:2016yrj}%
  \BibitemOpen
  \bibfield  {author} {\bibinfo {author} {\bibfnamefont {V.}~\bibnamefont
  {Skokov}}, \bibinfo {author} {\bibfnamefont {P.}~\bibnamefont {Sorensen}},
  \bibinfo {author} {\bibfnamefont {V.}~\bibnamefont {Koch}}, \bibinfo {author}
  {\bibfnamefont {S.}~\bibnamefont {Schlichting}}, \bibinfo {author}
  {\bibfnamefont {J.}~\bibnamefont {Thomas}}, \bibinfo {author} {\bibfnamefont
  {S.}~\bibnamefont {Voloshin}}, \bibinfo {author} {\bibfnamefont
  {G.}~\bibnamefont {Wang}},\ and\ \bibinfo {author} {\bibfnamefont {H.-U.}\
  \bibnamefont {Yee}},\ }\bibfield  {title} {\bibinfo {title} {{Chiral Magnetic
  Effect Task Force Report}},\ }\href
  {https://doi.org/10.1088/1674-1137/41/7/072001} {\bibfield  {journal}
  {\bibinfo  {journal} {Chin. Phys.}\ }\textbf {\bibinfo {volume} {C41}},\
  \bibinfo {pages} {072001} (\bibinfo {year} {2017})},\ \Eprint
  {https://arxiv.org/abs/1608.00982} {arXiv:1608.00982 [nucl-th]} \BibitemShut
  {NoStop}%
%%CITATION = ARXIV:1608.00982;%%
\bibitem [{\citenamefont {Hattori}\ and\ \citenamefont
  {Huang}(2017)}]{Hattori:2016emy}%
  \BibitemOpen
  \bibfield  {author} {\bibinfo {author} {\bibfnamefont {K.}~\bibnamefont
  {Hattori}}\ and\ \bibinfo {author} {\bibfnamefont {X.-G.}\ \bibnamefont
  {Huang}},\ }\bibfield  {title} {\bibinfo {title} {{Novel quantum phenomena
  induced by strong magnetic fields in heavy-ion collisions}},\ }\href
  {https://doi.org/10.1007/s41365-016-0178-3} {\bibfield  {journal} {\bibinfo
  {journal} {Nucl. Sci. Tech.}\ }\textbf {\bibinfo {volume} {28}},\ \bibinfo
  {pages} {26} (\bibinfo {year} {2017})},\ \Eprint
  {https://arxiv.org/abs/1609.00747} {arXiv:1609.00747 [nucl-th]} \BibitemShut
  {NoStop}%
\bibitem [{\citenamefont {Hernandez}\ and\ \citenamefont
  {Kovtun}(2017)}]{Hernandez:2017mch}%
  \BibitemOpen
  \bibfield  {author} {\bibinfo {author} {\bibfnamefont {J.}~\bibnamefont
  {Hernandez}}\ and\ \bibinfo {author} {\bibfnamefont {P.}~\bibnamefont
  {Kovtun}},\ }\bibfield  {title} {\bibinfo {title} {{Relativistic
  magnetohydrodynamics}},\ }\href {https://doi.org/10.1007/JHEP05(2017)001}
  {\bibfield  {journal} {\bibinfo  {journal} {JHEP}\ }\textbf {\bibinfo
  {volume} {05}},\ \bibinfo {pages} {001}},\ \Eprint
  {https://arxiv.org/abs/1703.08757} {arXiv:1703.08757 [hep-th]} \BibitemShut
  {NoStop}%
\bibitem [{\citenamefont {Fukushima}\ \emph {et~al.}(2008)\citenamefont
  {Fukushima}, \citenamefont {Kharzeev},\ and\ \citenamefont
  {Warringa}}]{Fukushima:2008xe}%
  \BibitemOpen
  \bibfield  {author} {\bibinfo {author} {\bibfnamefont {K.}~\bibnamefont
  {Fukushima}}, \bibinfo {author} {\bibfnamefont {D.~E.}\ \bibnamefont
  {Kharzeev}},\ and\ \bibinfo {author} {\bibfnamefont {H.~J.}\ \bibnamefont
  {Warringa}},\ }\bibfield  {title} {\bibinfo {title} {{The Chiral Magnetic
  Effect}},\ }\href {https://doi.org/10.1103/PhysRevD.78.074033} {\bibfield
  {journal} {\bibinfo  {journal} {Phys. Rev. D}\ }\textbf {\bibinfo {volume}
  {78}},\ \bibinfo {pages} {074033} (\bibinfo {year} {2008})},\ \Eprint
  {https://arxiv.org/abs/0808.3382} {arXiv:0808.3382 [hep-ph]} \BibitemShut
  {NoStop}%
\bibitem [{\citenamefont {Kharzeev}\ \emph {et~al.}(1998)\citenamefont
  {Kharzeev}, \citenamefont {Pisarski},\ and\ \citenamefont
  {Tytgat}}]{Kharzeev:1998kz}%
  \BibitemOpen
  \bibfield  {author} {\bibinfo {author} {\bibfnamefont {D.}~\bibnamefont
  {Kharzeev}}, \bibinfo {author} {\bibfnamefont {R.~D.}\ \bibnamefont
  {Pisarski}},\ and\ \bibinfo {author} {\bibfnamefont {M.~H.~G.}\ \bibnamefont
  {Tytgat}},\ }\bibfield  {title} {\bibinfo {title} {{Possibility of
  spontaneous parity violation in hot QCD}},\ }\href
  {https://doi.org/10.1103/PhysRevLett.81.512} {\bibfield  {journal} {\bibinfo
  {journal} {Phys. Rev. Lett.}\ }\textbf {\bibinfo {volume} {81}},\ \bibinfo
  {pages} {512} (\bibinfo {year} {1998})},\ \Eprint
  {https://arxiv.org/abs/hep-ph/9804221} {arXiv:hep-ph/9804221} \BibitemShut
  {NoStop}%
\bibitem [{\citenamefont {Kharzeev}(2006)}]{Kharzeev:2004ey}%
  \BibitemOpen
  \bibfield  {author} {\bibinfo {author} {\bibfnamefont {D.}~\bibnamefont
  {Kharzeev}},\ }\bibfield  {title} {\bibinfo {title} {{Parity violation in hot
  QCD: Why it can happen, and how to look for it}},\ }\href
  {https://doi.org/10.1016/j.physletb.2005.11.075} {\bibfield  {journal}
  {\bibinfo  {journal} {Phys. Lett. B}\ }\textbf {\bibinfo {volume} {633}},\
  \bibinfo {pages} {260} (\bibinfo {year} {2006})},\ \Eprint
  {https://arxiv.org/abs/hep-ph/0406125} {arXiv:hep-ph/0406125} \BibitemShut
  {NoStop}%
\bibitem [{\citenamefont {Kharzeev}\ \emph {et~al.}(2008)\citenamefont
  {Kharzeev}, \citenamefont {McLerran},\ and\ \citenamefont
  {Warringa}}]{Kharzeev:2007jp}%
  \BibitemOpen
  \bibfield  {author} {\bibinfo {author} {\bibfnamefont {D.~E.}\ \bibnamefont
  {Kharzeev}}, \bibinfo {author} {\bibfnamefont {L.~D.}\ \bibnamefont
  {McLerran}},\ and\ \bibinfo {author} {\bibfnamefont {H.~J.}\ \bibnamefont
  {Warringa}},\ }\bibfield  {title} {\bibinfo {title} {{The Effects of
  topological charge change in heavy ion collisions: 'Event by event P and CP
  violation'}},\ }\href {https://doi.org/10.1016/j.nuclphysa.2008.02.298}
  {\bibfield  {journal} {\bibinfo  {journal} {Nucl. Phys. A}\ }\textbf
  {\bibinfo {volume} {803}},\ \bibinfo {pages} {227} (\bibinfo {year}
  {2008})},\ \Eprint {https://arxiv.org/abs/0711.0950} {arXiv:0711.0950
  [hep-ph]} \BibitemShut {NoStop}%
\bibitem [{\citenamefont {Kharzeev}\ and\ \citenamefont
  {Yee}(2011)}]{Kharzeev:2010gd}%
  \BibitemOpen
  \bibfield  {author} {\bibinfo {author} {\bibfnamefont {D.~E.}\ \bibnamefont
  {Kharzeev}}\ and\ \bibinfo {author} {\bibfnamefont {H.-U.}\ \bibnamefont
  {Yee}},\ }\bibfield  {title} {\bibinfo {title} {{Chiral Magnetic Wave}},\
  }\href {https://doi.org/10.1103/PhysRevD.83.085007} {\bibfield  {journal}
  {\bibinfo  {journal} {Phys. Rev. D}\ }\textbf {\bibinfo {volume} {83}},\
  \bibinfo {pages} {085007} (\bibinfo {year} {2011})},\ \Eprint
  {https://arxiv.org/abs/1012.6026} {arXiv:1012.6026 [hep-th]} \BibitemShut
  {NoStop}%
\bibitem [{\citenamefont {Nakamura}\ \emph
  {et~al.}(2023{\natexlab{a}})\citenamefont {Nakamura}, \citenamefont
  {Miyoshi}, \citenamefont {Nonaka},\ and\ \citenamefont
  {Takahashi}}]{Nakamura:2022idq}%
  \BibitemOpen
  \bibfield  {author} {\bibinfo {author} {\bibfnamefont {K.}~\bibnamefont
  {Nakamura}}, \bibinfo {author} {\bibfnamefont {T.}~\bibnamefont {Miyoshi}},
  \bibinfo {author} {\bibfnamefont {C.}~\bibnamefont {Nonaka}},\ and\ \bibinfo
  {author} {\bibfnamefont {H.~R.}\ \bibnamefont {Takahashi}},\ }\bibfield
  {title} {\bibinfo {title} {{Directed flow in relativistic resistive
  magneto-hydrodynamic expansion for symmetric and asymmetric collision
  systems}},\ }\href {https://doi.org/10.1103/PhysRevC.107.014901} {\bibfield
  {journal} {\bibinfo  {journal} {Phys. Rev. C}\ }\textbf {\bibinfo {volume}
  {107}},\ \bibinfo {pages} {014901} (\bibinfo {year} {2023}{\natexlab{a}})},\
  \Eprint {https://arxiv.org/abs/2209.00323} {arXiv:2209.00323 [nucl-th]}
  \BibitemShut {NoStop}%
\bibitem [{\citenamefont {Nakamura}\ \emph
  {et~al.}(2023{\natexlab{b}})\citenamefont {Nakamura}, \citenamefont
  {Miyoshi}, \citenamefont {Nonaka},\ and\ \citenamefont
  {Takahashi}}]{Nakamura:2022wqr}%
  \BibitemOpen
  \bibfield  {author} {\bibinfo {author} {\bibfnamefont {K.}~\bibnamefont
  {Nakamura}}, \bibinfo {author} {\bibfnamefont {T.}~\bibnamefont {Miyoshi}},
  \bibinfo {author} {\bibfnamefont {C.}~\bibnamefont {Nonaka}},\ and\ \bibinfo
  {author} {\bibfnamefont {H.~R.}\ \bibnamefont {Takahashi}},\ }\bibfield
  {title} {\bibinfo {title} {{Relativistic resistive magneto-hydrodynamics code
  for high-energy heavy-ion collisions}},\ }\href
  {https://doi.org/10.1140/epjc/s10052-023-11343-y} {\bibfield  {journal}
  {\bibinfo  {journal} {Eur. Phys. J. C}\ }\textbf {\bibinfo {volume} {83}},\
  \bibinfo {pages} {229} (\bibinfo {year} {2023}{\natexlab{b}})},\ \Eprint
  {https://arxiv.org/abs/2211.02310} {arXiv:2211.02310 [nucl-th]} \BibitemShut
  {NoStop}%
\bibitem [{\citenamefont {Nakamura}\ \emph
  {et~al.}(2023{\natexlab{c}})\citenamefont {Nakamura}, \citenamefont
  {Miyoshi}, \citenamefont {Nonaka},\ and\ \citenamefont
  {Takahashi}}]{Nakamura:2022ssn}%
  \BibitemOpen
  \bibfield  {author} {\bibinfo {author} {\bibfnamefont {K.}~\bibnamefont
  {Nakamura}}, \bibinfo {author} {\bibfnamefont {T.}~\bibnamefont {Miyoshi}},
  \bibinfo {author} {\bibfnamefont {C.}~\bibnamefont {Nonaka}},\ and\ \bibinfo
  {author} {\bibfnamefont {H.~R.}\ \bibnamefont {Takahashi}},\ }\bibfield
  {title} {\bibinfo {title} {{Charge-dependent anisotropic flow in high-energy
  heavy-ion collisions from a relativistic resistive magneto-hydrodynamic
  expansion}},\ }\href {https://doi.org/10.1103/PhysRevC.107.034912} {\bibfield
   {journal} {\bibinfo  {journal} {Phys. Rev. C}\ }\textbf {\bibinfo {volume}
  {107}},\ \bibinfo {pages} {034912} (\bibinfo {year} {2023}{\natexlab{c}})},\
  \Eprint {https://arxiv.org/abs/2212.02124} {arXiv:2212.02124 [nucl-th]}
  \BibitemShut {NoStop}%
\bibitem [{\citenamefont {Mayer}\ \emph
  {et~al.}(2024{\natexlab{a}})\citenamefont {Mayer}, \citenamefont {Dash},
  \citenamefont {Inghirami}, \citenamefont {Elfner}, \citenamefont {Rezzolla},\
  and\ \citenamefont {Rischke}}]{Mayer:2024dze}%
  \BibitemOpen
  \bibfield  {author} {\bibinfo {author} {\bibfnamefont {M.}~\bibnamefont
  {Mayer}}, \bibinfo {author} {\bibfnamefont {A.}~\bibnamefont {Dash}},
  \bibinfo {author} {\bibfnamefont {G.}~\bibnamefont {Inghirami}}, \bibinfo
  {author} {\bibfnamefont {H.}~\bibnamefont {Elfner}}, \bibinfo {author}
  {\bibfnamefont {L.}~\bibnamefont {Rezzolla}},\ and\ \bibinfo {author}
  {\bibfnamefont {D.~H.}\ \bibnamefont {Rischke}},\ }\bibfield  {title}
  {\bibinfo {title} {{BHAC-QGP: three-dimensional MHD simulations of
  relativistic heavy-ion collisions, I. Methods and tests}},\ }\href@noop {} {\
   (\bibinfo {year} {2024}{\natexlab{a}})},\ \Eprint
  {https://arxiv.org/abs/2403.08668} {arXiv:2403.08668 [hep-ph]} \BibitemShut
  {NoStop}%
\bibitem [{\citenamefont {Mayer}\ \emph
  {et~al.}(2024{\natexlab{b}})\citenamefont {Mayer}, \citenamefont {Dash},
  \citenamefont {Inghirami}, \citenamefont {Elfner}, \citenamefont {Rezzolla},\
  and\ \citenamefont {Rischke}}]{Mayer:2024kkv}%
  \BibitemOpen
  \bibfield  {author} {\bibinfo {author} {\bibfnamefont {M.}~\bibnamefont
  {Mayer}}, \bibinfo {author} {\bibfnamefont {A.}~\bibnamefont {Dash}},
  \bibinfo {author} {\bibfnamefont {G.}~\bibnamefont {Inghirami}}, \bibinfo
  {author} {\bibfnamefont {H.}~\bibnamefont {Elfner}}, \bibinfo {author}
  {\bibfnamefont {L.}~\bibnamefont {Rezzolla}},\ and\ \bibinfo {author}
  {\bibfnamefont {D.~H.}\ \bibnamefont {Rischke}},\ }\bibfield  {title}
  {\bibinfo {title} {{BHAC-QGP: three-dimensional MHD simulations of
  relativistic heavy-ion collisions, II. Application to Au-Au collisions}},\
  }\href@noop {} {\  (\bibinfo {year} {2024}{\natexlab{b}})},\ \Eprint
  {https://arxiv.org/abs/2403.08669} {arXiv:2403.08669 [hep-ph]} \BibitemShut
  {NoStop}%
\bibitem [{\citenamefont {Xu}\ \emph {et~al.}(2022)\citenamefont {Xu},
  \citenamefont {Lin}, \citenamefont {Huang},\ and\ \citenamefont
  {Huang}}]{Xu:2022hql}%
  \BibitemOpen
  \bibfield  {author} {\bibinfo {author} {\bibfnamefont {K.}~\bibnamefont
  {Xu}}, \bibinfo {author} {\bibfnamefont {F.}~\bibnamefont {Lin}}, \bibinfo
  {author} {\bibfnamefont {A.}~\bibnamefont {Huang}},\ and\ \bibinfo {author}
  {\bibfnamefont {M.}~\bibnamefont {Huang}},\ }\bibfield  {title} {\bibinfo
  {title} {{\ensuremath{\Lambda}/\ensuremath{\Lambda}\textasciimacron{}
  polarization and splitting induced by rotation and magnetic field}},\ }\href
  {https://doi.org/10.1103/PhysRevD.106.L071502} {\bibfield  {journal}
  {\bibinfo  {journal} {Phys. Rev. D}\ }\textbf {\bibinfo {volume} {106}},\
  \bibinfo {pages} {L071502} (\bibinfo {year} {2022})},\ \Eprint
  {https://arxiv.org/abs/2205.02420} {arXiv:2205.02420 [hep-ph]} \BibitemShut
  {NoStop}%
\bibitem [{\citenamefont {Peng}\ \emph {et~al.}(2023)\citenamefont {Peng},
  \citenamefont {Wu}, \citenamefont {Wang}, \citenamefont {She},\ and\
  \citenamefont {Pu}}]{Peng:2022cya}%
  \BibitemOpen
  \bibfield  {author} {\bibinfo {author} {\bibfnamefont {H.-H.}\ \bibnamefont
  {Peng}}, \bibinfo {author} {\bibfnamefont {S.}~\bibnamefont {Wu}}, \bibinfo
  {author} {\bibfnamefont {R.-j.}\ \bibnamefont {Wang}}, \bibinfo {author}
  {\bibfnamefont {D.}~\bibnamefont {She}},\ and\ \bibinfo {author}
  {\bibfnamefont {S.}~\bibnamefont {Pu}},\ }\bibfield  {title} {\bibinfo
  {title} {{Anomalous magnetohydrodynamics with temperature-dependent electric
  conductivity and application to the global polarization}},\ }\href
  {https://doi.org/10.1103/PhysRevD.107.096010} {\bibfield  {journal} {\bibinfo
   {journal} {Phys. Rev. D}\ }\textbf {\bibinfo {volume} {107}},\ \bibinfo
  {pages} {096010} (\bibinfo {year} {2023})},\ \Eprint
  {https://arxiv.org/abs/2211.11286} {arXiv:2211.11286 [hep-ph]} \BibitemShut
  {NoStop}%
\bibitem [{\citenamefont {Buzzegoli}(2023)}]{Buzzegoli:2022qrr}%
  \BibitemOpen
  \bibfield  {author} {\bibinfo {author} {\bibfnamefont {M.}~\bibnamefont
  {Buzzegoli}},\ }\bibfield  {title} {\bibinfo {title} {{Spin polarization
  induced by magnetic field and the relativistic Barnett effect}},\ }\href
  {https://doi.org/10.1016/j.nuclphysa.2023.122674} {\bibfield  {journal}
  {\bibinfo  {journal} {Nucl. Phys. A}\ }\textbf {\bibinfo {volume} {1036}},\
  \bibinfo {pages} {122674} (\bibinfo {year} {2023})},\ \Eprint
  {https://arxiv.org/abs/2211.04549} {arXiv:2211.04549 [nucl-th]} \BibitemShut
  {NoStop}%
\bibitem [{\citenamefont {Sun}\ and\ \citenamefont {Yan}(2024)}]{Sun:2024isb}%
  \BibitemOpen
  \bibfield  {author} {\bibinfo {author} {\bibfnamefont {J.-A.}\ \bibnamefont
  {Sun}}\ and\ \bibinfo {author} {\bibfnamefont {L.}~\bibnamefont {Yan}},\
  }\bibfield  {title} {\bibinfo {title} {{Weak magnetic effect in quark-gluon
  plasma and local spin polarization}},\ }\href@noop {} {\  (\bibinfo {year}
  {2024})},\ \Eprint {https://arxiv.org/abs/2401.07458} {arXiv:2401.07458
  [nucl-th]} \BibitemShut {NoStop}%
\bibitem [{\citenamefont {Fang}\ \emph {et~al.}(2024)\citenamefont {Fang},
  \citenamefont {Hattori},\ and\ \citenamefont {Hu}}]{Fang:2024skm}%
  \BibitemOpen
  \bibfield  {author} {\bibinfo {author} {\bibfnamefont {Z.}~\bibnamefont
  {Fang}}, \bibinfo {author} {\bibfnamefont {K.}~\bibnamefont {Hattori}},\ and\
  \bibinfo {author} {\bibfnamefont {J.}~\bibnamefont {Hu}},\ }\bibfield
  {title} {\bibinfo {title} {{Analytic solutions for the linearized first-order
  magnetohydrodynamics and implications for causality and stability}},\ }\href
  {https://doi.org/10.1103/PhysRevD.110.056049} {\bibfield  {journal} {\bibinfo
   {journal} {Phys. Rev. D}\ }\textbf {\bibinfo {volume} {110}},\ \bibinfo
  {pages} {056049} (\bibinfo {year} {2024})},\ \Eprint
  {https://arxiv.org/abs/2402.18601} {arXiv:2402.18601 [physics.plasm-ph]}
  \BibitemShut {NoStop}%
\bibitem [{\citenamefont {Grozdanov}\ \emph {et~al.}(2017)\citenamefont
  {Grozdanov}, \citenamefont {Hofman},\ and\ \citenamefont
  {Iqbal}}]{Grozdanov:2016tdf}%
  \BibitemOpen
  \bibfield  {author} {\bibinfo {author} {\bibfnamefont {S.}~\bibnamefont
  {Grozdanov}}, \bibinfo {author} {\bibfnamefont {D.~M.}\ \bibnamefont
  {Hofman}},\ and\ \bibinfo {author} {\bibfnamefont {N.}~\bibnamefont
  {Iqbal}},\ }\bibfield  {title} {\bibinfo {title} {{Generalized global
  symmetries and dissipative magnetohydrodynamics}},\ }\href
  {https://doi.org/10.1103/PhysRevD.95.096003} {\bibfield  {journal} {\bibinfo
  {journal} {Phys. Rev. D}\ }\textbf {\bibinfo {volume} {95}},\ \bibinfo
  {pages} {096003} (\bibinfo {year} {2017})},\ \Eprint
  {https://arxiv.org/abs/1610.07392} {arXiv:1610.07392 [hep-th]} \BibitemShut
  {NoStop}%
\bibitem [{\citenamefont {Pu}\ \emph {et~al.}(2016)\citenamefont {Pu},
  \citenamefont {Roy}, \citenamefont {Rezzolla},\ and\ \citenamefont
  {Rischke}}]{Pu:2016ayh}%
  \BibitemOpen
  \bibfield  {author} {\bibinfo {author} {\bibfnamefont {S.}~\bibnamefont
  {Pu}}, \bibinfo {author} {\bibfnamefont {V.}~\bibnamefont {Roy}}, \bibinfo
  {author} {\bibfnamefont {L.}~\bibnamefont {Rezzolla}},\ and\ \bibinfo
  {author} {\bibfnamefont {D.~H.}\ \bibnamefont {Rischke}},\ }\bibfield
  {title} {\bibinfo {title} {{Bjorken flow in one-dimensional relativistic
  magnetohydrodynamics with magnetization}},\ }\href
  {https://doi.org/10.1103/PhysRevD.93.074022} {\bibfield  {journal} {\bibinfo
  {journal} {Phys. Rev. D}\ }\textbf {\bibinfo {volume} {93}},\ \bibinfo
  {pages} {074022} (\bibinfo {year} {2016})},\ \Eprint
  {https://arxiv.org/abs/1602.04953} {arXiv:1602.04953 [nucl-th]} \BibitemShut
  {NoStop}%
\bibitem [{\citenamefont {Hattori}\ \emph
  {et~al.}(2019{\natexlab{b}})\citenamefont {Hattori}, \citenamefont {Hirono},
  \citenamefont {Yee},\ and\ \citenamefont {Yin}}]{Hattori:2017usa}%
  \BibitemOpen
  \bibfield  {author} {\bibinfo {author} {\bibfnamefont {K.}~\bibnamefont
  {Hattori}}, \bibinfo {author} {\bibfnamefont {Y.}~\bibnamefont {Hirono}},
  \bibinfo {author} {\bibfnamefont {H.-U.}\ \bibnamefont {Yee}},\ and\ \bibinfo
  {author} {\bibfnamefont {Y.}~\bibnamefont {Yin}},\ }\bibfield  {title}
  {\bibinfo {title} {{MagnetoHydrodynamics with chiral anomaly: phases of
  collective excitations and instabilities}},\ }\href
  {https://doi.org/10.1103/PhysRevD.100.065023} {\bibfield  {journal} {\bibinfo
   {journal} {Phys. Rev. D}\ }\textbf {\bibinfo {volume} {100}},\ \bibinfo
  {pages} {065023} (\bibinfo {year} {2019}{\natexlab{b}})},\ \Eprint
  {https://arxiv.org/abs/1711.08450} {arXiv:1711.08450 [hep-th]} \BibitemShut
  {NoStop}%
\bibitem [{\citenamefont {Denicol}\ \emph {et~al.}(2018)\citenamefont
  {Denicol}, \citenamefont {Huang}, \citenamefont {Moln\'ar}, \citenamefont
  {Monteiro}, \citenamefont {Niemi}, \citenamefont {Noronha}, \citenamefont
  {Rischke},\ and\ \citenamefont {Wang}}]{Denicol:2018rbw}%
  \BibitemOpen
  \bibfield  {author} {\bibinfo {author} {\bibfnamefont {G.~S.}\ \bibnamefont
  {Denicol}}, \bibinfo {author} {\bibfnamefont {X.-G.}\ \bibnamefont {Huang}},
  \bibinfo {author} {\bibfnamefont {E.}~\bibnamefont {Moln\'ar}}, \bibinfo
  {author} {\bibfnamefont {G.~M.}\ \bibnamefont {Monteiro}}, \bibinfo {author}
  {\bibfnamefont {H.}~\bibnamefont {Niemi}}, \bibinfo {author} {\bibfnamefont
  {J.}~\bibnamefont {Noronha}}, \bibinfo {author} {\bibfnamefont {D.~H.}\
  \bibnamefont {Rischke}},\ and\ \bibinfo {author} {\bibfnamefont
  {Q.}~\bibnamefont {Wang}},\ }\bibfield  {title} {\bibinfo {title}
  {{Nonresistive dissipative magnetohydrodynamics from the Boltzmann equation
  in the 14-moment approximation}},\ }\href
  {https://doi.org/10.1103/PhysRevD.98.076009} {\bibfield  {journal} {\bibinfo
  {journal} {Phys. Rev. D}\ }\textbf {\bibinfo {volume} {98}},\ \bibinfo
  {pages} {076009} (\bibinfo {year} {2018})},\ \Eprint
  {https://arxiv.org/abs/1804.05210} {arXiv:1804.05210 [nucl-th]} \BibitemShut
  {NoStop}%
\bibitem [{\citenamefont {Denicol}\ \emph {et~al.}(2019)\citenamefont
  {Denicol}, \citenamefont {Moln\'ar}, \citenamefont {Niemi},\ and\
  \citenamefont {Rischke}}]{Denicol:2019iyh}%
  \BibitemOpen
  \bibfield  {author} {\bibinfo {author} {\bibfnamefont {G.~S.}\ \bibnamefont
  {Denicol}}, \bibinfo {author} {\bibfnamefont {E.}~\bibnamefont {Moln\'ar}},
  \bibinfo {author} {\bibfnamefont {H.}~\bibnamefont {Niemi}},\ and\ \bibinfo
  {author} {\bibfnamefont {D.~H.}\ \bibnamefont {Rischke}},\ }\bibfield
  {title} {\bibinfo {title} {{Resistive dissipative magnetohydrodynamics from
  the Boltzmann-Vlasov equation}},\ }\href
  {https://doi.org/10.1103/PhysRevD.99.056017} {\bibfield  {journal} {\bibinfo
  {journal} {Phys. Rev. D}\ }\textbf {\bibinfo {volume} {99}},\ \bibinfo
  {pages} {056017} (\bibinfo {year} {2019})},\ \Eprint
  {https://arxiv.org/abs/1902.01699} {arXiv:1902.01699 [nucl-th]} \BibitemShut
  {NoStop}%
\bibitem [{\citenamefont {Armas}\ and\ \citenamefont
  {Camilloni}(2022)}]{Armas:2022wvb}%
  \BibitemOpen
  \bibfield  {author} {\bibinfo {author} {\bibfnamefont {J.}~\bibnamefont
  {Armas}}\ and\ \bibinfo {author} {\bibfnamefont {F.}~\bibnamefont
  {Camilloni}},\ }\bibfield  {title} {\bibinfo {title} {{A stable and causal
  model of magnetohydrodynamics}},\ }\href
  {https://doi.org/10.1088/1475-7516/2022/10/039} {\bibfield  {journal}
  {\bibinfo  {journal} {JCAP}\ }\textbf {\bibinfo {volume} {10}},\ \bibinfo
  {pages} {039}},\ \Eprint {https://arxiv.org/abs/2201.06847} {arXiv:2201.06847
  [hep-th]} \BibitemShut {NoStop}%
\bibitem [{\citenamefont {Panda}\ \emph {et~al.}(2021)\citenamefont {Panda},
  \citenamefont {Dash}, \citenamefont {Biswas},\ and\ \citenamefont
  {Roy}}]{Panda:2021pvq}%
  \BibitemOpen
  \bibfield  {author} {\bibinfo {author} {\bibfnamefont {A.~K.}\ \bibnamefont
  {Panda}}, \bibinfo {author} {\bibfnamefont {A.}~\bibnamefont {Dash}},
  \bibinfo {author} {\bibfnamefont {R.}~\bibnamefont {Biswas}},\ and\ \bibinfo
  {author} {\bibfnamefont {V.}~\bibnamefont {Roy}},\ }\bibfield  {title}
  {\bibinfo {title} {{Relativistic resistive dissipative magnetohydrodynamics
  from the relaxation time approximation}},\ }\href
  {https://doi.org/10.1103/PhysRevD.104.054004} {\bibfield  {journal} {\bibinfo
   {journal} {Phys. Rev. D}\ }\textbf {\bibinfo {volume} {104}},\ \bibinfo
  {pages} {054004} (\bibinfo {year} {2021})},\ \Eprint
  {https://arxiv.org/abs/2104.12179} {arXiv:2104.12179 [nucl-th]} \BibitemShut
  {NoStop}%
\bibitem [{\citenamefont {Hattori}\ \emph {et~al.}(2022)\citenamefont
  {Hattori}, \citenamefont {Hongo},\ and\ \citenamefont
  {Huang}}]{Hattori:2022hyo}%
  \BibitemOpen
  \bibfield  {author} {\bibinfo {author} {\bibfnamefont {K.}~\bibnamefont
  {Hattori}}, \bibinfo {author} {\bibfnamefont {M.}~\bibnamefont {Hongo}},\
  and\ \bibinfo {author} {\bibfnamefont {X.-G.}\ \bibnamefont {Huang}},\
  }\bibfield  {title} {\bibinfo {title} {{New Developments in Relativistic
  Magnetohydrodynamics}},\ }\href {https://doi.org/10.3390/sym14091851}
  {\bibfield  {journal} {\bibinfo  {journal} {Symmetry}\ }\textbf {\bibinfo
  {volume} {14}},\ \bibinfo {pages} {1851} (\bibinfo {year} {2022})},\ \Eprint
  {https://arxiv.org/abs/2207.12794} {arXiv:2207.12794 [hep-th]} \BibitemShut
  {NoStop}%
\bibitem [{\citenamefont {Biswas}\ \emph {et~al.}(2020)\citenamefont {Biswas},
  \citenamefont {Dash}, \citenamefont {Haque}, \citenamefont {Pu},\ and\
  \citenamefont {Roy}}]{Biswas:2020rps}%
  \BibitemOpen
  \bibfield  {author} {\bibinfo {author} {\bibfnamefont {R.}~\bibnamefont
  {Biswas}}, \bibinfo {author} {\bibfnamefont {A.}~\bibnamefont {Dash}},
  \bibinfo {author} {\bibfnamefont {N.}~\bibnamefont {Haque}}, \bibinfo
  {author} {\bibfnamefont {S.}~\bibnamefont {Pu}},\ and\ \bibinfo {author}
  {\bibfnamefont {V.}~\bibnamefont {Roy}},\ }\bibfield  {title} {\bibinfo
  {title} {{Causality and stability in relativistic viscous non-resistive
  magneto-fluid dynamics}},\ }\href {https://doi.org/10.1007/JHEP10(2020)171}
  {\bibfield  {journal} {\bibinfo  {journal} {JHEP}\ }\textbf {\bibinfo
  {volume} {10}},\ \bibinfo {pages} {171}},\ \Eprint
  {https://arxiv.org/abs/2007.05431} {arXiv:2007.05431 [nucl-th]} \BibitemShut
  {NoStop}%
\bibitem [{\citenamefont {Bhadury}\ \emph {et~al.}(2022)\citenamefont
  {Bhadury}, \citenamefont {Florkowski}, \citenamefont {Jaiswal}, \citenamefont
  {Kumar},\ and\ \citenamefont {Ryblewski}}]{Bhadury:2022ulr}%
  \BibitemOpen
  \bibfield  {author} {\bibinfo {author} {\bibfnamefont {S.}~\bibnamefont
  {Bhadury}}, \bibinfo {author} {\bibfnamefont {W.}~\bibnamefont {Florkowski}},
  \bibinfo {author} {\bibfnamefont {A.}~\bibnamefont {Jaiswal}}, \bibinfo
  {author} {\bibfnamefont {A.}~\bibnamefont {Kumar}},\ and\ \bibinfo {author}
  {\bibfnamefont {R.}~\bibnamefont {Ryblewski}},\ }\bibfield  {title} {\bibinfo
  {title} {{Relativistic Spin Magnetohydrodynamics}},\ }\href
  {https://doi.org/10.1103/PhysRevLett.129.192301} {\bibfield  {journal}
  {\bibinfo  {journal} {Phys. Rev. Lett.}\ }\textbf {\bibinfo {volume} {129}},\
  \bibinfo {pages} {192301} (\bibinfo {year} {2022})},\ \Eprint
  {https://arxiv.org/abs/2204.01357} {arXiv:2204.01357 [nucl-th]} \BibitemShut
  {NoStop}%
\bibitem [{\citenamefont {Bhadury}\ \emph {et~al.}(2024)\citenamefont
  {Bhadury}, \citenamefont {Florkowski}, \citenamefont {Jaiswal}, \citenamefont
  {Kumar},\ and\ \citenamefont {Ryblewski}}]{Bhadury:2024whs}%
  \BibitemOpen
  \bibfield  {author} {\bibinfo {author} {\bibfnamefont {S.}~\bibnamefont
  {Bhadury}}, \bibinfo {author} {\bibfnamefont {W.}~\bibnamefont {Florkowski}},
  \bibinfo {author} {\bibfnamefont {A.}~\bibnamefont {Jaiswal}}, \bibinfo
  {author} {\bibfnamefont {A.}~\bibnamefont {Kumar}},\ and\ \bibinfo {author}
  {\bibfnamefont {R.}~\bibnamefont {Ryblewski}},\ }\bibfield  {title} {\bibinfo
  {title} {{Relativistic magnetohydrodynamics with spin}},\ }in\ \href@noop {}
  {\emph {\bibinfo {booktitle} {{25th International Spin Symposium}}}}\
  (\bibinfo {year} {2024})\ \Eprint {https://arxiv.org/abs/2401.16033}
  {arXiv:2401.16033 [hep-ph]} \BibitemShut {NoStop}%
\bibitem [{\citenamefont {Hongo}\ and\ \citenamefont
  {Hattori}(2021)}]{Hongo:2020qpv}%
  \BibitemOpen
  \bibfield  {author} {\bibinfo {author} {\bibfnamefont {M.}~\bibnamefont
  {Hongo}}\ and\ \bibinfo {author} {\bibfnamefont {K.}~\bibnamefont
  {Hattori}},\ }\bibfield  {title} {\bibinfo {title} {{Revisiting relativistic
  magnetohydrodynamics from quantum electrodynamics}},\ }\href
  {https://doi.org/10.1007/JHEP02(2021)011} {\bibfield  {journal} {\bibinfo
  {journal} {JHEP}\ }\textbf {\bibinfo {volume} {02}},\ \bibinfo {pages}
  {011}},\ \Eprint {https://arxiv.org/abs/2005.10239} {arXiv:2005.10239
  [hep-th]} \BibitemShut {NoStop}%
\bibitem [{\citenamefont {Belinfante}(1939)}]{Belinfante1939}%
  \BibitemOpen
  \bibfield  {author} {\bibinfo {author} {\bibfnamefont {F.}~\bibnamefont
  {Belinfante}},\ }\bibfield  {title} {\bibinfo {title} {On the spin angular
  momentum of mesons},\ }\href
  {https://doi.org/https://doi.org/10.1016/S0031-8914(39)90090-X} {\bibfield
  {journal} {\bibinfo  {journal} {Physica}\ }\textbf {\bibinfo {volume} {6}},\
  \bibinfo {pages} {887 } (\bibinfo {year} {1939})}\BibitemShut {NoStop}%
\bibitem [{\citenamefont {Belinfante}(1940)}]{Belinfante1940}%
  \BibitemOpen
  \bibfield  {author} {\bibinfo {author} {\bibfnamefont {F.}~\bibnamefont
  {Belinfante}},\ }\bibfield  {title} {\bibinfo {title} {On the current and the
  density of the electric charge, the energy, the linear momentum and the
  angular momentum of arbitrary fields},\ }\href
  {https://doi.org/https://doi.org/10.1016/S0031-8914(40)90091-X} {\bibfield
  {journal} {\bibinfo  {journal} {Physica}\ }\textbf {\bibinfo {volume} {7}},\
  \bibinfo {pages} {449 } (\bibinfo {year} {1940})}\BibitemShut {NoStop}%
\bibitem [{\citenamefont {Rosenfeld}(1940)}]{Rosenfeld1940}%
  \BibitemOpen
  \bibfield  {author} {\bibinfo {author} {\bibfnamefont {L.}~\bibnamefont
  {Rosenfeld}},\ }\bibfield  {title} {\bibinfo {title} {{On the current and the
  density of the electric charge, the energy, the linear momentum and the
  angular momentum of arbitrary fields}},\ }\href@noop {} {\bibfield  {journal}
  {\bibinfo  {journal} {Mem. Acad. Roy. Belg. Cl. Sc.}\ }\textbf {\bibinfo
  {volume} {18}},\ \bibinfo {pages} {1 } (\bibinfo {year} {1940})}\BibitemShut
  {NoStop}%
\bibitem [{\citenamefont {Israel}(1976)}]{Israel:1976tn}%
  \BibitemOpen
  \bibfield  {author} {\bibinfo {author} {\bibfnamefont {W.}~\bibnamefont
  {Israel}},\ }\bibfield  {title} {\bibinfo {title} {{Nonstationary
  irreversible thermodynamics: A Causal relativistic theory}},\ }\href
  {https://doi.org/10.1016/0003-4916(76)90064-6} {\bibfield  {journal}
  {\bibinfo  {journal} {Annals Phys.}\ }\textbf {\bibinfo {volume} {100}},\
  \bibinfo {pages} {310} (\bibinfo {year} {1976})}\BibitemShut {NoStop}%
%%CITATION = APNYA,100,310;%%
\bibitem [{\citenamefont {Israel}\ and\ \citenamefont
  {Stewart}(1976)}]{Israel:1976efz}%
  \BibitemOpen
  \bibfield  {author} {\bibinfo {author} {\bibfnamefont {W.}~\bibnamefont
  {Israel}}\ and\ \bibinfo {author} {\bibfnamefont {J.~M.}\ \bibnamefont
  {Stewart}},\ }\bibfield  {title} {\bibinfo {title} {{Thermodynamics of
  nonstationary and transient effects in a relativistic gas}},\ }\href
  {https://doi.org/10.1016/0375-9601(76)90075-X} {\bibfield  {journal}
  {\bibinfo  {journal} {Phys. Lett. A}\ }\textbf {\bibinfo {volume} {58}},\
  \bibinfo {pages} {213} (\bibinfo {year} {1976})}\BibitemShut {NoStop}%
\bibitem [{\citenamefont {Israel}\ and\ \citenamefont
  {Stewart}(1979)}]{Israel:1979wp}%
  \BibitemOpen
  \bibfield  {author} {\bibinfo {author} {\bibfnamefont {W.}~\bibnamefont
  {Israel}}\ and\ \bibinfo {author} {\bibfnamefont {J.~M.}\ \bibnamefont
  {Stewart}},\ }\bibfield  {title} {\bibinfo {title} {{Transient relativistic
  thermodynamics and kinetic theory}},\ }\href
  {https://doi.org/10.1016/0003-4916(79)90130-1} {\bibfield  {journal}
  {\bibinfo  {journal} {Annals Phys.}\ }\textbf {\bibinfo {volume} {118}},\
  \bibinfo {pages} {341} (\bibinfo {year} {1979})}\BibitemShut {NoStop}%
%%CITATION = APNYA,118,341;%%
\bibitem [{\citenamefont {Bemfica}\ \emph {et~al.}(2018)\citenamefont
  {Bemfica}, \citenamefont {Disconzi},\ and\ \citenamefont
  {Noronha}}]{Bemfica:2017wps}%
  \BibitemOpen
  \bibfield  {author} {\bibinfo {author} {\bibfnamefont {F.~S.}\ \bibnamefont
  {Bemfica}}, \bibinfo {author} {\bibfnamefont {M.~M.}\ \bibnamefont
  {Disconzi}},\ and\ \bibinfo {author} {\bibfnamefont {J.}~\bibnamefont
  {Noronha}},\ }\bibfield  {title} {\bibinfo {title} {{Causality and existence
  of solutions of relativistic viscous fluid dynamics with gravity}},\ }\href
  {https://doi.org/10.1103/PhysRevD.98.104064} {\bibfield  {journal} {\bibinfo
  {journal} {Phys. Rev. D}\ }\textbf {\bibinfo {volume} {98}},\ \bibinfo
  {pages} {104064} (\bibinfo {year} {2018})},\ \Eprint
  {https://arxiv.org/abs/1708.06255} {arXiv:1708.06255 [gr-qc]} \BibitemShut
  {NoStop}%
\bibitem [{\citenamefont {Bemfica}\ \emph {et~al.}(2019)\citenamefont
  {Bemfica}, \citenamefont {Disconzi},\ and\ \citenamefont
  {Noronha}}]{Bemfica:2019knx}%
  \BibitemOpen
  \bibfield  {author} {\bibinfo {author} {\bibfnamefont {F.~S.}\ \bibnamefont
  {Bemfica}}, \bibinfo {author} {\bibfnamefont {M.~M.}\ \bibnamefont
  {Disconzi}},\ and\ \bibinfo {author} {\bibfnamefont {J.}~\bibnamefont
  {Noronha}},\ }\bibfield  {title} {\bibinfo {title} {{Nonlinear Causality of
  General First-Order Relativistic Viscous Hydrodynamics}},\ }\href
  {https://doi.org/10.1103/PhysRevD.100.104020} {\bibfield  {journal} {\bibinfo
   {journal} {Phys. Rev. D}\ }\textbf {\bibinfo {volume} {100}},\ \bibinfo
  {pages} {104020} (\bibinfo {year} {2019})},\ \bibinfo {note} {[Erratum:
  Phys.Rev.D 105, 069902 (2022)]},\ \Eprint {https://arxiv.org/abs/1907.12695}
  {arXiv:1907.12695 [gr-qc]} \BibitemShut {NoStop}%
\bibitem [{\citenamefont {Bemfica}\ \emph {et~al.}(2022)\citenamefont
  {Bemfica}, \citenamefont {Disconzi},\ and\ \citenamefont
  {Noronha}}]{Bemfica:2020zjp}%
  \BibitemOpen
  \bibfield  {author} {\bibinfo {author} {\bibfnamefont {F.~S.}\ \bibnamefont
  {Bemfica}}, \bibinfo {author} {\bibfnamefont {M.~M.}\ \bibnamefont
  {Disconzi}},\ and\ \bibinfo {author} {\bibfnamefont {J.}~\bibnamefont
  {Noronha}},\ }\bibfield  {title} {\bibinfo {title} {{First-Order
  General-Relativistic Viscous Fluid Dynamics}},\ }\href
  {https://doi.org/10.1103/PhysRevX.12.021044} {\bibfield  {journal} {\bibinfo
  {journal} {Phys. Rev. X}\ }\textbf {\bibinfo {volume} {12}},\ \bibinfo
  {pages} {021044} (\bibinfo {year} {2022})},\ \Eprint
  {https://arxiv.org/abs/2009.11388} {arXiv:2009.11388 [gr-qc]} \BibitemShut
  {NoStop}%
\bibitem [{\citenamefont {Gavassino}(2022)}]{Gavassino:2021owo}%
  \BibitemOpen
  \bibfield  {author} {\bibinfo {author} {\bibfnamefont {L.}~\bibnamefont
  {Gavassino}},\ }\bibfield  {title} {\bibinfo {title} {{Can We Make Sense of
  Dissipation without Causality?}},\ }\href
  {https://doi.org/10.1103/PhysRevX.12.041001} {\bibfield  {journal} {\bibinfo
  {journal} {Phys. Rev. X}\ }\textbf {\bibinfo {volume} {12}},\ \bibinfo
  {pages} {041001} (\bibinfo {year} {2022})},\ \Eprint
  {https://arxiv.org/abs/2111.05254} {arXiv:2111.05254 [gr-qc]} \BibitemShut
  {NoStop}%
\bibitem [{\citenamefont {Heller}\ \emph {et~al.}(2023)\citenamefont {Heller},
  \citenamefont {Serantes}, \citenamefont {Spali\'nski},\ and\ \citenamefont
  {Withers}}]{Heller:2022ejw}%
  \BibitemOpen
  \bibfield  {author} {\bibinfo {author} {\bibfnamefont {M.~P.}\ \bibnamefont
  {Heller}}, \bibinfo {author} {\bibfnamefont {A.}~\bibnamefont {Serantes}},
  \bibinfo {author} {\bibfnamefont {M.}~\bibnamefont {Spali\'nski}},\ and\
  \bibinfo {author} {\bibfnamefont {B.}~\bibnamefont {Withers}},\ }\bibfield
  {title} {\bibinfo {title} {{Rigorous Bounds on Transport from Causality}},\
  }\href {https://doi.org/10.1103/PhysRevLett.130.261601} {\bibfield  {journal}
  {\bibinfo  {journal} {Phys. Rev. Lett.}\ }\textbf {\bibinfo {volume} {130}},\
  \bibinfo {pages} {261601} (\bibinfo {year} {2023})},\ \Eprint
  {https://arxiv.org/abs/2212.07434} {arXiv:2212.07434 [hep-th]} \BibitemShut
  {NoStop}%
\bibitem [{\citenamefont {Gavassino}(2023)}]{Gavassino:2023myj}%
  \BibitemOpen
  \bibfield  {author} {\bibinfo {author} {\bibfnamefont {L.}~\bibnamefont
  {Gavassino}},\ }\bibfield  {title} {\bibinfo {title} {{Bounds on transport
  from hydrodynamic stability}},\ }\href
  {https://doi.org/10.1016/j.physletb.2023.137854} {\bibfield  {journal}
  {\bibinfo  {journal} {Phys. Lett. B}\ }\textbf {\bibinfo {volume} {840}},\
  \bibinfo {pages} {137854} (\bibinfo {year} {2023})},\ \Eprint
  {https://arxiv.org/abs/2301.06651} {arXiv:2301.06651 [hep-th]} \BibitemShut
  {NoStop}%
\bibitem [{\citenamefont {Heller}\ \emph {et~al.}(2024)\citenamefont {Heller},
  \citenamefont {Serantes}, \citenamefont {Spali{\'n}ski},\ and\ \citenamefont
  {Withers}}]{Heller:2023jtd}%
  \BibitemOpen
  \bibfield  {author} {\bibinfo {author} {\bibfnamefont {M.~P.}\ \bibnamefont
  {Heller}}, \bibinfo {author} {\bibfnamefont {A.}~\bibnamefont {Serantes}},
  \bibinfo {author} {\bibfnamefont {M.}~\bibnamefont {Spali{\'n}ski}},\ and\
  \bibinfo {author} {\bibfnamefont {B.}~\bibnamefont {Withers}},\ }\bibfield
  {title} {\bibinfo {title} {{The space of transport coefficients allowed by
  causality}},\ }\href {https://doi.org/10.1038/s41567-024-02635-5} {\bibfield
  {journal} {\bibinfo  {journal} {Nature Phys.}\ }\textbf {\bibinfo {volume}
  {20}},\ \bibinfo {pages} {1948} (\bibinfo {year} {2024})},\ \Eprint
  {https://arxiv.org/abs/2305.07703} {arXiv:2305.07703 [hep-th]} \BibitemShut
  {NoStop}%
\bibitem [{\citenamefont {Denicol}\ \emph {et~al.}(2008)\citenamefont
  {Denicol}, \citenamefont {Kodama}, \citenamefont {Koide},\ and\ \citenamefont
  {Mota}}]{Denicol:2008ha}%
  \BibitemOpen
  \bibfield  {author} {\bibinfo {author} {\bibfnamefont {G.~S.}\ \bibnamefont
  {Denicol}}, \bibinfo {author} {\bibfnamefont {T.}~\bibnamefont {Kodama}},
  \bibinfo {author} {\bibfnamefont {T.}~\bibnamefont {Koide}},\ and\ \bibinfo
  {author} {\bibfnamefont {P.}~\bibnamefont {Mota}},\ }\bibfield  {title}
  {\bibinfo {title} {{Stability and Causality in relativistic dissipative
  hydrodynamics}},\ }\href {https://doi.org/10.1088/0954-3899/35/11/115102}
  {\bibfield  {journal} {\bibinfo  {journal} {J. Phys. G}\ }\textbf {\bibinfo
  {volume} {35}},\ \bibinfo {pages} {115102} (\bibinfo {year} {2008})},\
  \Eprint {https://arxiv.org/abs/0807.3120} {arXiv:0807.3120 [hep-ph]}
  \BibitemShut {NoStop}%
\bibitem [{\citenamefont {Biskamp}(1997)}]{biskamp1997nonlinear}%
  \BibitemOpen
  \bibfield  {author} {\bibinfo {author} {\bibfnamefont {D.}~\bibnamefont
  {Biskamp}},\ }\href@noop {} {\emph {\bibinfo {title} {Nonlinear
  magnetohydrodynamics}}},\ Vol.~\bibinfo {volume} {1}\ (\bibinfo  {publisher}
  {Cambridge University Press},\ \bibinfo {year} {1997})\BibitemShut {NoStop}%
\end{thebibliography}%

\end{document}